\documentclass[twocolumn]{aastex63}
\usepackage{amsmath} 
\usepackage[english]{babel}

\received{August 27, 2020}
\revised{November 17, 2020}
\accepted{November 24, 2020}
\shorttitle{Intracluster Light and Host Clusters}
\shortauthors{Kluge et al.}
\submitjournal{ApJS}

\begin{document}

\title{Photometric dissection of Intracluster Light and its correlations with host cluster properties}

\author[0000-0002-9618-2552]{M. Kluge}
\affil{Max Planck Institute for Extraterrestrial Physics, Giessenbachstrasse, D-85748 Garching, Germany}
\affil{University-Observatory, Ludwig-Maximilians-University, Scheinerstrasse 1, D-81679 Munich, Germany}

\author[0000-0001-7179-0626]{R. Bender}
\affil{University-Observatory, Ludwig-Maximilians-University, Scheinerstrasse 1, D-81679 Munich, Germany}
\affil{Max Planck Institute for Extraterrestrial Physics, Giessenbachstrasse, D-85748 Garching, Germany}

\author[0000-0002-5466-3892]{A. Riffeser}
\affil{University-Observatory, Ludwig-Maximilians-University, Scheinerstrasse 1, D-81679 Munich, Germany}

\author[0000-0002-2152-6277]{C. Goessl}
\affil{University-Observatory, Ludwig-Maximilians-University, Scheinerstrasse 1, D-81679 Munich, Germany}

\author[0000-0003-1008-225X]{U. Hopp}
\affil{University-Observatory, Ludwig-Maximilians-University, Scheinerstrasse 1, D-81679 Munich, Germany}
\affil{Max Planck Institute for Extraterrestrial Physics, Giessenbachstrasse, D-85748 Garching, Germany}

\author{M. Schmidt}
\affil{University-Observatory, Ludwig-Maximilians-University, Scheinerstrasse 1, D-81679 Munich, Germany}

\author{C. Ries}
\affil{University-Observatory, Ludwig-Maximilians-University, Scheinerstrasse 1, D-81679 Munich, Germany}

\begin{abstract}

We explore several ways to dissect Brightest Cluster Galaxies (BCGs) and their surrounding Intracluster Light (ICL) using a surface brightness cut, a luminosity cut, excess light above a de Vaucouleurs profile, or a double S\'ersic decomposition. Assuming that all light above $M<-21.85~g'~\rm{mag}$ is attributable to the ICL, we find an average ICL fraction of $f^{\rm MT}_{\rm ICL}=71\pm22\%$ of all diffuse light centered on the BCG to belong to the ICL. Likewise, if we assume all light fainter than $\rm{SB}>27$ $g'$ mag arcsec$^{-2}$ to belong to the ICL, the average ICL fraction is $f^{\rm SB27}_{\rm ICL}=34\pm19\%$. After fitting a de Vaucouleurs profile to the inner parts of the SB profile, we detect excess light at large radii, corresponding to an average ICL fraction of $f^{\rm DV}_{\rm ICL}=48\pm20\%$. Finally, by decomposing the SB profile into two S\'ersic functions, we find an average ICL fraction of $f^{\rm S\times}_{\rm ICL}=52\pm21\%$ associated with the outer S\'ersic component. Our measured ICL and BCG+ICL luminosities agree well with predictions from high-resolution simulations where the outer S\'ersic component traces the unrelaxed, accreted stellar material.

BCG and ICL properties defined in this way are correlated with cluster parameters to study the co-evolution of BCGs, ICL, and their host clusters. We find positive correlations between BCG+ICL brightness and cluster mass, cluster velocity dispersion, cluster radius, and integrated satellite brightness, confirming that BCG/ICL growth is indeed coupled with cluster growth.

On average, the ICL is better aligned than the BCG with the host cluster in terms of position angle, ellipticity, and centering. That makes it a potential Dark Matter tracer.

\end{abstract}


\section{Introduction}

Near the center of a galaxy cluster often resides an exceptionally extended and luminous elliptical galaxy, a so called Brightest Cluster Galaxy (BCG). In these galaxies, uncommon for ellipticals, the stellar velocity dispersion often rises outwards (\citealt{Dressler1979,Carter1981,Ventimiglia2010,Toledo2011,Arnaboldi2012,Melnick2012,Murphy2014,Bender2015,Barbosa2018,Loubser2018,Spiniello2018,Gu2020}), in some cases reaching the cluster velocity dispersion. These observations show that BCGs are generally surrounded by a dynamically hot, host-cluster-bound stellar component, the so-called Intracluster Light (ICL) (for reviews, see e.g., \citealt{Mihos2016,Montes2019}). 

Moreover, the broken slope in the scaling relations for non-BCG Ellipticals toward BCGs \citep{Kluge2020} shows that BCGs have excess light at large radii compared to simply up-scaled versions of non-BCG ellipticals. This raises the question how much ICL exists in galaxy clusters and how it is spatially distributed. Unfortunately, more often than not, the BCG's stellar light merges smoothly into the ICL, meaning the transition leaves no trace in the smooth surface brightness (SB) profiles (e.g., \citealt{Bender2015,Kluge2020}). 

Even in the cases that show a distinct upward change in slope, it is not clear whether the excess light traces the ICL. Changes in position angle, ellipticity and color occur in NGC 1399, the BCG in the Fornax cluster, around the transition radius between the two S\'ersic profile fits at $\sim$13\,kpc semimajor-axis radius (\citealt{Iodice2016,Spavone2017}). This could possibly be the signature of the ICL associated with the outer S\'ersic component. However, there is no systematic study so far, neither observationally nor numerically, that verifies the association of a photometrically distinct component with the kinematically confirmed ICL.

The determination of the total amount of ICL per cluster depends on how it is separated from the BCG in the inner regions and how the SB profile is extrapolated beyond the outermost measurable radius. The uncertainty of the extrapolation was examined in \cite{Kluge2020}. In this paper, we focus on four methods to attempt a photometric separation between BCG and ICL. They encompass SB cuts (\citealt{Feldmeier2004,Rudick2011,Burke2012,Cui2014,Cooper2015}) or the fitting of double \cite{DeVaucouleurs1948}, double \cite{Sersic1968} or similar functions to the SB profiles (\citealt{Gonzalez2005a,Seigar2007,Puchwein2010,Donzelli2011,Cooper2015,Spavone2020}) or fitting only the inner SB profile and defining the excess luminosity in the outskirts as ICL (\citealt{Schombert1986,Zibetti2005}).

A different approach is to consider stellar velocities. It is motivated by the radially rising velocity dispersion profiles that approach the cluster velocity dispersion, that is, the relative velocities of the cluster galaxies. The ICL is hereby the dynamically hot component, which is kinematically controlled by the gravitational potential of the whole cluster, that is, unbound from the BCG. \cite{Bender2015} have applied a simplified approach to observational data of NGC 6166 by assuming constant velocity dispersions for both components.

In a more complex form, the kinematic approach is often applied in numerical simulations where full phase-space information of the particles is accessible. A BCG+ICL system is decomposed by fitting a double Maxwell distribution to the particle velocities. The component with the higher characteristic velocity is called the diffuse stellar component (DSC, e.g., \citealt{Dolag2010}). Contrary to expectation, the "photometrically" determined ICL does not necessarily resemble the DSC (\citealt{Dolag2010,Puchwein2010,Rudick2011,Cui2014,Remus2017}). A different set of components alternative to the bound / unbound criterion are in-situ formed / accreted stars \citep{Cooper2013,Cooper2015}. The in-situ stars were formed directly from the cluster cooling flow whereas accreted stars have been stripped from satellite galaxies. \cite{Cooper2015} showed that in their used N-body simulations, 80--95\% of stellar mass found below ${\rm SB}\gtrsim 26.5~V$ mag arcsec$^{-2}$ is associated with accreted stars. The question whether the outer photometric component traces the DSC and/or the accreted stellar mass or none of them is a matter of on-going research and will be addressed in this paper.

The second aim of this paper is to find correlations between BCG/ICL- and host cluster parameters. Since the ICL is dynamically bound to the host cluster and therefore co-evolves with it, we expect to find correlations between their structural and kinematic parameters. An argument for the co-evolution is that more massive clusters host more luminous BCGs \citep{Lin2004,Yang2005,Zheng2007,Popesso2007,Brough2008,Hansen2009,SampaioSantos2020}. Furthermore, BCGs align their position angle (PA) with that of their host clusters \citep{Sastry1968,Dressler1978,Binggeli1982,Struble1990,Kim2002,Yang2006,Niederste-Ostholt2010,Huang2016,West2017,Okabe2020}. Such an alignment is found in numerical simulations, too \citep{Faltenbacher2008,Okabe2020}. It persists out to at least 6 virial radii in the Horizon-AGN simulations \citep{Faltenbacher2008}. At such large clustrocentric distances, the filamentary cosmic structure becomes apparent. The BCG/ICL's alignment with its host cluster is therefore likely an imprint of the infalling direction of matter along these filaments during cluster formation (e.g., \citealt{Dubinski1998}).

Since the ICL's dynamics are regulated by the overall cluster gravitational potential, we examine in this paper whether the ICL aligns its PA and isophotal centers even better than the BCG with its host cluster. Both effects are expected to be small (\citealt{Gonzalez2005a,Kluge2020}) but possibly detectable.

This paper is organized as follows. In Section \ref{sec:introiclfrac}, we mention our methods to calculate fiducial ICL luminosities and in Section \ref{sec:clusterprops} we define the host cluster parameters. The details of the ICL decomposition methods and the results of the fiducial ICL luminosities are presented in Section \ref{sec:iclfrac} and the results regarding the correlations with host cluster parameters are presented in Section \ref{sec:corrhosttext}. A discussion about the reliability of the photometric ICL decomposition methods and the potential to consider the ICL as a dark matter tracer can be found in Section \ref{sec:discussion}. All results are summarized in Section \ref{sec:summary}.

Throughout the paper, we assume a flat cosmology with $H_0=69.6$ km s$^{-1}$ Mpc$^{-1}$ and $\Omega_{\rm m}=0.286$ \citep{Bennett2014}. Distances and angular scales were calculated using the web tool from \cite{Wright2006}. Virgo infall is not considered. Three types of flux corrections were applied: (1) dust extinction using the maps from \cite{Schlafly2011}, (2) K corrections following \cite{Chilingarian2010} and \cite{Chilingarian2012}, and (3) cosmic $(1+z)^4$ dimming. Magnitudes are always given in the AB system.

\begin{figure*}
	\includegraphics[width=\linewidth]{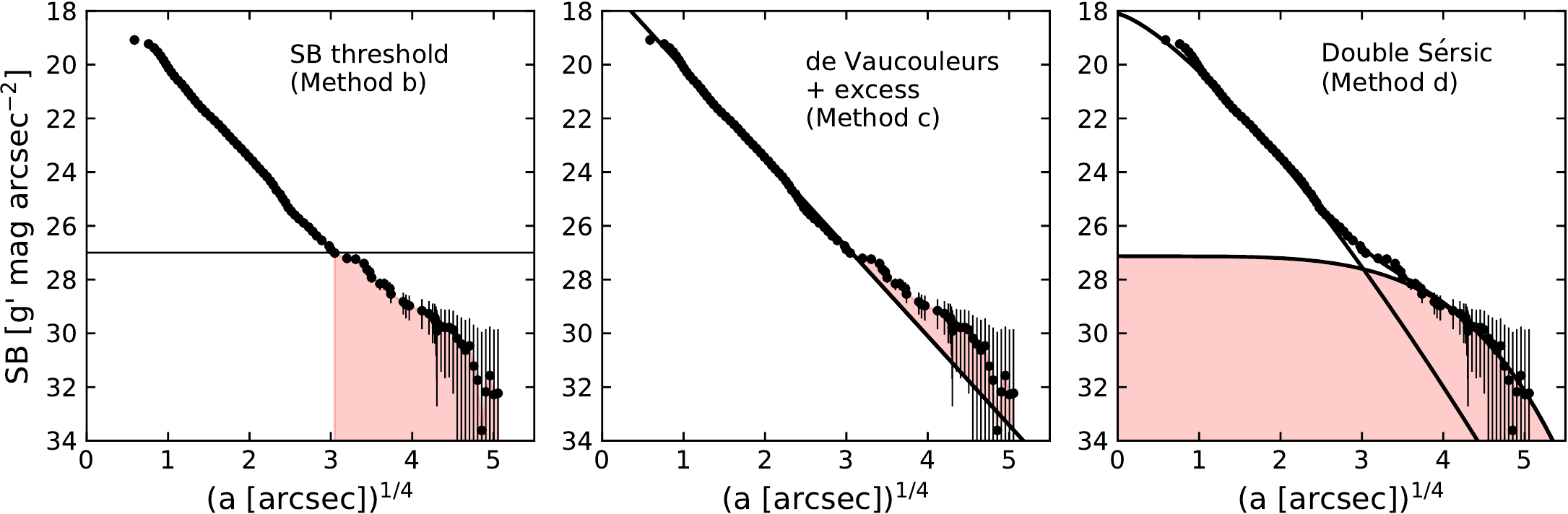}
	\caption{ICL decomposition methods, which are based on SB profile decompositions are illustrated on the SB profile of the A1668 BCG (black dots). The red shaded region is assigned to the ICL. Method b (\textit{left panel}) is the SB threshold, in this case at 27 $g'$ mag arcsec$^{-2}$ (horizontal line). All light at fainter SBs is defined as ICL. Method c (\textit{middle panel}) is the de Vaucouleurs plus excess light decomposition. A de Vaucouleurs profile (black line) is fitted to the inner region (${\rm SB}<23~g'$ mag arcsec$^{-2}$) and all light above that profile is counted as ICL. Method d (\textit{right panel}) is the double S\'ersic decomposition (two black lines), where all light in the outer S\'ersic component (black line above the shaded region) is defined as ICL. \label{fig:icldecomposition}}
\end{figure*}

\section{Sample}

Our full observational dataset and data reduction pipeline are detailed in \cite{Kluge2020}. Here, we briefly summarize the sample selection and most important data reduction steps.
	
We have observed 170 low-redshift ($z\lesssim0.08$) galaxy clusters in the northern hemisphere ($Decl.\gtrsim5\degr$) with the 2\,m Fraunhofer telescope at the Wendelstein Observatory, Germany. The sample is mostly selected from the ACO catalog \citep{Abell1989} with 13 additional clusters selected from the \cite{Linden2007}, \cite{AWM1977}, and \cite{MKW1975} catalogs. Clusters in the projected vicinity of bright foreground stars were rejected. The completeness is about $\sim80\%$. For a subsample of 50 clusters, satellite galaxy catalogs of sufficient quality are available to calculate gravitational masses $\log(M_{\rm g}[\rm{M}_{\odot}])=14.75\pm0.25$.

The observations were taken in the $g'$ band with the Wendelstein Wide Field Imager. Its large field of view with $27.6\arcmin\times28.9\arcmin$ in combination with a large dither pattern provides sufficient sky coverage to model and subtract the sky background using night-sky flats. Ghosts and the extended point-spread function (PSF) wings are subtracted from bright foreground stars. Large-scale PSF broadening of the BCG+ICL by the extended PSF wings is corrected by subtracting the scattered light due to the BCG's nucleus from the images. The central resolution is increased with archival \textit{Hubble Space Telescope} imaging data or Richardson--Lucy-deconvolved WWFI data.

Semimajor axis SB profiles of all BCG+ICLs are measured by fitting ellipses to the isophotes with the code {\tt ellfitn} \citep{Bender1987}. All isophotal parameters besides the semimajor-axis radius are kept fixed beyond the largest plausible radius. Single or double S\'ersic functions are fitted to the SB profiles based on a nonlinear least squares method using the Levenberg--Marquardt algorithm. SB uncertainties are set to 0.18 $g'$ mag arcsec$^{-2}$, which is on the order of typical intrinsic deviations from best-fit S\'ersic profiles. Additionally, we add a linear background uncertainty of $\Delta {\rm BG}=\pm1$ count arcsec$^{-2}$ for a photometric zero point of ${\rm ZP}=30~g'$ mag. Upper error bars are mirrored downwards because of their asymmetry in logarithmic SB units. Light below our limiting SB of ${\rm SB}_{\rm lim}=30~g'$ mag arcsec$^{-2}$ has a considerable impact on the BCG+ICL total brightnesses. We estimate this by extrapolating the best-fit single or double S\'ersic profiles out to infinite radius. We finally take the average value of the total brightnesses, determined by integrating the SB profiles once to ${\rm SB}_{\rm lim}=30~g'$ mag arcsec$^{-2}$ and once to infinite radius. The uncertainty of the total brightnesses is determined by using these different integration limits is added quadratically to the propagated uncertainty of the best-fit parameters to estimate the total uncertainty of the BCG+ICL brightnesses and of the ICL fractions.

\section{ICL fractions by photometric decomposition}\label{sec:introiclfrac}

For all galaxy clusters that were observed by \cite{Kluge2020}, we calculate a fiducial ICL fraction
\begin{equation}
	f_{\rm ICL} = L_{\rm ICL}/(L_{\rm BCG}+L_{\rm ICL}) \label{eq:ficl}
\end{equation}
as the luminosity of the photometric component that we define as ICL, relative to the total luminosity of the combined BCG+ICL system. We stress that this probably includes at least part of the stellar halos of the BCGs.

The ICL brightness is then calculated as
\begin{equation}
	M_{\rm ICL} = M_{\rm BCG+ICL} - 2.5 \log(f_{\rm ICL}).
\end{equation}

To separate the ICL from the BCG, we apply a simple integrated brightness cut (a) and three methods that are commonly used in the literature (b), (c) and (d). The ICL is either defined as

\begin{enumerate}
	\item[(a)] all stellar light above a given integrated brightness,
	\item[(b)] all stellar light below a given SB threshold,
	\item[(c)] the excess light above a de Vaucouleurs profile or as
	\item[(d)] the outer component determined by an SB profile decomposition into two S\'ersic functions.
\end{enumerate}

The last method is only applied for double S\'ersic (DS) BCGs (see Section 4.4 in \citealt{Kluge2020}). Methods (b), (c), and (d) are based on SB profile decompositions. They are visualized in Figure \ref{fig:icldecomposition} for the SB profile of the A1668 BCG. The red area corresponds to the fiducial ICL as defined using each method.

Whether or not these methods actually dissect the real, dynamically hot ICL from the BCG is an ongoing debate. We join that discussion in Section \ref{sec:discussionicl}.

\section{Host cluster properties}\label{sec:clusterprops}

\subsection{Physical parameters} \label{sec:clusterparams}

In order to understand the connection between BCG/ICL and cluster formation and evolution, we calculate and measure various parameters that describe the current evolutionary state of the host clusters of our BCG sample.

The cluster velocity dispersions $\sigma_{\rm C}$ is the dispersion of line-of-sight velocities of the cluster satellite galaxies. They are taken from \cite{Lauer2014}. The satellite galaxies' positions were retrieved from the {\tt SIMBAD} database. As a search radius, we chose $r_{\rm max}=2\,$Mpc in transversal and $v_{\rm BCG}\pm 3000\,$km\,s$^{-1}$ (using spectroscopic or photometric redshifts) along the line of sight direction around the BCG. The satellite galaxy samples are inhomogeneous with respect to the detection thresholds. However, we assume that this adds only a statistical error to our inferred correlations.

Following \cite{Tully2015}, we define the projected gravitational radius $r_{\rm g}$ as

\begin{equation}
r_{\rm g}(r_{\rm max} = 2 \rm\,Mpc) = \frac{S^2}{\sum_{i<j} 1 / r_{ij}} \label{eq:rg}
\end{equation}

where $S$ is the total number of {\tt SIMBAD} satellite galaxies inside $r_{\rm max} = 2\,$Mpc, that is, the cluster richness, and $r_{ij}$ is the projected distance between galaxy pairs. The gravitational radius resembles the characteristic separation between two satellite galaxies. We estimate the error of $r_{\rm g}$ by using two different radial boundaries on the satellite galaxy samples:

\begin{equation}
\delta r_{\rm g} = |r_{\rm g}(r_{\rm max} = 2.5 \rm\,Mpc) - r_{\rm g}(r_{\rm max} = 1.5\,\rm Mpc)|
\end{equation}

\begin{figure*}
	\includegraphics[width=\linewidth]{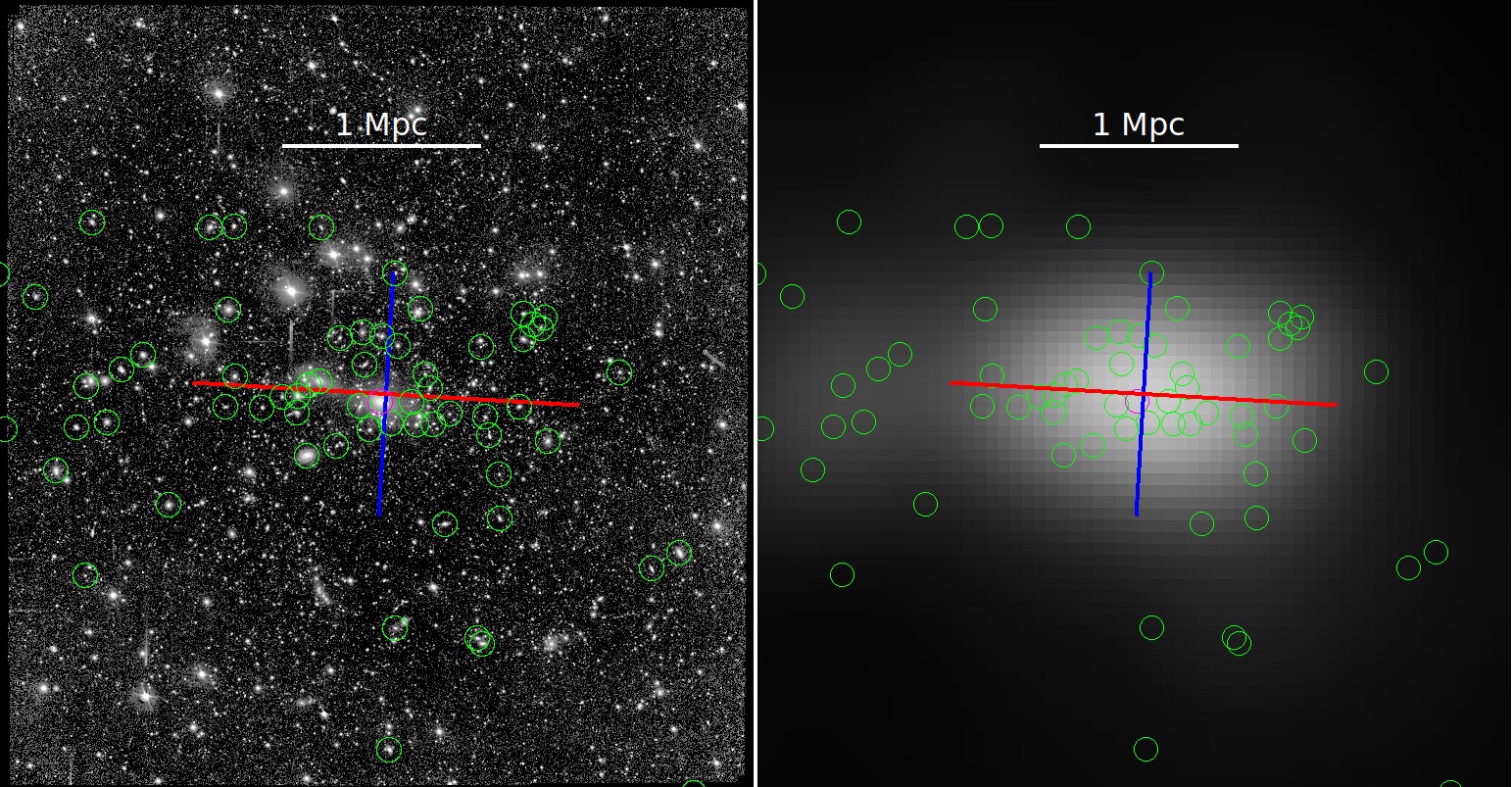}
	\caption{{\it Left panel:} stacked WWFI image of A1668. {\it Right panel:} smoothed galaxy density distribution obtained by Voronoi binning the cluster galaxies (green circles). The magenta circle marks the BCG. The red (blue) line indicates the major (minor) axis. They cross at the cluster center. \label{fig:alignmentA1668}}
\end{figure*}

\begin{figure*}
	\centering
	\includegraphics[width=\linewidth]{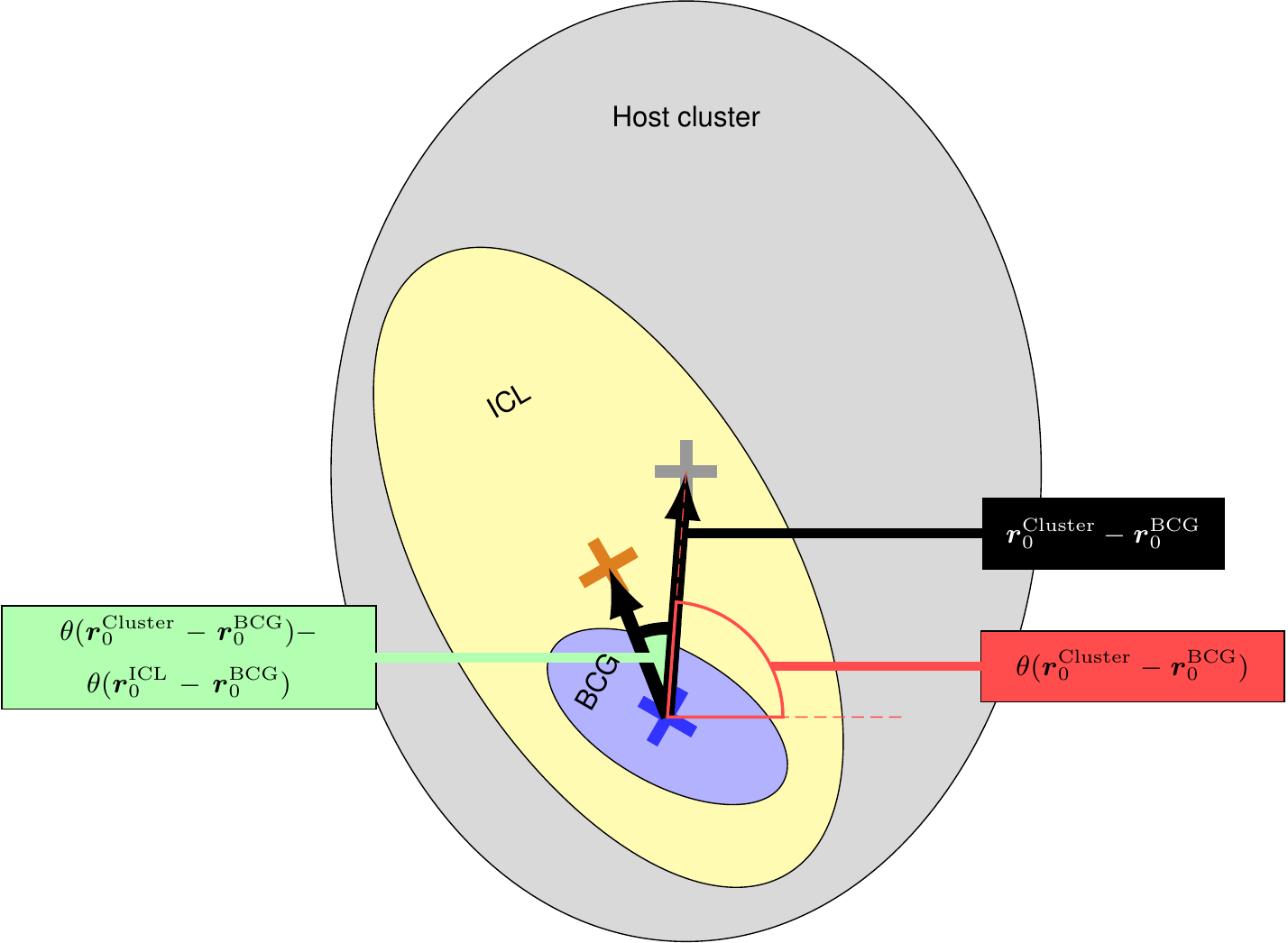}
	\caption{Schematic alignment between BCG (blue), ICL (yellow) and host cluster (gray). All three components have different position angles $PA$, which are measured independently from the offset direction. The two black arrows point from the BCG center toward the ICL center or host-cluster center, each marked with a cross. The angle $\theta$ of these vectors is counted anti-clockwise from the horizontal, red dashed line. However, only the absolute difference (green) between the two angles is of interest because it is the offset direction of the ICL with respect to the BCG. If the ICL is offset toward the host cluster center, then $|\theta(\boldsymbol{r}_0^{\rm Cluster} - \boldsymbol{r}_0^{\rm BCG}) - \theta(\boldsymbol{r}_0^{\rm ICL} - \boldsymbol{r}_0^{\rm BCG})|=0\degr$. \label{fig:alignment}}
\end{figure*}

The gravitational mass $\mathcal{M_{\rm g}}$ is given by

\begin{equation}
\mathcal{M_{\rm g}} = \alpha \sigma^2_{\rm C} \times (\pi/2) r_{\rm g} / G
\end{equation}	

where $G$ is the gravitational constant. The factors $(\pi/2)$ and $\alpha$ arise from the deprojection of the galaxy positions and velocity dispersion, respectively. The value of $\alpha=2.4$ is adopted from \cite{Mamon2010} for an anisotropy model \citep{Mamon2005} that is a good fit to $\Lambda$CDM halos.

We further use the gravitational cluster volume \mbox{$V_{\rm g}=(4/3 \pi r_{\rm g}^3)$} to calculate the cluster mass density

\begin{equation}
\rho = \mathcal{M_{\rm g}} / V_{\rm g},
\end{equation}

the satellite galaxy number density

\begin{equation}
s = S / V_{\rm g},
\end{equation}

the mass phase space density

\begin{equation}
f_{\mathcal{M_{\rm g}}} = \rho / \sigma_{\rm C}^3,
\end{equation}

and the galaxy number phase space density

\begin{equation}
f_{\rm s} = S / (V_{\rm g} \cdot \sigma_{\rm C}^3).
\end{equation}

The integrated brightness of all satellite galaxies $M_{\rm sat}$ is measured directly from the wide-field CCD images (WWFI imager) as presented by \cite{Kluge2020} and is independent of the {\tt SIMBAD} sample. The average field of view spans 1.64$\pm$0.49\,Mpc in radius centered on the BCG. The procedure is the following: we mask all foreground stars that are listed in the Tycho-2 \citep{Tycho2000} and Pan-STARRS PV3 \citep{Flewelling2020} catalogs, as well as in the source catalogs created with {\tt SExtractor} from the WWFI images. We then multiply the masks which were created to measure the isophotal flux of the BCG (see Section 3.2.2. in \citealt{Kluge2020}) onto the WWFI stacks. That product image is then subtracted from the WWFI stacks so that only galaxies (excluding the BCG+ICL) remain in the difference image. The remaining flux is then measured in circular apertures centered around the BCG. The background is subtracted from the light profiles analogous to the procedure described in Section 4.2 in \cite{Kluge2020}. These light profiles are then integrated and the uncertainties in the flux are estimated from the uncertainty of the background level choice.

All derived cluster properties are listed in Appendix \ref{sec:hostclusterparams}.

\subsection{Alignment}

Since the BCG, the ICL and the cluster galaxy density distribution are all to first order elliptical (or triaxial in 3D), we can measure the centers $\boldsymbol{r}_0$ and position angles $PA$ for all three components by fitting ellipses to the isophotes or isodensity contours.

For the clusters, we use the {\tt SIMBAD} satellite galaxy catalogs that are described in Section \ref{sec:clusterparams} to measure the galaxy density distribution. An example is shown in Figure \ref{fig:alignmentA1668}. The green circles mark the cluster galaxies in A1668. Firstly, we Voronoi bin the projected galaxy positions. Each voronoi cell is then divided by its surface area to obtain a galaxy density map. These maps are then smoothed using a Gaussian kernel with a standard deviation of 4\,kpc $ < \sigma_{\rm K} < $ 80\,kpc which is optimized for each cluster by hand. The result for A1668 is shown in Figure \ref{fig:alignmentA1668}, right panel. The isodensity contours of the smoothed galaxy density maps are then fitted with ellipses using {\tt ellfitn} (see Section 4.1 in \citealt{Kluge2020}). Since substructures deform the isodensity contours, we manually select the fitted contour which resembles best the overall cluster center and PA. The red (blue) line in Figure \ref{fig:alignmentA1668} indicates the major (minor) axis of that chosen ellipse and the cluster center is located at the crossing of these two lines. No further analysis of the ellipticity is performed because of its strong dependency on the smoothing scale.

The projected offset between the center of one component $i$ with respect to a second component $j$ is given as

\begin{equation}
\boldsymbol{r}_0^{i} - \boldsymbol{r}_0^{j} = \binom{\Delta R.A. \times \cos(Dec)}{\Delta Dec},
\end{equation}

where $i$ and $j$ are either the BCG, ICL or the cluster galaxy density distribution. The angle of that resulting vector is

\begin{equation}
\theta(\boldsymbol{r}_0^{i} - \boldsymbol{r}_0^{j}) = \arctan \left(\frac{\Delta Dec}{\Delta R.A. \times \cos(Dec)}\right).
\end{equation}

A schematic illustration for these quantities is shown in Figure \ref{fig:alignment}. Both, $\boldsymbol{r}_0$ and $PA$ have measurement uncertainties which are especially large at large radii. In order to obtain a high S/N measurement for the BCG and ICL, we average $PA$ and $\boldsymbol{r}_0$ below a major-axis radius $a < 30$\,kpc for the BCG and beyond $a > 30$\,kpc for the ICL. That radius is technically motivated so that roughly the same amount of data points are averaged in each interval. No averaging was done for the galaxy density distribution; the isophote which resembles the cluster $PA$ best is chosen by hand.

\section{Results: BCG/ICL decomposition} \label{sec:iclfrac}

\subsection{Integrated brightness threshold}

The brightness where the slope of the size--luminosity relation breaks (see Figure 16 in \citealt{Kluge2020}) separates regular ellipticals from BCGs quite well. We therefore assume that there is a maximum brightness for regular ellipticals at the knee $M_{\rm max} = -21.85~g'$ mag. All light above that brightness is defined as ICL:

\begin{equation}
	M_{\rm ICL}^{\rm MT} = -2.5 \log( 10^{-0.4 \times M_{\rm BCG+ICL}} - 10^{-0.4 \times M_{\rm max}} )
\end{equation}

\begin{figure}
	\includegraphics[width=\linewidth]{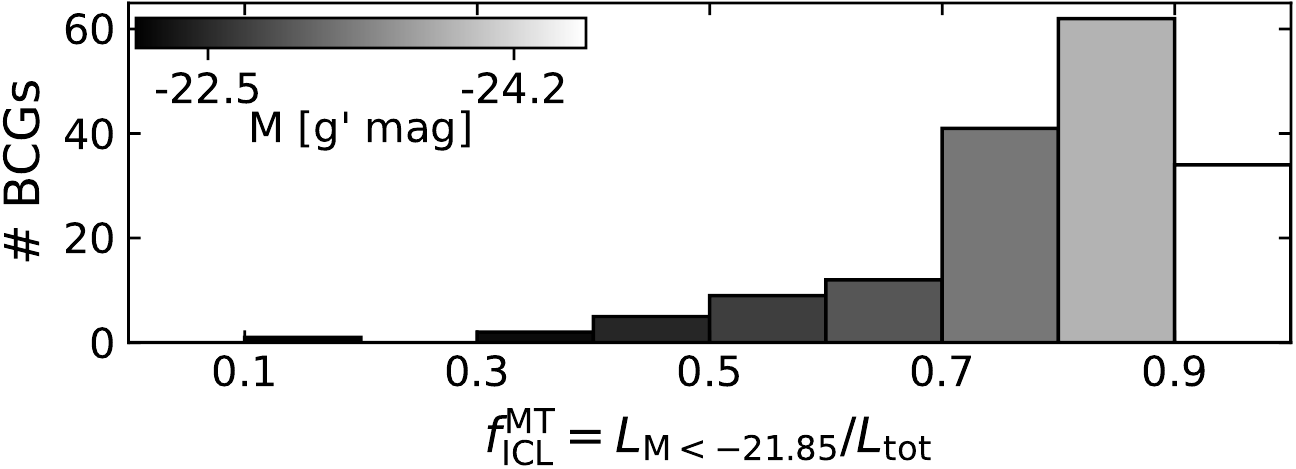}
	\caption{Method a: Luminosity fraction of the light brighter than $M_{\rm BCG+ICL}<-21.85~g'$ in relation to the total BCG+ICL luminosity $L_{\rm tot}$. The average absolute BCG+ICL brightness in the bins is coded in the gray-shading of each bar.\label{fig:fmtotfrac}}
\end{figure}

We calculate an average ICL fraction of $f_{\rm ICL}^{\rm MT} = 71\pm22\%$. The distribution is shown in Figure \ref{fig:fmtotfrac}. By definition, the ICL fraction rises monotonically with increasing BCG+ICL brightness.

\subsection{Surface brightness threshold}\label{sec:iclmethod_sbthresh}

We apply a surface brightness threshold on the light profiles. The faint light below this threshold is defined as the ICL in this context. We calculate the ICL fraction by integrating the light profiles numerically while considering the radially varying ellipticity. The results are strongly sensitive toward the choice of the brightness threshold (see Table \ref{tab:iclfrac}). So which threshold separates the ICL from the BCG most accurately? \cite{Cooper2015} have shown that an SB threshold of ${\rm SB_{cut}}=26.5~V$ mag arcsec$^{-2}$ attributes 80 -- 95 \% of the accreted stars to the ICL in their $N$-body simulations. We transform this $V$-band magnitude into a $g'$-band magnitude using a color transformation derived for NGC6166. By matching the $V$-band SB profile measured by \cite{Bender2015} to our $g'$-band SB profile, we get $g' \simeq V + 0.45$ mag.

The threshold in the $g'$-band is therefore set to ${\rm SB_{cut}}=27~g'$ mag arcsec$^{-2}$, for which we calculate an average ICL fraction $f^{\rm SB27}_{\rm ICL} = 34\pm19\%$. That agrees well with the prediction of $\sim30\%$ by \cite{Cooper2015}, but it is slightly higher than the prediction of $19-31\%$ from \cite{Cui2014} for the hydrodynamical simulations that include AGN feedback. The comparison with \cite{Cui2014} has to be taken with caution because the combined BCG+ICL systems are too massive in the simulations (see column 5 in Table \ref{tab:iclfrac} and discussion in Section \ref{sec:comparesim}).

The results of \cite{Feldmeier2004} for five BCGs are between $f^{\rm SB26}_{\rm ICL} = 28\%$ for ${\rm SB_{cut}}=26$ V mag arcsec$^{-2}$ and $f^{\rm SB27.5}_{\rm ICL}=2\%$ for ${\rm SB_{cut}}=27.5$ V mag arcsec$^{-2}$. Both values are lower than ours, likely because Feldmeier et al. studied only non-cD clusters.

\begin{figure}
	\includegraphics[width=\linewidth]{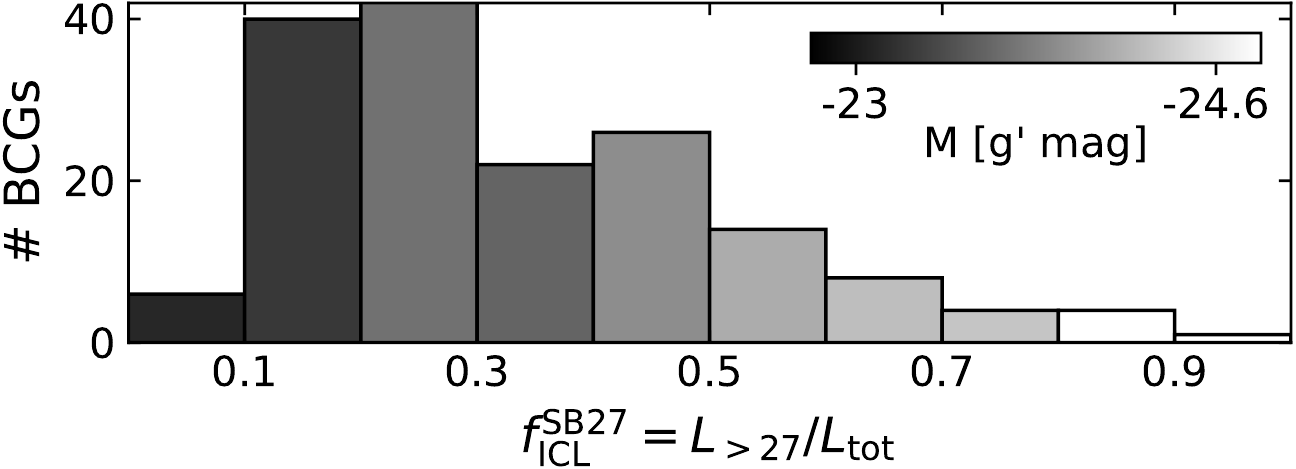}
	\caption{Method b: Luminosity fraction of the light below ${\rm SB}>27~g'$ mag arcsec$^{-2}$ in relation to the total BCG+ICL luminosity $L_{\rm tot}$. The average absolute BCG+ICL brightness in the bins is coded in the gray-shading of each bar.\label{fig:f27frac}}
\end{figure}

We discover a trend that brighter BCGs have larger ICL fractions (see Figure \ref{fig:f27frac}) with a relatively high absolute value of the Pearson coefficient $R=-0.53 \pm 0.07$. The uncertainty is based on 10\,000 bootstraps. That indicates that the recent growth of BCGs happens predominantly by accreting stellar material in their low-SB outskirts.

\subsection{Excess light above the inner de Vaucouleurs profile}

The S\'ersic indices $n$ for BCGs whose SB profiles are well fitted by a single S\'ersic function (SS BCGs) are often (83\%) significantly larger than $n>4$ (see Figure 17 in \citealt{Kluge2020}). That is about the transition value between lower mass \citep{Faber1997,Lauer2007} disky--extralight--rotating ($n<4$) and higher mass boxy--core--nonrotating ($n>4$) ellipticals in the Virgo cluster \citep{Kormendy2009}. It is also known that the velocity dispersion profiles of large ellipticals flatten out toward larger radii whereas they decrease for lower-mass ellipticals \citep{Veale2017}. One could therefore hypothesize that large S\'ersic indices are due to the presence of intragroup- or ICL (e.g., \citealt{Bender2015}).

By assuming that the BCG has a de Vaucouleurs \mbox{($n=4$)} SB profile, we define the ICL in this approach as the excess light above the outwards extrapolation of that profile. The de Vaucouleurs profile is fitted to the inner SB profile of all BCGs below ${\rm SB}<23~g'$ mag arcsec$^{-2}$. A larger fitting range would lead to large errors $\Delta {\rm SB} \gtrsim 0.1~g'$ mag arcsec$^{-2}$ in the inner regions due to the $n>4$ curvature of the SB profiles. As for the S\'ersic fits, we exclude the cores from the fitting.

\begin{figure}
	\includegraphics[width=\linewidth]{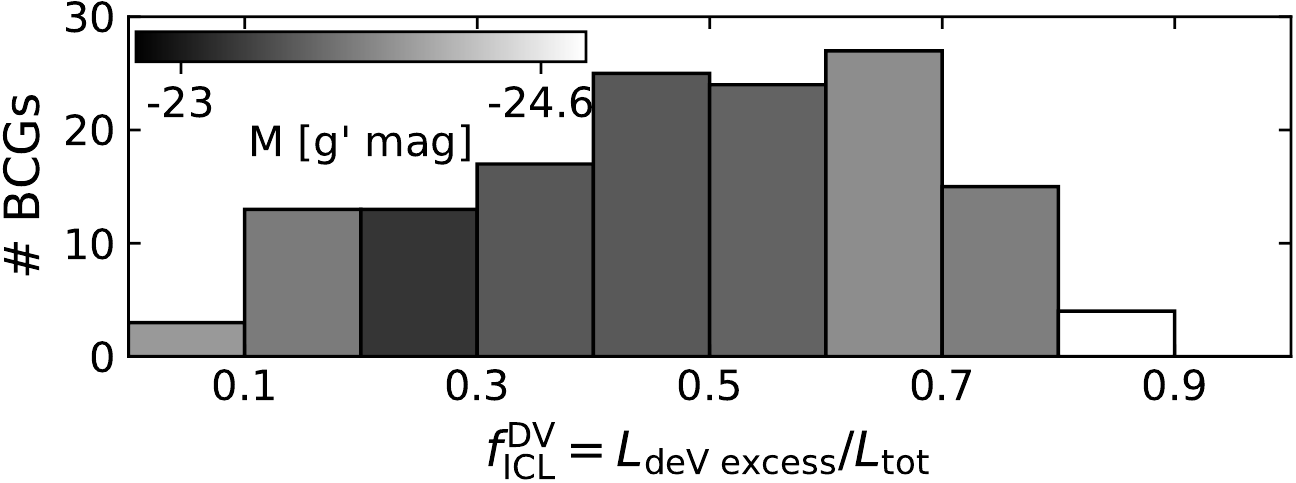}
	\caption{Method c: Luminosity fraction of the light above the de Vaucouleurs profile, which is fitted to the inner SB profile $L_{\rm deV~excess}$, in relation to the total BCG+ICL luminosity $L_{\rm tot}$. The average absolute BCG+ICL brightness in the bins is coded in the gray-shading of each bar.\label{fig:fdevfrac}}
\end{figure}

In Figure \ref{fig:fdevfrac}, we show a histogram of the ICL fractions, measured with the de Vaucouleurs plus excess light method. The average value of $f_{\rm ICL}^{\rm DV} = 48\pm20\%$ is not directly comparable to the result from \cite{Zibetti2005}. The authors measured the SB profile in an averaged SDSS-DR1 image of 683 BCG+ICLs. The de Vaucouleurs fit to the inner $\sim 15 - 80$\,kpc gives $r_{\rm e} = 19.29$\,kpc and ${\rm SB_e}=23.39~g'$ mag arcsec$^{-2}$ (after K-, color- and cosmic dimming correction). They calculate an average ICL fraction of $f^{\rm DV}_{\rm ICL} = 33\pm6\%$. The fit to our (in fixed SB bins) averaged SB profile (see Section 6.2 in \citealt{Kluge2020}) along the effective axis in the same radial interval gives $r_{\rm e} = 35.44\pm0.24$\,kpc and ${\rm SB_e} = 24.61\pm0.01~g'$ mag arcsec$^{-2}$ and the ICL fraction is $f^{\rm DV}_{\rm ICL} = 21\pm12\%$. The uncertainty of $f^{\rm DV}_{\rm ICL}$ is estimated from the SB uncertainty of the averaged SB profile. For that, we assume a lower flux uncertainty than that for individual SB profiles because statistical errors average out. Our choice corresponds to an SB of 31 $g'$ mag arcsec$^{-2}$.

Our calculated average ICL fraction is consistent within the measurement uncertainties with the average ICL fraction for the Zibetti et al. sample. The decomposition is much clearer for the SB profile measured by Zibetti et al. (see Figure 15 in \citealt{Kluge2020}, top left panel) because its shape is much closer to a double de Vaucouleurs profile. Note that the sample of Zibetti et al. is at a higher average redshift: $\bar{z}_{\rm Z}=0.25$ compared to the sample of this work $\bar{z}_{\rm K}=0.06$ and therefore 2.16 Gyrs younger. As discussed in Section 6.2 in \cite{Kluge2020}, the different average shapes might be the result of a time-evolution in which the SB profiles have evolved to become smoother. The increase in effective radius of the inner de Vaucouleurs component by 84\% and the large fraction (71\%) of smooth SS BCGs today are presumably a direct consequence of that.

\subsection{Double S\'ersic decomposition}

\begin{figure}
	\includegraphics[width=\linewidth]{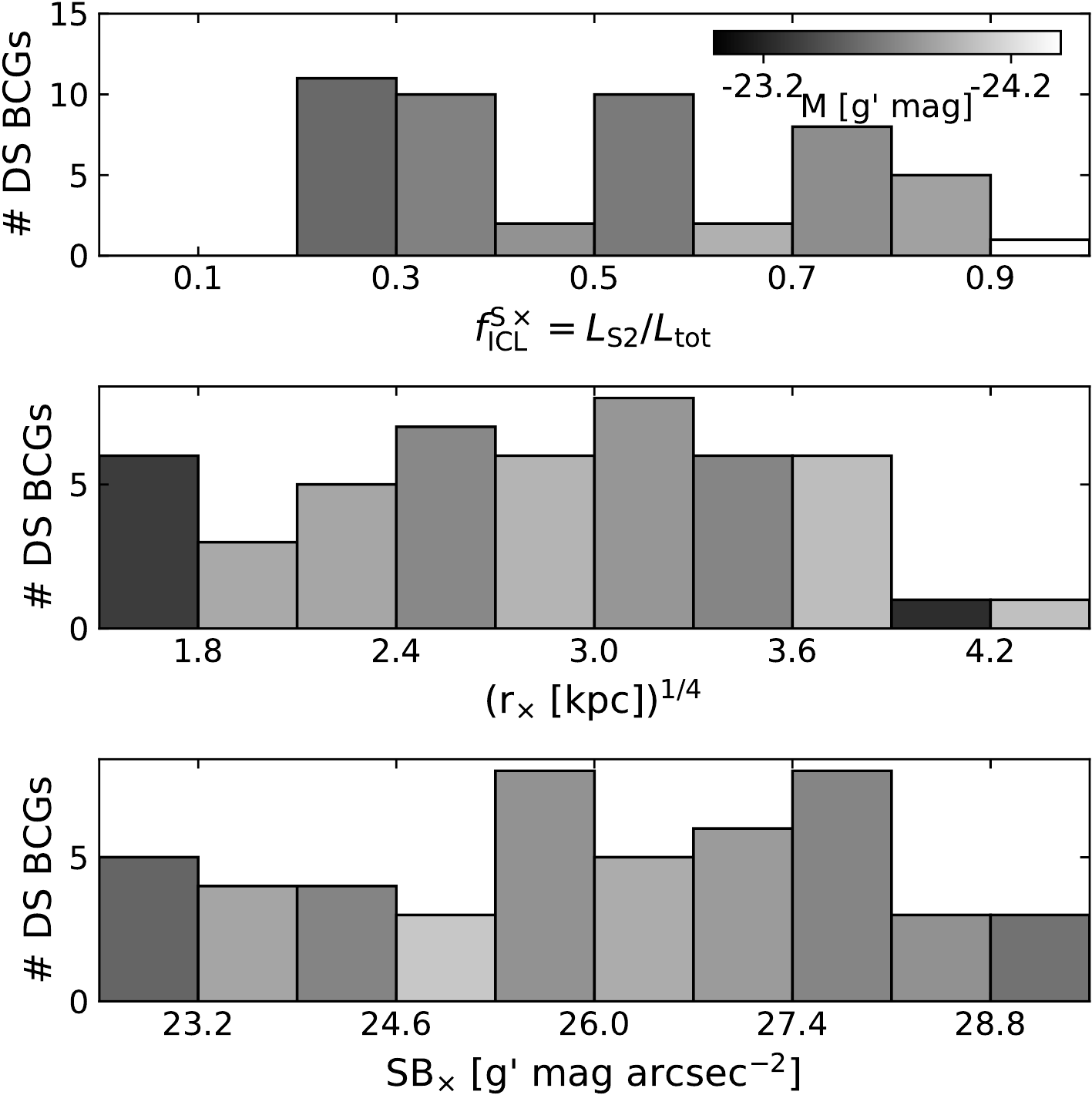}
	\caption{Method d: \textit{Top panel:} Luminosity fraction of the outer S\'ersic component $L_{\rm S2}$ in relation to the total BCG+ICL luminosity $L_{\rm tot}$ of 49 double-S\'ersic BCGs. That is column (8) in Table 4 in \cite{Kluge2020}. \textit{Middle panel:} Radius $r_{\times}$ beyond which the outer S\'ersic component is brighter than the inner. \textit{Bottom panel:} SB of the transition point ${\rm SB}(r_{\times})$. The average absolute BCG+ICL brightness in the bins is coded in the gray-shading of each bar.\label{fig:S2frac}}
\end{figure}

\begin{table*}
	\centering
	\resizebox{\linewidth}{!}{%
		\begin{tabular}{lccccccc}
			\multicolumn{1}{c}{Author} & Method & $f_{\rm ICL}=L_{\rm ICL}/(L_{\rm BCG+ICL})$ [\%] & $L_{\rm ICL}/L_{\rm Cluster}$ [\%] & $(L_{\rm BCG+ICL})/L_{\rm Cluster}$ [\%] & Lim. Mag & Filter & Lim. Mag ($g'$) \\
			\hline
			\hline
			\textit{OBSERVATIONS:}\\
			~~this work (method a) & $M<-21.85$ & $\mathbf{71\pm22}$ & $20\pm12$ & $28\pm17$ & 30 & $g'$ & \ldots \\
			~~this work & ${\rm SB_{cut}}=25$   & $52\pm17$ & $16\pm14$ & $28\pm17$ & 30 & $g'$ & \ldots \\
			~~this work & ${\rm SB_{cut}}=26$   & $43\pm19$ & $13\pm13$ & $28\pm17$ & 30 & $g'$ & \ldots \\
			~~this work (method b) & ${\rm SB_{cut}}=27$   & $\mathbf{34\pm19}$ & $10\pm12$  & $28\pm17$ & 30 & $g'$ & \ldots \\
			~~this work & ${\rm SB_{cut}}=28$   & $26\pm18$ & $8\pm12$  & $28\pm17$ & 30 & $g'$ & \ldots \\
			~~this work & ${\rm SB_{cut}}=29$   & $19\pm17$ & $7\pm11$  & $28\pm17$ & 30 & $g'$ & \ldots \\
			~~this work & ${\rm SB_{cut}}=30$   & $15\pm16$ & $5\pm11$  & $28\pm17$ & 30 & $g'$ & \ldots \\
			~~\cite{Feldmeier2004} & ${\rm SB_{cut}}=26$ & $\sim20$ & \ldots & \ldots & 26.5 & $V$ & 27 \\
			~~\cite{Feldmeier2004} & ${\rm SB_{cut}}=27.5$ & $\sim2$ & \ldots & \ldots & 26.5 & $V$ & 27 \\
			~~this work (method c) & dV+excess & $\mathbf{48\pm20}$ & $13\pm9$ & $28\pm17$ & 30 & $g'$ & \ldots \\
			~~\cite{Zibetti2005} & dV+excess & $33\pm6$ & $10.9\pm5.0$ & $33\pm16$ & 32 & $r+i$ & 31.5 \\
			~~\cite{Zhang2019} & \ldots & \ldots & \ldots & $44\pm17$ & 30 & $r$ & 29.5 \\
			~~\cite{SampaioSantos2020} & \ldots & \ldots & \ldots & $26\pm19$ & 30 & $r$ & 29.5\\
			~~this work (method d) & DS & $\mathbf{52\pm21}$ & $18\pm17$ & $28\pm17$ & 30 & $g'$ & \ldots \\
			~~\cite{Seigar2007} & DS & 59 -- 98 & \ldots & \ldots & 26.5 & $R$ & 26.0 \\
			~~\cite{Donzelli2011} & S+Exp & $40\pm14$ & \ldots & \ldots & 24.5 & $R$ & 25.7 \\
			~~\cite{Gonzalez2005a} & 2dV & 40 -- 90 & \ldots & \ldots & 28.4 & $I$ & 30\\
			~~\cite{Gonzalez2007} & \ldots & \ldots & \ldots & $26\pm8$ & 24.5 & $R$ & 25.7 \\

			~~\cite{Kravtsov2018} & \ldots & \ldots & \ldots & $29\pm7$ & \ldots & mass & \ldots \\

			~~\cite{Mihos2017} (Virgo) & $\Sigma$ streams & \ldots & $7-15$ & \ldots & 28.5 & $V$ & 29 \\
			~~\cite{Spavone2020} (Fornax) & DS/TS (all) & $\sim65$ & $34\pm20$ & $65\pm30$ & 29 & $r$ & 29.7 \\
			~~various studies (Coma) & \ldots & \ldots & $25-50$ & \ldots & \ldots & \ldots & \ldots \\

			\hline
			\textit{SIMULATIONS:}\\
			~~\cite{Puchwein2010}\\
			~~~~~~ (w/ AGN feedback) & 2dV & 91 -- 97 & 50 -- 54 & 51 -- 59 & \ldots & mass & \ldots\\
			~~~~~~ (w/o AGN feedback) & 2dV & $\sim$ 91 & 41 -- 49 & 45 -- 54 & \ldots & mass & \ldots\\
			
			~~~~~~ (w/ AGN feedback) & 2 Maxwell & 72 -- 79 & 40 -- 43 & 51 -- 59 & \ldots & mass & \ldots\\
			~~~~~~ (w/o AGN feedback) & 2 Maxwell & 71 -- 88 & 38 -- 40 & 45 -- 54 & \ldots & mass & \ldots\\
			
			~~\cite{Rudick2011} & ${\rm SB_{cut}}=26.5$ & \ldots & 11 & \ldots & \ldots & $V$ & \ldots\\
			~~\cite{Rudick2011} & 2 Maxwell & \ldots & 16 & \ldots & \ldots & $V$ & \ldots\\
			~~\cite{Cui2014}\\
			~~~~~~ (w/ AGN feedback) & ${\rm SB_{cut}}=26.5$ & $\sim$ 19 -- 31 & $\sim$ 15 -- 25 & $\sim$ 80 & \ldots & $V$ & \ldots\\
			~~~~~~ (w/o AGN feedback) & ${\rm SB_{cut}}=26.5$ & $\sim$ 15 -- 22 & $\sim$ 10 -- 15 & $\sim$ 65 -- 70 & \ldots & $V$ & \ldots\\

			~~~~~~ (w/ AGN feedback) & 2 Maxwell & $\sim$ 61 -- 76 & $\sim$ 50 -- 60 & $\sim$ 75 -- 85 & \ldots & $V$ & \ldots\\
			~~~~~~ (w/o AGN feedback) & 2 Maxwell & $\sim$ 59 -- 72 & $\sim$ 40 -- 45 & $\sim$ 60 -- 70 & \ldots & $V$ & \ldots\\

			~~\cite{Contini2014}\\			
			~~~~~~ (Disruption) & \ldots & $\sim$ 54 -- 88 & 18 -- 23 & 23 -- 35 & \ldots & mass & \ldots\\
			~~~~~~ (Disruption + Mergers) & \ldots & $\sim$ 60 -- 90 & 22 -- 27 & 27 -- 38 & \ldots & mass & \ldots\\
			~~~~~~ (Tidal Radius) & \ldots & $\sim$ 60 -- 86 & 26 -- 31 & 33 -- 45 & \ldots & mass & \ldots\\
			~~~~~~ (Tidal Radius + Mergers) & \ldots & $\sim$ 65 -- 92 & 29 -- 36 & 36 -- 47 & \ldots & mass & \ldots\\
			~~~~~~ (Continuous Stripping) & \ldots & $\sim$ 34 -- 54 & 17 -- 23 & 38 -- 53 & \ldots & mass & \ldots\\
			~~~~~~ (Continuous Stripping + Mergers) & \ldots & $\sim$ 50 -- 74 & 29 -- 37 & 45 -- 61 & \ldots & mass & \ldots\\
			~~\cite{Cooper2015} & DS & $66\pm27$ & $17\pm8$ & $26\pm5$ & \ldots & $V$ & \ldots\\
			~~\cite{Cooper2015} & ${\rm SB_{cut}}=26.5$ & $\sim30$ & $\sim8$ & $26\pm5$ & \ldots & $V$ & \ldots\\
			~~\cite{Pillepich2018} & $r>30$\,kpc & $47-63$ & $21-45$ & $30-56$ & \ldots & mass & \ldots\\
			~~\cite{Pillepich2018} & $r>100$\,kpc & $72-84$ & $12-31$ & $30-56$ & \ldots & mass & \ldots\\
		\end{tabular}
	}
	\caption{Comparisons of ICL fractions with published values from the literature. The methods to dissect the ICL from the BCG are either by applying a surface brightness cut, where the SB threshold in the filterband (5) is given in the second column (2); by decomposing the light profiles into two S\'ersic functions (DS), two de Vaucouleurs function (2dV), an inner S\'ersic and outer exponential function (S+Exp) or by fitting the inner light profile with a de Vaucouleurs function and counting the excess light above that (dV+excess); by applying a brightness cut at $M_{\rm BCG+ICL} < -21.85~g'$ mag; or by applying a circular radius cut. The limiting magnitudes given by the authors (6) are converted to $g'$-band (8) by matching the photometric zero-points of individual light profiles to our data (for \citealt{Seigar2007} and \citealt{Donzelli2011}), by converting $g' = V + 0.45$ mag (for \citealt{Feldmeier2004}), by applying the color transformations derived by Lupton (2005, \url{https://www.sdss.org/dr12/algorithms/sdssUBVRITransform/}) (for \citealt{Gonzalez2005a}) or by using the multiband light profiles measured by the authors and applying K- and for cosmic dimming correction (for \citealt{Zibetti2005}, \citealt{Zhang2019}, and \citealt{SampaioSantos2020}). The cluster luminosities in Zibetti et al. are calculated inside 500\,kpc around the BCG, in Zhang et al. and Sampaio-Santos et al. inside 1\,Mpc around the BCG, in \cite{Gonzalez2007} inside $r_{200}$, and the cluster masses in \cite{Kravtsov2018} inside $r_{500}$. \cite{Mihos2017} calculated the summed-up luminosity of all stellar streams found in the Virgo cluster and assumed that they make up 10\% of the total ICL luminosity \citep{Rudick2011}. \cite{Spavone2020} performed double S\'ersic (DS) or triple S\'ersic (TS) decompositions of bright ($<15$ $B$ mag) galaxies inside the virial radius of the Fornax cluster. Following \cite{Cooper2013,Cooper2015}, they assume that the ICL is stored in the outer second and third S\'ersic components. Their total ICL luminosity also includes the ICL stored in satellite galaxies. For the Coma cluster, we adopt the ICL estimates compiled by \cite{Spavone2020}. They refer to \cite{Melnick1977,Thuan1977,Bernstein1995,Adami2005}, and \cite{JimenezTeja2019}. The simulation results from \cite{Puchwein2010} and \cite{Cui2014} are calculated inside $r_{500}$, from \cite{Cooper2015} inside $r_{200}$, and from \cite{Pillepich2018} inside the virial radius. The $L_{\rm ICL}/(L_{\rm BCG+ICL})$ fractions from \cite{Cooper2015} were inverted in case the effective radius of the outer second component was smaller than that of the first component. The $L_{\rm ICL}/L_{\rm Cluster}$ fraction is calculated for \cite{Contini2014} inside $r_{200}$ and for $(L_{\rm BCG+ICL})/L_{\rm Cluster}$ inside $r_{500}$. For the comparisons with \cite{Gonzalez2007}, \cite{Kravtsov2018}, and the simulation data, we only consider clusters in our mass range $\log(\mathcal{M}_{\rm g}[{\rm M}_{\odot}]) = 14.75\pm0.25$. \label{tab:iclfrac}}
\end{table*}

The fourth approach to determine ICL fractions is by decomposing the SB profiles into two S\'ersic functions. Both S\'ersic components are independently integrated from the 2D isophote models. In some galaxies, there is evidence that the transition to the outer component correlates with changes in ellipticity, position angle and color \citep{Gonzalez2005a,Iodice2016,Spavone2017}. This would empirically motivate fixing different ellipticities for each component for integration. However, ellipticity profiles $\epsilon(r)$ are often not monotonic \citep{Iodice2016,Kluge2020} and therefore not well reproduced by weighting two transitioning S\'ersic profiles. We therefore choose the simplest option of not fixing any ellipticities and instead, using the same ellipticity profile for both components.

We find that the outer S\'ersic component encompasses $f_{\rm ICL}^{\rm S\times}=52\pm21\%$ of the total light. The intrinsic scatter is huge. We show a more detailed histogram of the distribution in the top panel of Figure \ref{fig:S2frac}. The integrated brightness $M$ of the BCG+ICL is coded in the gray-scaling of the bars. Only a weak correlation is found between $f_{\rm ICL}$ and $M$ (Pearson $R=-0.26\pm0.16$).

The other two histograms show the transition point where the S\'ersic components intersect. Only a very weak correlation with BCG+ICL brightness is found for the transition radii $r_{\times}$ (Pearson $R=-0.17\pm0.15$) or no correlation at all for the transition surface brightnesses ${\rm SB}_{\times}$ (Pearson $R=0.00\pm0.16$). The transition SBs between the BCGs and the DSCs (Diffuse Stellar Component = kinematically confirmed ICL) in the simulations used by \cite{Cui2014} have similar scatter around ${\rm SB}_{\times}\sim25~g'$ mag arcsec$^{-2}$, but the non-negligible fraction of ${\rm SB}_{\times}>27~g'$ mag arcsec$^{-2}$ is not found there.

We now compare our results with those from previously published work by other authors. An overview of the derived ICL fractions and the limiting depths of the corresponding surveys can be found in Table \ref{tab:iclfrac}.

The largest sample so far was compiled by \cite{Donzelli2011}. They derived 430 BCGs light profiles from data taken between 1989 and 1995 and fitted them using either one S\'ersic function or using an inner S\'ersic functions plus an outer exponential profile ($n_2=1$). Our average outer S\'ersic index $n_2=1.16\pm1.28$ is also consistent with being exponential but with significant scatter. Our average ICL fraction $f_{\rm ICL}^{\rm S\times}=52\% \pm 21\%$ agrees with the value of $f^{\rm S\times}_{\rm ICL}=40\% \pm 14\%$, calculated from the $S/e$ column Table 2 of \cite{Donzelli2011}. However, closer examination of the results for individual clusters reveals large discrepancies. The limiting depth of their used survey is relatively shallow with ${\rm SB_{lim}} = 24.5$ $R$ mag arcsec$^{-2}$. S\'ersic fits to SB profiles that were derived from shallow data are unconstrained at large radii. The goodness of fit can sometimes be improved significantly in the inner regions by choosing a different set of parameters. They however often fail to describe the shape of the outer ICL halos.

\cite{Gonzalez2005a} observed 24 galaxy clusters in drift-scan mode similar to SDSS. Background inhomogeneities due to hardware are largely averaged out by that technique. The limiting magnitude of the survey corresponds to an equivalent of ${\rm SB_{lim}}=30~g'$ mag arcsec$^{-2}$ in the $g'$-band and is therefore as deep as our survey. They found a typical ICL contribution of $\sim80-90\%$ for the 24 BCGs by decomposing their light profiles into two de Vaucouleurs profiles. That is larger than our result, but still consistent with our large error bars. We cannot compare individual galaxies because none of their observed clusters are visible in the northern sky and therefore, there is no sample overlap.

\cite{Seigar2007} calculated between 59\% and 98\% ICL fractions for the extrapolated DS profiles of four BCGs which we will discuss now individually. \mbox{NGC 6173} in A2197 is classified by both, Seigar et al. and us as an SS BCG. For NGC 3551 in A1177, both results ($f_{\rm ICL}^{\rm S\times}=53\pm14\%$, Seigar et al.; $f^{\rm S\times}_{\rm ICL}=55\pm4\%$, this work) are also consistent within the uncertainties. However, there is disagreement for the NGC 4874 in A1656 (Coma cluster). Seigar et al. found $f^{\rm S\times}_{\rm ICL}=98\pm1\%$ whereas we chose to fit it with an SS function because the transition radius would be close to the nucleus (see Section 4.4 and 7.2 in \citealt{Kluge2020}). The results for UGC 9799 in A2052 also disagree. Even though the galaxy has a relaxed morphology at first glance, there is strong intrinsic scatter in the radial light profile. The addition of a second S\'ersic component improves the fit at small radii but provides no significant gain considering the overall, almost linear $r^{1/4}$ profile shape. \cite{Donzelli2011} agree with our perception because they also classify UGC 9799 as an SS BCG.

The ICL fractions calculated from the hydrodynamical simulations of \cite{Puchwein2010} $f^{\rm S\times}_{\rm ICL} \sim 91 - 97\%$ are higher than what we determined from our observations (see Table \ref{tab:iclfrac}). The authors decomposed the projected $r$-band light profiles into two de Vaucouleurs profiles instead of two S\'ersic profiles like we did. The comparison is therefore not entirely fair. Furthermore, the authors note that their simulated BCG+ICLs are too massive. The simulation results from \cite{Cooper2015} $f^{\rm S\times}_{\rm ICL} 66\pm27\%$ agree well with our measurements ($52\pm21\%$) and the BCG+ICL masses are also consistent.

The large intrinsic scatter of photometrically determined ICL fractions is reproduced in the Magneticum simulation. \cite{Remus2017} decomposed the mass profiles of simulated BCGs into two S\'ersic functions. They discovered that there is no correlation between perceived ICL fractions that are determined by mass profile decompositions and the true ICL fractions derived by kinematic separation of the particles. That evokes skepticism on the validity of the physical interpretation that the outer S\'ersic component traces the ICL.

\subsection{Comparisons to numerical simulations}\label{sec:comparesim}

In this section, we compare our photometrically determined ICL luminosity fractions with predictions from three numerical simulations.

\cite{Cui2014} used 29 galaxy clusters from an $N$-body cosmological simulation, which were resimulated using a hydrodynamical code with merger trees and a comparably low mass resolution of $8.47\times10^8 h^{-1} {\rm M}_{\odot}$ per dark matter particle. They compare two methods of ICL identification: the SB threshold and a kinematic approach. Our $L_{\rm ICL}/(L_{\rm BCG+ICL})$ fractions from method b (${\rm SB_{cut}} = 27~g'$ mag arcsec$^{-2}$; see Section \ref{sec:iclmethod_sbthresh}) agree well with \cite{Cui2014}. The authors used the same SB cut (when converted to $g'$ band) on synthetic photometry of their projected, simulated clusters. However, their $(L_{\rm BCG+ICL})/L_{\rm Cluster}$ fractions are with $60\%-85\%$ significantly higher than our observational results of $28\%\pm17\%$. \cite{Cui2014} realize that this is too high and speculate that two effects might be responsible. The limited numerical resolution makes simulated galaxies too fragile and therefore easily disrupted in the cluster environment. Moreover, the assumption of a constant mass-to-light ratio for the whole cluster galaxy population can underestimate the BCG stellar mass relative to the rest of the cluster stellar mass.

In order to examine the validity of the SB threshold method to dissect the real, dynamically hot ICL from the dynamically colder BCG, \cite{Cui2014} compared the simple SB threshold approach to a more robust kinematic approach. The kinematic method \citep{Dolag2010} decomposes the stellar particle velocity histogram into two Maxwell distributions. The component with the slower characteristic velocity is assigned to the BCG and the component with the faster characteristic velocity to the ICL. Compared to the SB threshold, this more robust method yields significantly smaller transition radii between BCG and ICL, and consequently, $2.3-3.5$ times larger $L_{\rm ICL}/(L_{\rm BCG+ICL})$ fractions in our cluster mass range. \cite{Dolag2010} already note that the SB threshold method underestimates the dynamically hot ICL. A much brighter surface brightness threshold of ${\rm SB_{cut}}\simeq$ 24.75 ($\simeq$ 23) $V$ mag arcsec$^{-2}$ would yield the correct ICL fractions for the simulations in \cite{Cui2014} with (without) AGN feedback, although with large object-by-object scatter. \cite{Cui2014} stress, that this threshold should not be blindly applied to observational data because the BCG+ICL systems are too massive in the simulation (see Table \ref{tab:iclfrac}).

\cite{Cooper2015} studied the usefulness of double S\'ersic SB profile decompositions. They used dark-matter-only 'zoomed-in' $N$-body simulations of 9 massive clusters with a high resolution between $6.1\times10^6 {\rm M}_{\odot}$ and $2.5\times10^7 {\rm M}_{\odot}$ per particle. The cluster mass range $14.9<\log(\mathcal{M}_{200}[10^{14} {\rm M}_{\odot}])<15.5$ corresponds to the high-mass end of our sample. Semi-analytic models of galaxy formation are applied to the simulations. The authors used a particle tagging technique, assigning selected dark matter particles a stellar mass, based on star formation recipes. Thereby, in-situ formed stars and accreted stars from different progenitors are identified in the present-day SB profiles. They find that a double S\'ersic fit to the total SB profile does {\it not} separate the in-situ and accreted stellar components because the latter ones dominate at all radii. The same result was found for lower mass clusters $\log(\mathcal{M}_{200}) > 13$ \citep{Cooper2013}. However, the inner and outer fitted S\'ersic profiles trace well the relaxed and unrelaxed accreted material, respectively. The relaxed material is centered around the BCG and roughly symmetric, whereas the unrelaxed material is more diffuse and includes the distended envelopes of other bright cluster galaxies. This decomposition is alternative and not equivalent to a BCG/ICL decomposition when the ICL is defined as the high-velocity-dispersion stellar component in the surroundings of a BCG.

In our observational data, we confirm that visible accretion signatures and less relaxed morphology are more common in BCG+ICLs which need two S\'ersic functions, instead of one, to fit the SB profiles well \citep{Kluge2020}. The outer S\'ersic component encompasses $L_{\rm ICL}/(L_{\rm BCG+ICL})=52\pm21\%$ of the total BCG+ICL luminosity in our data. This agrees with the value of $66\pm27\%$ by \cite{Cooper2015}. Furthermore, the shape of the outer component is more or less exponential in both our work (S\'ersic index $n_2=1.16\pm1.28$) and \cite{Cooper2015} ($n_2=1.89\pm1.01$). A simple SB threshold at $SB_{\rm cut}=26.5~V$ mag arcsec$^{-2}$ recovers $\sim30\%$ of the BCG+ICL stellar mass in the simulations. This is very consistent with our luminosity fraction $L_{\rm ICL}/(L_{\rm BCG+ICL})=34\pm19\%$ for the corresponding $g'$ band threshold at ${\rm SB_{cut}}=27~g'$ mag arcsec$^{-2}$.

\begin{figure*}
	\centering
	\includegraphics[width=\linewidth]{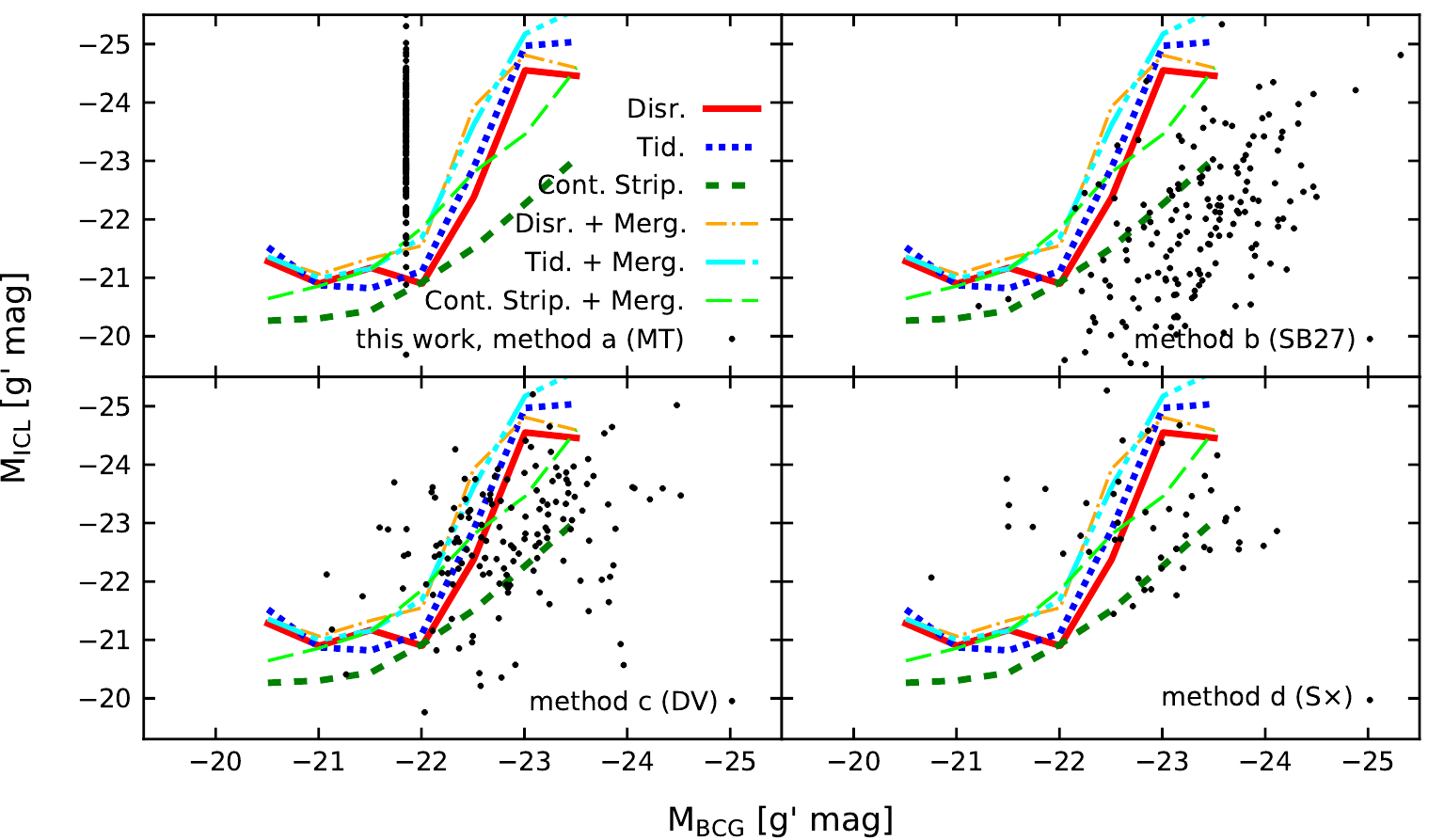}
	\caption{Comparison between our photometrically determined BCG vs. ICL brightnesses (black points) and predictions from semi-analytic models \citep{Contini2014}: disruption model (red line), tidal radius model (blue dotted line), continuous stripping model (green dashed line), and the same models including the merger channel. All models are summarized in Section \ref{sec:comparesim}. The different BCG/ICL decomposition methods for our observational data are: a) integrated brightness threshold, b) is the surface brightness threshold at ${\rm SB_{cut}}=27~g'$ mag arcsec$^{-2}$, c) is the de Vaucouleurs fit + excess light and d) double S\'ersic decomposition. The methods are detailed in Section \ref{sec:iclfrac}. The sample from \cite{Contini2014} also includes lower mass clusters ($\mathcal{M}_{200}>10^{13} {\rm M}_{\odot}$), located toward the left, which we do not test with our data. In order to convert the stellar masses from \cite{Contini2014} to $g'$ band luminosities, we assume a constant mass-to-light ratio of 4.5 times solar in the $g'$-band \citep{Tang2018}. \label{fig:contini}}
\end{figure*}

We now perform a detailed comparison between our BCG/ICL decompositions and the results of semi-analytic models from \cite{Contini2014}. The authors used 27 dark-matter only 'zoomed-in' $N$-body simulations with a mass resolution of $10^8 h^{-1} {\rm M}_{\odot}$ per dark matter particle. Their semi-analytic models of mass assembly trace the ICL build up. They are based on liberation of stellar mass from satellite galaxies in the cluster tidal field. We only consider the cluster mass range in the simulations that is consistent with our observational sample ($\log({\mathcal M}_{\rm g}[{\rm M}_{\odot}])=14.75\pm0.25$). Table \ref{tab:iclfrac} summarizes the resulting ICL fractions and Figure \ref{fig:contini} shows a detailed comparison with our results. \cite{Contini2014} follow three different ICL formation models: disruption, tidal stripping, and continuous stripping. Each model has an optional merger channel.

In the disruption model, a satellite galaxy is disrupted and its stellar material is added to the ICL when its Dark Matter halo is 1) stripped below the resolution limit of the simulation and 2) the satellite stellar mass density is lower than the local cluster Dark Matter density. Additionally, a type 0 galaxy, that is, the central galaxy of the main halo, hosting its own ICL, can fall into a larger system and become a type 1 galaxy, that is, a satellite galaxy in the larger cluster. When its Dark Matter halo is stripped below the resolution limit, its ICL is transferred to the ICL of the new type 0 galaxy.

The predicted $(L_{\rm BCG+ICL})/L_{\rm Cluster}$ fraction by the disruption model is consistent with our results. In other words, for a given cluster luminosity, the disruption model from \cite{Contini2014} predicts correct BCG+ICL luminosities. However, the $L_{\rm ICL}/L_{\rm Cluster}$ fractions are slightly higher than those derived using our photometric decompositions (see Table \ref{tab:iclfrac}). Therefore, for a given BCG+ICL brightness, the BCGs must become fainter on average compared to our decompositions. Indeed, Figure \ref{fig:contini} shows that our results for decomposition methods b, c, and d (black data points) are located below the model predictions (colored lines) from \cite{Contini2014}. For our method a, the BCG brightness is always fixed to $M_{\rm BCG}=-21.85~g'$ mag. The BCG vs. ICL brightnesses calculated using method d (double S\'ersic decomposition) agree on average best with the model predictions (see Table \ref{tab:iclfrac}) but the scatter is also the largest in this case. For method b, brightening the surface brightness threshold from 27 to 25 $g'$ mag arcsec$^{-2}$ shifts our data points closer to the disruption and tidal radius model predictions by $~0.5~g'$ mag arcsec$^{-2}$, but they remain lower.

The second model from \cite{Contini2014} is the tidal radius model. Here, the satellite Dark Matter halo cannot fully retain the stellar material. It can be stripped when the half-mass radius of its parent Dark Matter subhalo is smaller than the half-mass radius of the galaxy's disk. Disruption occurs when the tidal radius for the satellite galaxy is smaller than its bulge radius. If the tidal radius is larger than bulge radius but smaller than the disk radius, all stellar material beyond the tidal radius is stripped and added to the ICL of the central galaxy.

The predicted ICL fractions by the tidal radius model are, again, higher than those derived using our photometric decompositions (see Table \ref{tab:iclfrac}). It is important to mention that the results for both, the disruption and tidal radius models depend on the numerical resolution and must be considered as upper limits. Hence, these models are not inconsistent with our photometric decompositions.

The third model from \cite{Contini2014} is the continuous stripping model. Using a different set of simulations, the authors derived a fitting formula for the stripped stellar mass from a satellite, depending on its orbit. This stripped stellar mass is added to the ICL. The $L_{\rm ICL}/(L_{\rm BCG+ICL})$ fractions are most consistent with our photometric decompositions. However, this model predicts $60\%-90\%$ higher $(L_{\rm BCG+ICL})/L_{\rm Cluster}$ fractions than our observational measurements, especially when the merger channel is activated. The authors explain that this overprediction of the BCG+ICL masses is due to much shorter merging time-scales than in the other two models. The BCGs grow too fast.

All three models have the option to add a merger channel. In case of a merging event, 20\% of the stellar material from the satellite galaxy gets unbound and is added to the ICL. That increases the resulting ICL fractions by a few percent for the disruption and tidal radius models and by $\sim15\%$ for the continuous stripping model. Consequently, the predicted ICL fractions are higher and less consistent with our photometric decompositions (see Figure \ref{fig:contini}). The merger channel turns out to not provide the dominant contribution to the ICL formation. However, by increasing the liberated stellar mass fraction in a merger event from 20\% to 50\%, while fully neglecting tidal stripping, brings the predicted results in agreement with the tidal radius + merging model \citep{Contini2018}.

We cannot distinguish whether the models from \cite{Contini2014} overpredict the ICL fractions or whether our photometric decompositions underpredict them. Kinematic data from spatially resolved stellar spectra will help answering this question by providing more robust measurements of the ICL fractions. We conclude that the disruption and tidal radius models predict consistent upper boundaries for the BCG+ICL luminosities while the continuous stripping model predictions are too high, especially since the latter model does not directly depend on numerical resolution. In the disruption and tidal radius models, the bulk of the ICL comes from galaxies with stellar masses $\sim10^{11} {\rm M}_{\odot}$, while dwarf galaxies contribute very little. The authors interpret this with the effects of dynamical friction: \textit{"the most massive satellites decay through dynamical friction to the inner regions of the halo on shorter timescales than their lower mass counterparts. Tidal forces are stronger closer to the halo centre".}

In summary, we find our $(L_{\rm BCG+ICL})/L_{\rm Cluster}$ fractions to be consistent with previous observational surveys. Simulations often overpredict the BCG+ICL luminosities and the discrepancy gets worse with decreasing numerical resolution. The likely cause is that satellite galaxies get disrupted too easily and their debris builds up the ICL too efficiently. All of our measured ICL and BCG+ICL fractions agree well with \cite{Cooper2015}, where the outer S\'ersic component traces the unrelaxed, accreted stellar material. Whether a double S\'ersic decomposition also separates the dynamically hot ICL from the BCG remains unanswered, because it has not yet been tested rigorously, neither in observations nor in simulations. An SB threshold of ${\rm SB_{cut}}=27~g'$ mag arcsec$^{-2}$ is too faint to separate the dynamically hot ICL from the BCG \citep{Dolag2010,Cui2014}.

\begin{figure*}
	\centering
	\includegraphics[width=\linewidth]{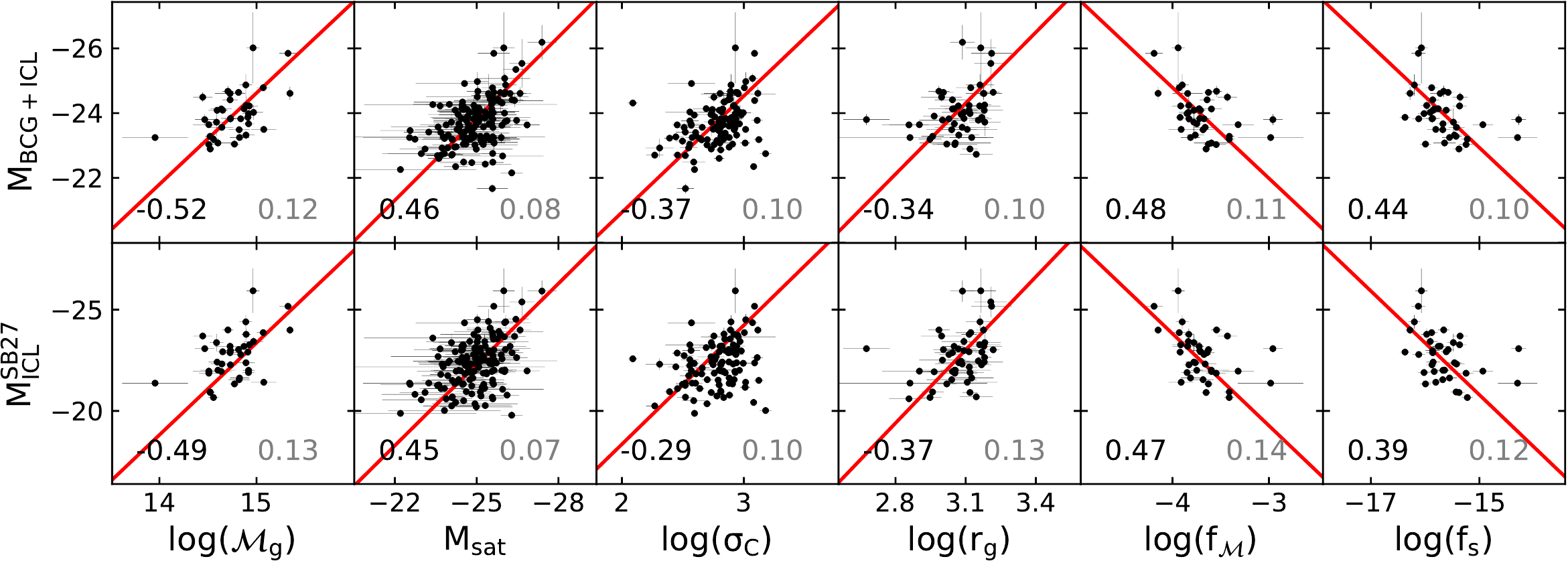}
	\caption{The best correlations between absolute BCG+ICL brightness $M_{\rm BCG+ICL}$ [$g'$ mag] (upper panels) or the absolute brightness of the ICL only $M_{\rm ICL}$ [$g'$ mag], dissected via the surface brightness threshold of 27 $g'$ mag $M_{\rm ICL}^{\rm SB27}$ (lower panels; method b) and host cluster parameters. Each row shows a different host cluster parameter: the (1) gravitational mass $\mathcal{M_{\rm g}} [{\rm M }_{\odot}$], (2) integrated absolute brightness of all satellite galaxies (excluding the BCG+ICL) $M_{\rm sat}$ [$g'$ mag], (3) the velocity dispersion of the satellite galaxies $\sigma_{\rm C}$ [km s$^{-1}$] (taken from \citealt{Lauer2014}), (4) gravitational radius $r_{\rm g}$ [kpc], (5) mass phase space density $f_{\mathcal{M_{\rm g}}} [{\rm M}_{\odot}$ kpc$^{-3}$ km$^{-3}$ s$^3$], and (6) number phase space density of the satellite galaxies $f_{\rm s}$ [kpc$^{-3}$ km$^{-3}$ s$^3$]. The Pearson coefficient for each correlation is given as a black label and its error (calculated using 10\,000 bootstraps) as a gray label. The full collection of plots are shown in Appendix \ref{sec:corrhost} and the best-fit parameters are listed in Appendix \ref{sec:correlateall}.
		\label{fig:correlatebest}}
\end{figure*}

\section{Results: Correlations between BCG/ICL and host cluster properties}\label{sec:corrhosttext}

\subsection{Structural parameters} \label{sec:galdistribution}

The widely hypothesized two-phase formation scenario of BCGs and ICL states that the ICL is made mostly of stellar material that has been accreted from cluster satellite galaxies. Consequently, we expect to find correlations between the satellite galaxies distribution of the host clusters and ICL properties. A selection of the strongest correlations is shown in Figure \ref{fig:correlatebest}. The full collection of plots are shown in Appendix \ref{sec:corrhost} and the best-fit parameters are listed in Appendix \ref{sec:correlateall}.

After standardizing the variables, we fit only the slope and convert the result back to the non-standardized form. The Pearson coefficients, which give a measure for the strength of a linear correlation, are overplotted as a black text label in each subplot. The measurement errors are neglected for the fitting of the parameter correlations. Otherwise, a significant amount of data points had almost zero weight due to the high inhomogeneity of the errors, especially for the brightnesses. Errors of the correlation coefficients are calculated by the standard deviation of the correlation coefficients calculated using 10\,000 bootstraps of the data. They are plotted as the gray text labels in each subplot.

We distinguish between direct and indirect observables. Indirect observables (gravitational mass $\mathcal{M_{\rm g}}$, mass density $\rho$, satellite galaxy number density $s$, mass phase space density $f_{\mathcal{M_{\rm g}}}$ and galaxy number phase space density $f_{\rm s}$) are constructed from a combination of direct observables (cluster velocity dispersion $\sigma_{\rm C}$, richness $S$, gravitational radius $r_{\rm g}$ and brightness of all satellite galaxies $M_{\rm sat}$). They are defined in Section \ref{sec:clusterparams} and the values for each cluster are listed in Table \ref{tab:clusterparams}.

The first row in Figure \ref{fig:correlatebest} or the first columns in Figures \ref{fig:correlateall} and \ref{fig:correlateall2} show the integrated brightnesses of the BCGs+ICL. The BCG luminosity is known to increase with cluster mass \citep{Lin2004,Yang2005,Zheng2007,Popesso2007,Brough2008,Hansen2009}. \cite{Hansen2009} found that the $i$-band luminosity scales $L_{\rm BCG} \propto \mathcal{M}_{200}^{0.30\pm0.01}$ for 13\,823 SDSS clusters at mean redshift $\bar{z}_{\rm H09}=0.25$ (see their Figure 13). We find a significantly steeper slope in our $g'$-band data at $\bar{z}_{\rm K20}=0.06$: $L_{\rm BCG+ICL} \propto \mathcal{M}_{\rm g}^{1.14\pm0.24}$. Under the assumption that the gravitational mass scales linearly with $\mathcal{M}_{200}$, the discrepancy must arise from the measured BCG luminosities. A previously underestimated ICL contribution is likely the cause. We show in Figure \ref{fig:f27frac} that the luminosity fraction below ${\rm SB}>27~g'$ mag arcsec$^{-2}$ increases with total BCG+ICL luminosity. Therefore, the luminosities of the brightest BCGs will be underestimated the most from shallow data. That will consequently lead to a shallower slope in the BCG luminosity -- cluster mass relation. Alternatively, redshift evolution or color effects cannot be excluded as a possible explanation for the discrepancy.

The BCG+ICL (and also solely ICL) brightness correlates positively with the gravitational mass $\mathcal{M}_{\rm g}$ and satellite brightness $M_{\rm sat}$ of the host cluster. That indicates that the growth of the ICL is indeed coupled to the growth of the cluster. Growth is also quantified in cluster gravitational radius $r_{\rm g}$, cluster richness $S$ and cluster velocity dispersion $\sigma_{\rm C}$.

We confirm that BCGs+ICL grow slower in brightness than their host cluster satellite brightness: $M_{\rm sat} = 1.19(\pm0.12)~M_{\rm BCG+ICL}$ + const. (cf. Appendix \ref{sec:correlateall} and e.g., Figure 14 in \citealt{Hansen2009}). But we find a stronger increase in ICL brightness at low surface brightnesses: $M_{\rm sat} = 0.75(\pm0.07)~M^{\rm SB27}_{\rm ICL}$ + const. That is another way of quantifying that BCGs grow predominantly in their outskirts at present epoch.

This also implies that the ICL brightness should correlate with other host cluster parameters, which is indeed shown in the lower panels of Figure \ref{fig:correlatebest}. The fact that the ICL--host cluster correlation strengths are as strong as the BCG+ICL--host cluster correlation strengths does not imply that the correlations in the top panels are solely driven by the ICL. Both, the BCG \textit{and} the ICL correlate with host cluster properties. It might indicate that the virialization time scales for the accreted stellar material are relatively short.

The bottom four rows of subplots in Figures \ref{fig:correlateall2} and \ref{fig:correlateall3} show expressions for densities. Strong correlations are found between the ICL brightness and the phase-space densities $f_{\mathcal{M_{\rm g}}}$ and $f_{\rm s}$ (see also Figure \ref{fig:correlatebest}). The stripping efficiency is proportional to the phase space density because tidal forces are strongest at short separations and more material can be accumulated when interaction time scales are long. Surprisingly, instead of an expected positive correlation, we find a negative correlation. The reason is possibly that we do not see the former host cluster properties but rather the effect that the phase space density decreases after energy is deposited into the clusters by mergers.

\subsection{Alignment} \label{sec:alignment}

Another quantity which is related to the connection between BCG/ICL and cluster formation/evolution is their alignment. Many studies have found strong correlations between the alignment of the BCG and the cluster galaxies distribution \citep{Sastry1968,Dressler1978,Binggeli1982,Struble1990,Kim2002,Yang2006,Niederste-Ostholt2010,Huang2016,West2017,Okabe2020}. Our deep PA profiles allow us not only to investigate the alignment but also whether it improves with radius. Furthermore, we investigate whether the ICL is offset from the BCG toward the cluster center. For these analyses, we consider only clusters where a center and position angle could be reliably determined from the satellite galaxy samples retrieved from the {\tt SIMBAD} database. We have selected 50 out of the 170 clusters from our sample that fulfill these two criteria sufficiently well.

\begin{figure}
	\includegraphics[width=0.95\linewidth]{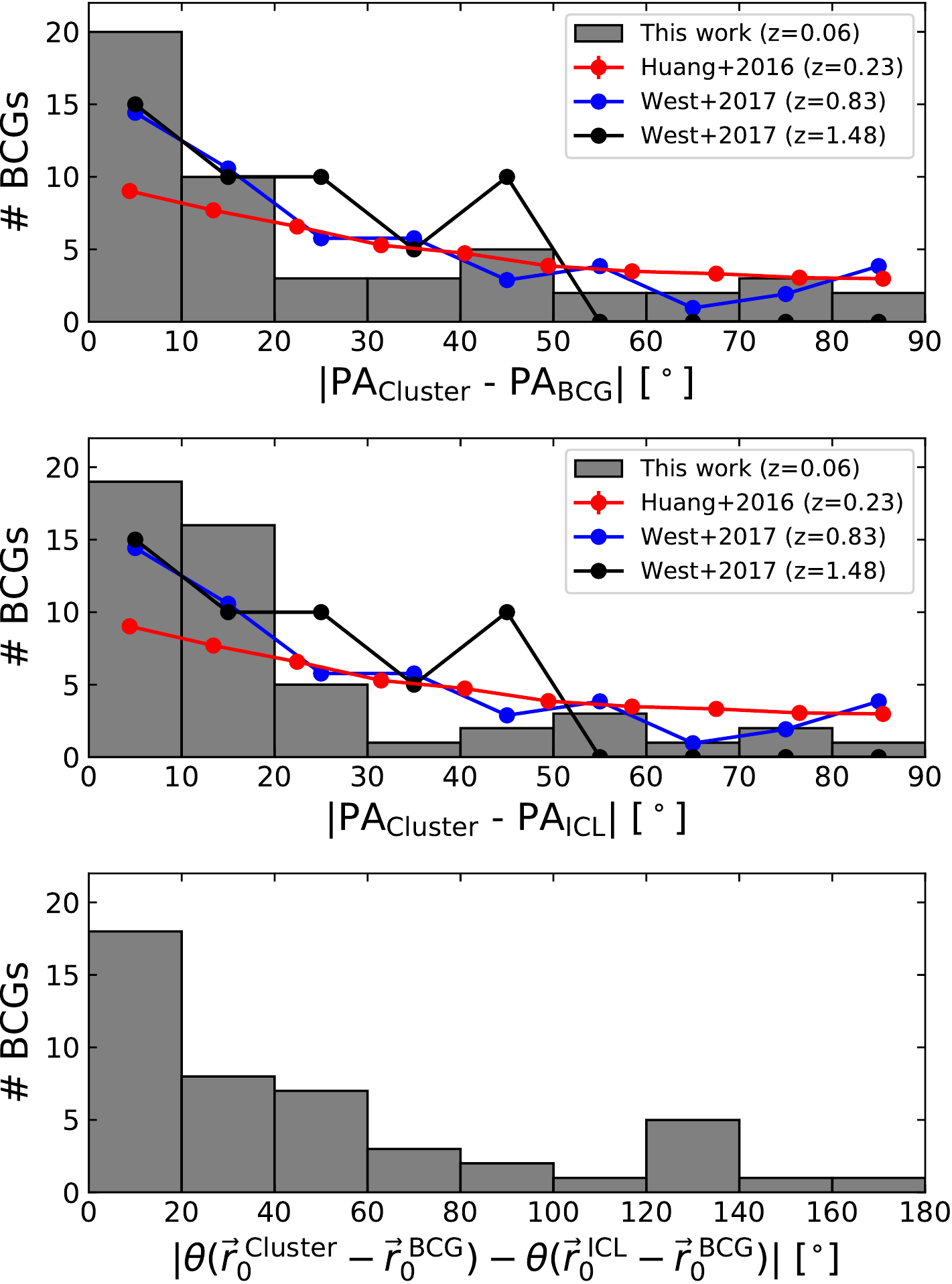}
	\caption{{\it Top and middle panels:} alignment between the BCG (top panel) or ICL (middle panel) with the cluster galaxies orientation. The data from \cite{Huang2016} (red) and \cite{West2017} (blue and black) are renormalized and overplotted for comparison. {\it Bottom panel:} direction of the ICL offset compared to the direction of the cluster galaxies offset with respect to the BCG. A value of $0\degr$ means that the ICL and the cluster galaxies (on average) are offset in the same direction. \label{fig:pa}}
\end{figure}

The results are shown in Figure \ref{fig:pa}. We confirm that both, the BCG and the ICL are strongly aligned with the host cluster orientation, as traced by the satellite galaxies distribution. Moreover, the ICL is aligned even better. Its PA is aligned to less than $PA<30\degr$ with its host cluster in 40/50 = 80\% cases (Figure \ref{fig:pa}, middle panel) compared to the BCGs who are aligned in only 33/50 = 66\% of the cases with their host clusters (Figure \ref{fig:pa}, top panel). The expectation value for a random distribution would be 33\%. Our results show an overall better alignment than the results from \cite{West2017} (32/52 = 62\% with $\Delta PA<30\degr$). That might be an effect of relaxation over time. Their analyzed clusters are at significantly higher redshifts than ours and there is also an improvement in the alignment visible from their higher redshift sample to their lower redshift sample. The results from \cite{Huang2016} show a weaker alignment. A possible explanation is related to our visual optimization of the smoothing kernels for the galaxy density distributions: in case of isodensity contour twists (e.g., in A2029; possibly due to triaxiality of the Dark Matter halo) we favor the inner PA (that is, closer to the radii where we measure the ICL PA) instead of the average PA of the galaxies density distribution. We expect the intrinsic alignment to be stronger where cluster- and ICL-radii are more similar.

Furthermore, the direction of the ICL offset compared to the direction of the cluster galaxies number density peak, both viewed from the BCG center, is aligned better than $<60\degr$ in $33/46 = 72\%$ of the clusters. The expectation value for a random distribution would be again 33\%. Four outlier clusters are discarded from our subsample because these clusters are offsets far ($>400$\,kpc) from their BCG. Our measurements for the ICL offsets from the BCG and the galaxy number density peak offsets from the BCG are independent from each other and a strong correlation between them is unlikely to occur by chance. Contrary to our results, \cite{Gonzalez2005a} found no significant ICL offsets from the BCG (except in one case) for their sample of 24 BCGs.

\subsection{Systemic velocity- and X-ray offsets}

\begin{figure}
	\includegraphics[width=\linewidth]{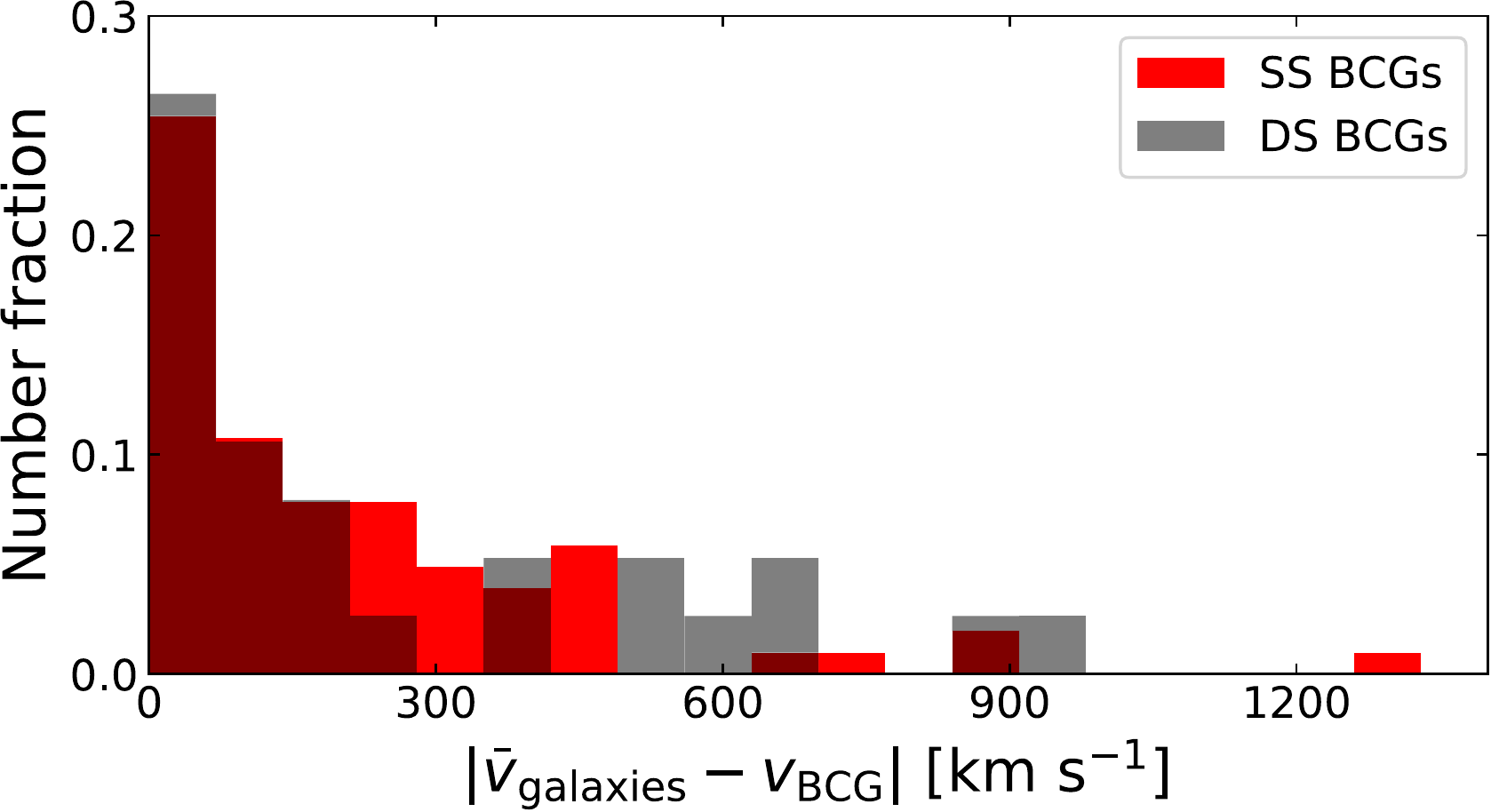}
	\includegraphics[width=\linewidth]{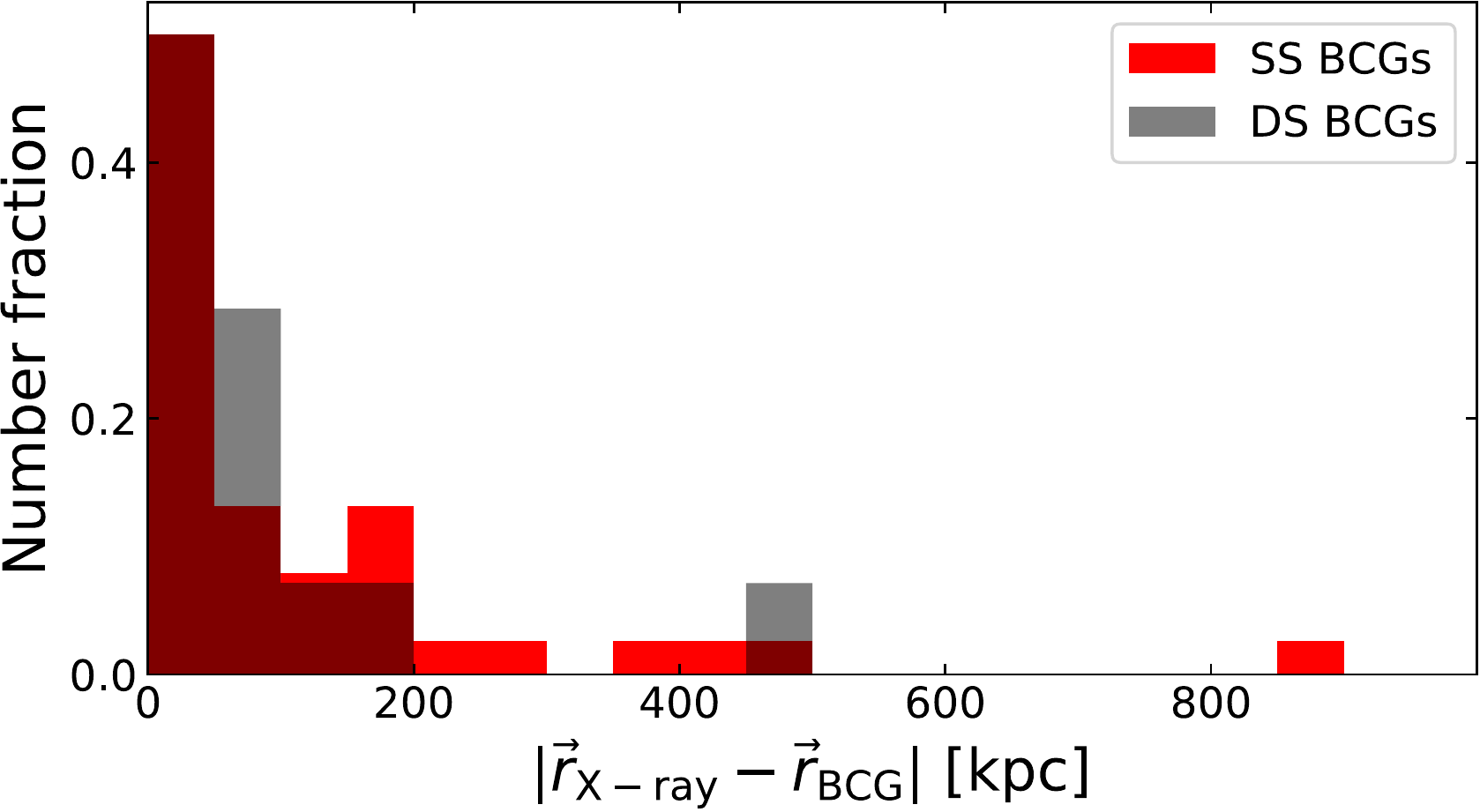}
	\caption{Normalized histograms of systemic velocity offsets (top panel) and X-ray offsets (bottom panel) from the BCG. Single-S\'ersic BCGs are shown in red whereas double-S\'ersic BCGs are shown in gray. Data are taken from \cite{Lauer2014}. Only BCGs that overlap with the Lauer et al. sample were considered.\label{fig:peculiarv}}
\end{figure}

The discovery that DS BCGs show slightly more disturbed morphologies than SS BCGs \citep{Kluge2020} lets us presume that this less relaxed state manifests further in higher systemic velocity offsets and larger X-ray offsets for DS BCGs. A systemic velocity offset is defined as the line of sight velocity difference between the average velocity of the cluster galaxies and the one of the BCG's core $|\bar{v}_{\rm galaxies} - v_{\rm BCG}|$ \citep{Oegerle2001}. The X-ray offset is the analogous measurement in the two other spatial dimensions. It is the projected radial distance between the peak of the X-ray emission which traces the cluster center and the center of the BCG. We use published data from \cite{Lauer2014} for this analysis.

Figure \ref{fig:peculiarv} shows overlapping histograms of absolute systemic velocity offsets (top panel) and absolute X-ray offsets (bottom panel). The numbers for each S\'ersic type add up to one. The normalization allows for a fair comparison since the two types of BCGs have different subsample sizes in this study (73 SS BCGs vs.\ 27 DS BCGs for the systemic velocity offsets and 39 SS BCGs vs.\ 14 DS BCGs for the X-ray offsets). The average error is 52 ${\rm km~s}^{-1}$ for the systemic velocity offsets and on the order of a few tens of kpc for the X-ray offsets.

A Kolmogorov--Smirnov test gives a 43\% probability that the SS and DS BCG samples are drawn from the same systemic velocity offset distribution. The test for the X-ray offsets gives 52\% probability. These numbers do not allow us to draw any conclusion here.

\section{Discussion}\label{sec:discussion}

\subsection{Does the outer S\'ersic component trace the ICL?}\label{sec:discussionicl}

This discussion was started around the pioneering work of \cite{Schombert1986} and the thesis is supported by recent simulations \citep{Cooper2015}. First of all, we find that 121 (71\%) out of our 170 observed BCGs are of SS type whereas 49 (29\%) are of DS type. The bare existence of SS BCGs is problematic in this context. For these not uncommon cases, the transition between inner and outer S\'ersic component is smooth so that any photometric decomposition is strongly degenerate \citep{Bender2015}.

The number fraction of DS BCGs in our work is lower by 19\% points than the 48\% found by \cite{Donzelli2011}. A lower number fraction is expected because we require a minimum transition radius between the two S\'ersic components. The number fraction in the Magneticum simulation is with 58\% \citep{Remus2017} also higher than ours.  Similarly to us, \cite{Remus2017} classify BCGs with very large ICL fractions $f^{\rm S\times}_{\rm ICL}$ as SS type. However, all of their simulated BCGs with $0.1>f^{\rm S\times}_{\rm ICL}>0.9$ are classified as DS, whereas we favor the SS classification if the SB profile is similarly well fit by only one S\'ersic function. Moreover, their maximum fitting radius is with $\sim$\,2\,Mpc about six times larger than our maximum radius and, hence, DS transitions at radii $r_{\times}\gtrsim$ 200\,kpc are rarely detected by us (Figure \ref{fig:S2frac}, middle panel). Both considerations will increase the DS number fraction in the simulation.

Nevertheless, we do confirm the large scatter in $f^{\rm S\times}_{\rm ICL}$ as predicted by the Magneticum simulation (compare Figure \ref{fig:S2frac} (top panel) in this paper to Figure 6 (left panel) in \citealt{Remus2017}). That agreement strengthens the conclusions drawn from this simulation on the inner structure and dynamics of BCGs+ICL. By decomposing the stellar velocity distributions into two Maxwellians, they isolated the ICL as the dynamically hot component. This approach is motivated by the observed rises in velocity dispersion profiles toward the ICL (\citealt{Dressler1979,Carter1981,Ventimiglia2010,Toledo2011,Arnaboldi2012,Melnick2012,Murphy2014,Bender2015,Barbosa2018,Loubser2018,Spiniello2018,Gu2020}) and currently consensus for numerical simulation (\citealt{Dolag2010,Puchwein2010,Rudick2011,Cui2014}). In their publication, \cite{Remus2017} state that they find \textit{"no clear correlation between the presence of a second component in the velocity distribution and the presence of a second component in the radial density profile"}.

We now examine DS parameters that could possibly depend on the total integrated brightnesses $M_{\rm BCG+ICL}$ of the BCG+ICL systems. We use $M_{\rm BCG+ICL}$ as a proxy for the evolvedness of the system: more evolved BCGs have had more time to accrete stellar debris onto their ICL halos and have grown since in total  brightness. If the hypothesis was correct that the outer photometric component traces the ICL, then at least one of the following four relations is necessary to emerge:

\begin{enumerate}
	\item larger fraction of DS BCGs to all BCGs $f_{\rm DS}$ with increasing BCG+ICL luminosity: $f_{\rm DS}=26\pm6\%$ for the faint half sample and $f_{\rm DS}=32\pm7\%$ for the bright half sample. Errors are determined from Poisson statistics.
	\item increasing light fraction encompassed in the outer S\'ersic component $f^{\rm S\times}_{\rm ICL}$ with BCG+ICL luminosity if the BCG is unaffected by the accretion (Pearson $R=0.26\pm0.16$),
	\item larger transition radii $r_{\times}$ for higher luminosity if the components mix (Pearson $R=0.17\pm0.15$),
	\item a vertical size--luminosity relation if stars are accreted homogeneously over all radii.
\end{enumerate}

In order to discuss option four, we refer to Table 3 in \cite{Kluge2020}. The size--luminosity relation for BCGs is $\log(r_{\rm e}) \propto 1.41 (\pm0.08) \log(L_{\rm BCG+ICL})$. That means fractional BCG growth is 41\% larger in radius than in luminosity. That argument plus our finding that the light fraction at low SBs increases with BCG+ICL brightness (see Section \ref{sec:iclmethod_sbthresh}) disprove the fourth option in the list above. For the remaining options, we give the Pearson coefficients in the brackets. The absence of relations one and two are the strongest indicators that the two-component structure of the light profiles might be nothing more than a result of the recent accretion events and a photometric decomposition into two S\'ersic functions is likely to be unphysical. Final conclusions can only be drawn from additional velocity information. We will explore whether the transition between the two components coincides with a distinct rise in velocity dispersion for a small subsample of BCGs in a subsequent paper.

\subsection{ICL as a Dark Matter tracer}\label{sec:discussiondm}

ICL is the dynamically hot stellar component that was assembled by tidally stripping stars from cluster galaxies. These stars move freely in the cluster potential and, when virialized, should trace the overall mass distribution. This expectation was recently confirmed for six clusters from the Hubble Frontier Fields \citep{MontesTrujillo2019}. The ICL in these clusters traces the overall Dark Matter distribution including substructure better than the hot X-ray gas, because it is less perturbed by mergers than the dissipative gas. The alignment between ICL and Dark Matter contours was also confirmed in numerical simulations \citep{AlonsoAsensio2020}. Besides being more concentrated, the averaged radial ICL profile is comparable to averaged Dark Matter profiles when expressed in annular differential surface density \citep{SampaioSantos2020}.

We examine four criteria that potentially qualify ICL as a good Dark Matter tracer: (1) the ICL PA alignment with the cluster PA, (2) the offset from the BCG toward the cluster center, (3) the ellipticity, and (4) the line-of-sight velocity. We have selected a subset of 50 clusters from our dataset with the most reliable cluster PA and cluster center measurements. The satellite galaxies are used as test particles for the underlying Dark Matter distribution (e.g., \citealt{Shin2018}).

(1) We begin our discussion with the PA alignment between the BCG, ICL and their host clusters.

It is well known that BCGs are well aligned with their host clusters (see Section \ref{sec:alignment}). Our results show that the alignment is even better for the ICL. The difference in PA is $\Delta PA < 30\degr$ in 80\% of the clusters whereas that is only the case for 66\% of the BCGs (see Figure \ref{fig:pa}). The expectation value for a random distribution is only 33\%.

(2) Criterion two is investigating the offset between BCG, ICL and their host clusters. We begin with the direction of the offsets and then discuss their amplitudes.

The direction of the ICL offset from the BCG coincides in 72\% of the cases to better than $60\degr$ with the direction of the cluster offset from the BCG. A random distribution would have only 33\% matches. We conclude that the ICL is generally more at rest in the cluster potential than the BCG.

The amplitude of the ICL offset from the BCG radially increases (see Figure 15 in \citealt{Kluge2020}, central panel). At 200\,kpc circular radius, the isophotes are shifted on average by 17\%, that is, 37\,kpc with 34\,kpc intrinsic scatter. For a subsample of 31 clusters, in which the cluster and ICL offset directions agree by $<60\degr$, and the cluster offset is less than $<400$\,kpc, we find an amplitude of $93\pm62$\,kpc of the cluster offsets, compared to $9.3\pm9.7$\,kpc for the ICL offsets.

We now compare our results to ICL offsets, X-ray gas offsets and Dark Matter centroid offsets with respect to the BCGs, which have been published in the literature.

Similar to our results, ICL-to-BCG offsets exist in Hydra (12\,kpc, \citealt{Arnaboldi2012}) and A1651 (15\,kpc, \citealt{Gonzalez2005a}). However, 23 out of 24 BCGs are consistent with having no ICL offsets from the BCG in the study by \cite{Gonzalez2005a}. By comparing the projected center of the satellite galaxy distribution with the projected location of BCGs, \cite{Zitrin2012} found typical BCG offsets of around 12\,kpc in 10\,000 SDSS clusters. \cite{Oguri2010} determined mass profiles of 25 clusters using weak lensing methods from high-quality Subaru/Suprime-Cam imaging data. They find that most of the centroids coincide with the location of the BCG within their measurement uncertainty of 35\,kpc. However, a non-negligible number of clusters shows large offsets of up to 500\,kpc. The intracluster medium (hot, X-ray emitting gas) is a good tracer for the total mass distribution because it can be assumed to be in hydrostatic equilibrium. \cite{Umetsu2014} found for 20 CLASH clusters a median offset between the BCG and the X-ray peak of 7\,kpc with 21\,kpc intrinsic scatter. Using high-resolution Chandra data, \cite{Lauer2014} found a median offset of 10\,kpc but $\sim15\%$ of the BCGs have offsets larger than 100\,kpc (see also Figure \ref{fig:peculiarv}, bottom panel).

We conclude that the average amplitude our measured ICL offsets from the BCG is consistent with the Dark Matter offsets from the BCG as measured by other authors.

(3) A third quality, that is required for ICL to be a good Dark Matter tracer, is that its average ellipticity $\epsilon$ must be similar to the average ellipticity of Dark Matter halos. For the ICL, we have measured an axis ratio $(b/a)_{\rm ICL} = 1- \epsilon \approx 0.5 - 0.6$ at the largest circular radius $r=\sqrt{ab}\approx200$\,kpc. The average ellipticity of Dark Matter halos can be measured by using satellite galaxies as test particles or using stacked weak lensing measurements. Both methods have been applied by \cite{Shin2018} for 10\,428 SDSS clusters. They found an average axis ratio $(b/a)_{\rm DM} = 0.573\pm0.002$(stat)$\pm0.039$(sys) using the satellite galaxy method or $(b/a)_{\rm DM} = 0.56\pm0.09$(stat)$\pm0.03$(sys) using the stacked weak lensing method. The results agree well with our measured average ICL ellipticity.

(4) By measuring line-of-sight velocities, \cite{Bender2015} found the mean ICL line-of-sight velocity around the BCG NGC 6166 to be shifted toward the mean line-of-sight velocity of the cluster galaxies. Therefore, the ICL is more at rest with respect to the cluster as a whole, and subsequently, also with the Dark Matter. We plan to measure ICL velocity offsets for a subsample of the BCGs presented here in a subsequent paper.

From all of the considerations above, we conclude that the ICL is better aligned than the BCG in position, velocity, ellipticity and position angle with respect to the cluster galaxies and consequently, with respect to the Dark Matter distribution.

\section{Summary and Conclusions}\label{sec:summary}

Using the deep SB profiles of 170 local BCGs published in \cite{Kluge2020}, we have applied four photometric methods to attempt BCG/ICL decompositions. We emphasize that we do not distinguish between stellar envelope, stellar halo, and ICL because they are probably indistinguishable with photometric data alone. At least part of the stellar halos of the BCGs are probably included in the ICL fractions. The results are summarized as follows:

\begin{enumerate}
	\item[A1.] The large intrinsic scatter of the ICL fraction $f^{\rm S\times}_{\rm ICL}$ is well reproduced by the Magneticum simulation where \cite{Remus2017} found no correlation between the existence of a two-component nature of the stellar velocities which represent the BCG and dynamically hot ICL and the existence of a two-component nature of the SB profiles. In our observational data, the BCG+ICL brightness also correlates not at all or only very weakly with $f^{\rm S\times}_{\rm ICL}$ (Pearson $R=-0.26\pm0.16$), with the transition radius $r_{\times}$ between the two S\'ersic components ($R=-0.17\pm0.15$) and with the transition surface brightness ${\rm SB}_{\times}$ ($R=0.00\pm0.16$). Both results suggest that the separation between BCG and dynamically hot ICL is not possible based on photometric decompositions of their light profiles. That is in agreement with the photometric and kinematic results of a case study of NGC 6166 \citep{Bender2015}.
	
	\item[A2.] All of our ICL fractions agree very well with predictions by \cite{Cooper2015}. In their simulations, the accreted stellar material in BCG+ICL systems dominates at all radii over the in-situ formed stellar material. The double S\'ersic decomposition separates well the relaxed, inner and unrelaxed, outer accreted stellar material.
	
	\item[A3.] Our BCG/ICL fractions agree best with the continuous stripping model by \cite{Contini2014}. However, their BCG+ICLs are $60-90\%$ too luminous in the cluster mass range of our sample. The BCG+ICL luminosities predicted by their disruption and tidal radius models are closer to our measured values, but these models predict higher ICL fractions.
			
	\item[A4.] The fiducial ICL/(BCG+ICL) luminosity fraction above an integrated brightness of \mbox{$M<-21.85$} g' mag is $f^{\rm MT}_{\rm ICL}=71\pm22\%$. The corresponding ICL/cluster luminosity fraction is $20\%\pm12\%$.
	
	\item[A5.] The fiducial ICL/(BCG+ICL) luminosity fraction below ${\rm SB} > 27$ g' mag arcsec$^{-2}$ is $f^{\rm SB27}_{\rm ICL}=34\%\pm19\%$. It increases with total BCG+ICL brightness (Pearson $R=-0.53 \pm 0.07$), indicating that BCGs grow predominantly by accretion in their low-SB outskirts. The corresponding ICL/cluster luminosity fraction is $10\%\pm12\%$.
	
	\item[A6.] The fiducial ICL/(BCG+ICL) luminosity fraction above an inner de Vaucouleurs function is $f^{\rm DV}_{\rm ICL}=48\%\pm20\%$. The corresponding ICL/cluster luminosity fraction is $13\%\pm9\%$.
		
	\item[A7.] The fiducial ICL/(BCG+ICL) luminosity fraction inferred via the DS profile decomposition method is $f^{\rm S\times}_{\rm ICL}=52\%\pm21\%$. The corresponding ICL/cluster luminosity fraction is $18\%\pm17\%$.
	
	\item[A8.] The (BCG+ICL)/cluster luminosity fraction is $28\%\pm17\%$.
\end{enumerate}

Our calculated ICL fractions agree with most observational works that apply the same methods. However, the large scatter among the various decomposition methods ($f_{\rm ICL}=34\%-71\%$) underlines that the ICL dissection cannot be done unambiguously with photometric data alone. Future work based on spatially resolved kinematic data and multi-band color information will allow a critical test of the reliability of the various decomposition methods. We plan to make these tests for a subsample of BCGs in a subsequent paper.

Furthermore, we find that the light fraction at faint SB levels below ${\rm SB}>27~g'$ mag arcsec$^{-2}$ increases with total BCG+ICL brightness. That confirms that BCG+ICLs grow predominantly in their outskirts at present epoch, in agreement with the two-phase formation scenario. The growth proceeds up to the point where the BCG+ICL sizes and luminosities become similar to those of whole galaxy clusters.

The second aim of this paper was to find correlations between BCG/ICL parameters and host cluster parameters. Our results are:

\begin{enumerate}
	\item[B1.] We find positive correlations between BCG+ICL brightness and cluster mass ($R=-0.52\pm0.12$), cluster velocity dispersion ($R=-0.37\pm0.10$), cluster radius ($R=-0.34\pm0.10$), and integrated satellite brightness ($R=0.46\pm0.08$). That confirms that the growth of BCGs is connected to the growth of their host clusters.
	
	\item[B2.] The correlation coefficients for the photometrically dissected ICL are no stronger than those for the total BCG+ICL luminosities. This does not imply that the correlations are solely driven by the ICL. Both, the BCG \textit{and} the ICL correlate with host cluster properties. It might indicate that the virialization time scales for the accreted stellar material are relatively short.

	\item[B3.] The BCG position angles (PAs) are aligned to better than $\Delta PA<30\degr$ in $33/50 = 66\%$ with the PAs of their host clusters. The alignment between the ICL and their host clusters is even stronger: $41/50 = 82\%$ are better aligned than $\Delta PA<30\degr$.
	
	\item[B4.] The ICL offset with respect to the BCG at 200\,kpc circular radius is 37\,kpc with 34\,kpc intrinsic scatter. That is consistent with the offsets between the BCG and X-ray gas centroids or Dark Matter mass centroids.
	
	\item[B5.] The direction of the ICL offsets agrees to better than $60\degr$ with the direction of the cluster galaxies number density peak offset in 33/46 = 72\% of the clusters.
	
	\item[B6.] The ICL ellipticity increases with radius and reaches $\epsilon = 0.4 - 0.5$ at a circular radius of $r \approx 200$\,kpc. That is consistent with the ellipticity of cluster Dark Matter halos \citep{Shin2018}.
\end{enumerate}

The correlations between BCG+ICL brightness and various host cluster parameters show that BCG+ICL formation is tightly linked to the growth of their host clusters. The position angle, ellipticity and spatial offset alignment with the host clusters (Results B3 -- B6) qualify ICL as a potential Dark Matter tracer.

We are grateful to Stella Seitz, Rhea-Silvia Remus, Klaus Dolag, John Kormendy, and Walter Dehnen for helpful conversations. We also wish to thank the anonymous referee for his or her comments and suggestions that allowed us to significantly improve the paper.

The 2m telescope project was funded by the Bavarian government and by the German Federal government through a common funding process. Part of the 2m instrumentation including some of the upgrades for the infrastructure and the 40cm telescope housing were funded by the Cluster of Excellence "Origin of the Universe" of the German Science foundation DFG. The 40cm telescope was funded by Ludwig-Maximilians-University, Munich.

This work made use of data products based on observations made with the NASA/ESA \textit{Hubble Space Telescope}, and obtained from the Hubble Legacy Archive, which is a collaboration between the Space Telescope Science Institute (STScI/NASA), the Space Telescope European Coordinating Facility (ST-ECF/ESA), and the Canadian Astronomy Data Centre (CADC/NRC/CSA).

This work would not have been practical without extensive use of NASA's Astrophysics Data System Bibliographic Services and the SIMBAD database, operated at CDS, Strasbourg, France.

We also used the image display tool SAOImage DS9 developed by Smithsonian Astrophysical Observatory and the image display tool Fitsedit, developed by Johannes Koppenhoefer.

This research made use of Astropy, a community-developed core Python package for Astronomy \citep{Astropy2013}.

\facilities{WO:2m (Wide-field camera), \textit{HST} (WFPC2, ACS), Planck (HFI)}

\newpage

\appendix

\section{Host Cluster Parameters}\label{sec:hostclusterparams}

The host cluster properties are compiled in Table \ref{tab:clusterparams}. All parameters are defined in Section \ref{sec:clusterparams}.

\begin{longrotatetable}
	\begin{deluxetable}{lccccccccccc}
		\tablewidth{0pt}
		\tabletypesize{\footnotesize}
		
		\tablecaption{Host cluster parameters. \label{tab:clusterparams}}
		\tablehead{
			\colhead{Cluster} & \colhead{$\sigma_{\rm C}$} & \colhead{$S$} &  \colhead{$r_{\rm g}$} & \colhead{$M_{\rm sat}$} & \colhead{$\log(\mathcal{M}_{\rm g})$} & \colhead{$v_{\rm syst}$} & \colhead{$r_{\rm syst}$} & \colhead{$\log(\rho)$} & \colhead{$\log(s)$} & \colhead{$\log(f_{\mathcal{M_{\rm g}}})$} & \colhead{$\log(f_{\rm s})$} \\
			\colhead{} & \colhead{(km s$^{-1}$)} & \colhead{} &  \colhead{(kpc)} & \colhead{($g'$ mag)} & \colhead{($\log({\rm M}_{\odot})$)} & \colhead{(km s$^{-1}$)} & \colhead{(kpc)} & \colhead{($\log({\rm M}_{\odot} kpc^{-3})$)} & \colhead{($\log({\rm kpc^{-3}})$)} & \colhead{($\log({\rm M}_{\odot}$} & \colhead{($\log({\rm kpc^{-3}}$} \\
			\colhead{} & \colhead{} & \colhead{} &  \colhead{} & \colhead{} & \colhead{} & \colhead{} & \colhead{} & \colhead{} & \colhead{} & \colhead{${\rm kpc^{-3}~km^{-3}~s^3)}$)} & \colhead{${\rm km^{-3}~s^3)}$)} \\
		}
		\colnumbers
		\startdata
		A76    &  491 $\pm$  120 &  \ldots &  \ldots & -24.94 $\pm$ 0.40 & \ldots & 686 $\pm$ 149.0 & 467 & \ldots & \ldots & \ldots & \ldots \\
		A85    & 1009 $\pm$   31 &  314 & 1324 $\pm$  195 & -25.58 $\pm$ 0.40 & 15.07 $\pm$ 0.07 & 41 $\pm$ 83.5 & 1 & 5.08 $\pm$ 0.07 & -6.87 $\pm$ 0.11 & -3.93 $\pm$ 0.07 & -15.88 $\pm$ 0.11 \\
		A150   &  664 $\pm$  151 &  \ldots &  \ldots & -25.59 $\pm$ 0.40 & \ldots & 4 $\pm$ 194.7 & 34 & \ldots & \ldots & \ldots & \ldots \\
		A152   &  844 $\pm$   59 &  \ldots &  \ldots & -24.90 $\pm$ 1.18 & \ldots & 120 $\pm$ 105.0 & \ldots & \ldots & \ldots & \ldots & \ldots \\
		A154   &  988 $\pm$  146 &  \ldots &  \ldots & -25.61 $\pm$ 0.40 & \ldots & \ldots & \ldots & \ldots & \ldots & \ldots & \ldots \\
		A158   &  \ldots &  \ldots &  \ldots & -25.43 $\pm$ 0.55 & \ldots & \ldots & \ldots & \ldots & \ldots & \ldots & \ldots \\
		A160   &  \ldots &  122 &  923 $\pm$  100 & -24.67 $\pm$ 0.40 & \ldots & \ldots & \ldots & \ldots & -6.81 $\pm$ 0.10 & \ldots & \ldots \\
		A161   &  \ldots &  \ldots &  \ldots & -25.49 $\pm$ 0.86 & \ldots & 0 $\pm$ 63.6 & \ldots & \ldots & \ldots & \ldots & \ldots \\
		A171   &  \ldots &  \ldots &  \ldots & -24.00 $\pm$ 1.33 & \ldots & 0 $\pm$ 63.6 & \ldots & \ldots & \ldots & \ldots & \ldots \\
		A174   &  \ldots &  \ldots &  \ldots & -25.57 $\pm$ 0.70 & \ldots & 0 $\pm$ 60.8 & \ldots & \ldots & \ldots & \ldots & \ldots \\
		A179   &  284 $\pm$  100 &  \ldots &  \ldots & -24.98 $\pm$ 0.53 & \ldots & \ldots & \ldots & \ldots & \ldots & \ldots & \ldots \\
		A193   &  776 $\pm$   62 &   93 &  968 $\pm$  100 & -25.04 $\pm$ 0.40 & 14.71 $\pm$ 0.06 & 135 $\pm$ 89.1 & 1 & 5.13 $\pm$ 0.06 & -6.99 $\pm$ 0.10 & -3.54 $\pm$ 0.06 & -15.66 $\pm$ 0.11 \\
		A225   &  660 $\pm$  272 &  \ldots &  \ldots & -25.31 $\pm$ 0.81 & \ldots & 13 $\pm$ 272.1 & \ldots & \ldots & \ldots & \ldots & \ldots \\
		A240   &  \ldots &  \ldots &  \ldots & -25.07 $\pm$ 0.41 & \ldots & 0 $\pm$ 50.9 & \ldots & \ldots & \ldots & \ldots & \ldots \\
		A245   &  \ldots &  \ldots &  \ldots & -25.31 $\pm$ 0.96 & \ldots & \ldots & \ldots & \ldots & \ldots & \ldots & \ldots \\
		A257   &  499 $\pm$   42 &  \ldots &  \ldots & -24.94 $\pm$ 1.57 & \ldots & 89 $\pm$ 83.4 & \ldots & \ldots & \ldots & \ldots & \ldots \\
		A260   &  754 $\pm$   74 &  \ldots &  \ldots & -24.38 $\pm$ 1.15 & \ldots & 289 $\pm$ 92.0 & \ldots & \ldots & \ldots & \ldots & \ldots \\
		A262   &  540 $\pm$   38 &  150 & 1313 $\pm$  100 & -23.11 $\pm$ 0.93 & 14.52 $\pm$ 0.05 & 47 $\pm$ 60.1 & 2 & 4.55 $\pm$ 0.05 & -7.18 $\pm$ 0.09 & -3.65 $\pm$ 0.05 & -15.38 $\pm$ 0.10 \\
		A292   &  \ldots &  \ldots &  \ldots & -24.96 $\pm$ 0.81 & \ldots & 0 $\pm$ 67.9 & \ldots & \ldots & \ldots & \ldots & \ldots \\
		A347   &  627 $\pm$   61 &  \ldots &  \ldots & -23.61 $\pm$ 0.40 & \ldots & 394 $\pm$ 81.2 & \ldots & \ldots & \ldots & \ldots & \ldots \\
		A376   &  830 $\pm$   59 &  135 & 1303 $\pm$  100 & -24.90 $\pm$ 1.23 & 14.89 $\pm$ 0.05 & 166 $\pm$ 80.5 & 75 & 4.93 $\pm$ 0.05 & -7.21 $\pm$ 0.09 & -3.83 $\pm$ 0.05 & -15.97 $\pm$ 0.10 \\
		A397   &  638 $\pm$   85 &  \ldots &  \ldots & -24.41 $\pm$ 1.09 & \ldots & 461 $\pm$ 112.3 & \ldots & \ldots & \ldots & \ldots & \ldots \\
		A399   & 1224 $\pm$   62 &  \ldots &  \ldots & -25.73 $\pm$ 0.51 & \ldots & 135 $\pm$ 120.4 & 26 & \ldots & \ldots & \ldots & \ldots \\
		A400   &  683 $\pm$   39 &  133 &  887 $\pm$  100 & -24.10 $\pm$ 0.49 & 14.56 $\pm$ 0.06 & 31 $\pm$ 63.8 & 5 & 5.09 $\pm$ 0.06 & -6.72 $\pm$ 0.10 & -3.41 $\pm$ 0.06 & -15.22 $\pm$ 0.10 \\
		A407   &  762 $\pm$   62 &  \ldots &  \ldots & -26.11 $\pm$ 0.44 & \ldots & 187 $\pm$ 139.0 & \ldots & \ldots & \ldots & \ldots & \ldots \\
		A426   &  \ldots &  254 &  721 $\pm$  139 & -23.88 $\pm$ 0.96 & \ldots & \ldots & \ldots & \ldots & -6.17 $\pm$ 0.12 & \ldots & \ldots \\
		A498   &  \ldots &  \ldots &  \ldots & -24.64 $\pm$ 0.56 & \ldots & 0 $\pm$ 77.8 & \ldots & \ldots & \ldots & \ldots & \ldots \\
		A505   &  \ldots &  \ldots &  \ldots & -26.02 $\pm$ 0.40 & \ldots & 0 $\pm$ 66.5 & \ldots & \ldots & \ldots & \ldots & \ldots \\
		A539   &  833 $\pm$   40 &  161 & 1097 $\pm$  265 & -24.34 $\pm$ 1.01 & 14.82 $\pm$ 0.11 & \ldots & \ldots & 5.08 $\pm$ 0.11 & -6.91 $\pm$ 0.14 & -3.68 $\pm$ 0.11 & -15.68 $\pm$ 0.14 \\
		A553   &  \ldots &  \ldots &  \ldots & -26.44 $\pm$ 0.47 & \ldots & \ldots & \ldots & \ldots & \ldots & \ldots & \ldots \\
		A559   &  \ldots &  \ldots &  \ldots & -25.62 $\pm$ 0.48 & \ldots & \ldots & \ldots & \ldots & \ldots & \ldots & \ldots \\
		A568   &  687 $\pm$  100 &  \ldots &  \ldots & -25.87 $\pm$ 0.74 & \ldots & 471 $\pm$ 263.4 & \ldots & \ldots & \ldots & \ldots & \ldots \\
		A569   &  394 $\pm$   25 &  \ldots &  \ldots & -22.20 $\pm$ 2.78 & \ldots & \ldots & \ldots & \ldots & \ldots & \ldots & \ldots \\
		A582   &  324 $\pm$   56 &  \ldots &  \ldots & -25.60 $\pm$ 1.02 & \ldots & 182 $\pm$ 74.2 & \ldots & \ldots & \ldots & \ldots & \ldots \\
		A592   &  123 $\pm$  100 &  \ldots &  \ldots & -25.93 $\pm$ 0.40 & \ldots & 149 $\pm$ 65.2 & \ldots & \ldots & \ldots & \ldots & \ldots \\
		A595   &  601 $\pm$   56 &  \ldots &  \ldots & -25.16 $\pm$ 1.09 & \ldots & \ldots & \ldots & \ldots & \ldots & \ldots & \ldots \\
		A600   &  \ldots &  \ldots &  \ldots & -25.25 $\pm$ 1.00 & \ldots & 0 $\pm$ 48.1 & \ldots & \ldots & \ldots & \ldots & \ldots \\
		A602   &  796 $\pm$   61 &   69 & 1476 $\pm$  100 & -25.14 $\pm$ 0.75 & 14.91 $\pm$ 0.04 & \ldots & \ldots & 4.78 $\pm$ 0.04 & -7.67 $\pm$ 0.09 & -3.92 $\pm$ 0.04 & -16.37 $\pm$ 0.10 \\
		A607   &  \ldots &  \ldots &  \ldots & -25.14 $\pm$ 1.84 & \ldots & \ldots & \ldots & \ldots & \ldots & \ldots & \ldots \\
		A612   &  \ldots &  \ldots &  \ldots & -25.62 $\pm$ 0.89 & \ldots & \ldots & \ldots & \ldots & \ldots & \ldots & \ldots \\
		A634   &  331 $\pm$   25 &  \ldots &  \ldots & -23.56 $\pm$ 0.40 & \ldots & 218 $\pm$ 40.2 & \ldots & \ldots & \ldots & \ldots & \ldots \\
		A671   &  850 $\pm$   33 &  133 & 1314 $\pm$  181 & -25.36 $\pm$ 0.40 & 14.92 $\pm$ 0.06 & 93 $\pm$ 141.5 & 111 & 4.94 $\pm$ 0.06 & -7.23 $\pm$ 0.11 & -3.85 $\pm$ 0.06 & -16.02 $\pm$ 0.11 \\
		A688   &  \ldots &  \ldots &  \ldots & -25.54 $\pm$ 2.41 & \ldots & \ldots & \ldots & \ldots & \ldots & \ldots & \ldots \\
		A690   &  546 $\pm$   46 &   79 & 1512 $\pm$  275 & -25.11 $\pm$ 0.63 & 14.59 $\pm$ 0.09 & 295 $\pm$ 258.6 & \ldots & 4.43 $\pm$ 0.09 & -7.64 $\pm$ 0.12 & -3.78 $\pm$ 0.09 & -15.85 $\pm$ 0.13 \\
		A695   &  402 $\pm$   52 &  \ldots &  \ldots & \ldots & \ldots & 278 $\pm$ 116.0 & \ldots & \ldots & \ldots & \ldots & \ldots \\
		A734   &  \ldots &  \ldots &  \ldots & -25.29 $\pm$ 0.92 & \ldots & \ldots & \ldots & \ldots & \ldots & \ldots & \ldots \\
		A744   &  445 $\pm$  920 &  \ldots &  \ldots & -25.02 $\pm$ 1.40 & \ldots & 21 $\pm$ 80.0 & 2 & \ldots & \ldots & \ldots & \ldots \\
		A757   &  360 $\pm$   32 &  \ldots &  \ldots & -24.60 $\pm$ 0.74 & \ldots & 5 $\pm$ 46.0 & \ldots & \ldots & \ldots & \ldots & \ldots \\
		A834   &  392 $\pm$  100 &  \ldots &  \ldots & \ldots & \ldots & 397 $\pm$ 84.4 & \ldots & \ldots & \ldots & \ldots & \ldots \\
		A883  &  \ldots &  \ldots &  \ldots & \ldots & \ldots & \ldots & \ldots & \ldots & \ldots & \ldots & \ldots \\
		A999   &  286 $\pm$   25 &  \ldots &  \ldots & -23.69 $\pm$ 0.78 & \ldots & 129 $\pm$ 39.3 & \ldots & \ldots & \ldots & \ldots & \ldots \\
		A1003  &  501 $\pm$   50 &  \ldots &  \ldots & -24.75 $\pm$ 0.83 & \ldots & 427 $\pm$ 82.4 & \ldots & \ldots & \ldots & \ldots & \ldots \\
		A1016  &  204 $\pm$   53 &  \ldots &  \ldots & -23.73 $\pm$ 0.84 & \ldots & 18 $\pm$ 30.4 & \ldots & \ldots & \ldots & \ldots & \ldots \\
		A1020  &  314 $\pm$   41 &  \ldots &  \ldots & -25.26 $\pm$ 0.81 & \ldots & 13 $\pm$ 68.7 & 481 & \ldots & \ldots & \ldots & \ldots \\
		A1056  &  \ldots &  \ldots &  \ldots & -26.85 $\pm$ 0.46 & \ldots & \ldots & \ldots & \ldots & \ldots & \ldots & \ldots \\
		A1066  &  817 $\pm$   55 &  \ldots &  \ldots & -25.55 $\pm$ 0.40 & \ldots & 498 $\pm$ 79.3 & \ldots & \ldots & \ldots & \ldots & \ldots \\
		A1100  &  451 $\pm$  100 &  \ldots &  \ldots & -24.57 $\pm$ 0.40 & \ldots & 34 $\pm$ 58.6 & \ldots & \ldots & \ldots & \ldots & \ldots \\
		A1108  &  \ldots &  \ldots &  \ldots & \ldots & \ldots & \ldots & \ldots & \ldots & \ldots & \ldots & \ldots \\
		A1142  &  757 $\pm$   44 &  \ldots &  \ldots & -24.19 $\pm$ 0.40 & \ldots & 457 $\pm$ 88.1 & 52 & \ldots & \ldots & \ldots & \ldots \\
		A1155  &  277 $\pm$   41 &  \ldots &  \ldots & -23.77 $\pm$ 3.73 & \ldots & 76 $\pm$ 108.3 & \ldots & \ldots & \ldots & \ldots & \ldots \\
		A1173  &  \ldots &  \ldots &  \ldots & -23.39 $\pm$ 2.52 & \ldots & \ldots & \ldots & \ldots & \ldots & \ldots & \ldots \\
		A1177  &  331 $\pm$   59 &  \ldots &  \ldots & -23.24 $\pm$ 0.92 & \ldots & 36 $\pm$ 55.5 & \ldots & \ldots & \ldots & \ldots & \ldots \\
		A1185  &  758 $\pm$   54 &  292 & 1335 $\pm$  283 & -24.86 $\pm$ 0.40 & 14.83 $\pm$ 0.10 & 730 $\pm$ 63.5 & 161 & 4.83 $\pm$ 0.10 & -6.91 $\pm$ 0.13 & -3.81 $\pm$ 0.10 & -15.55 $\pm$ 0.13 \\
		A1187  &  952 $\pm$   55 &  \ldots &  \ldots & -25.04 $\pm$ 1.43 & \ldots & 1317 $\pm$ 118.0 & 428 & \ldots & \ldots & \ldots & \ldots \\
		A1190  &  671 $\pm$   43 &   60 & 1106 $\pm$  260 & -25.55 $\pm$ 0.60 & 14.64 $\pm$ 0.11 & 905 $\pm$ 73.4 & \ldots & 4.88 $\pm$ 0.11 & -7.35 $\pm$ 0.14 & -3.60 $\pm$ 0.11 & -15.83 $\pm$ 0.14 \\
		A1203  &  552 $\pm$   36 &  \ldots &  \ldots & -24.51 $\pm$ 0.68 & \ldots & 90 $\pm$ 66.9 & \ldots & \ldots & \ldots & \ldots & \ldots \\
		A1213  &  572 $\pm$   43 &  154 & 1132 $\pm$  246 & -25.24 $\pm$ 0.40 & 14.51 $\pm$ 0.10 & 516 $\pm$ 61.9 & \ldots & 4.73 $\pm$ 0.10 & -6.97 $\pm$ 0.13 & -3.55 $\pm$ 0.10 & -15.25 $\pm$ 0.14 \\
		A1218  &  \ldots &  \ldots &  \ldots & -25.90 $\pm$ 0.57 & \ldots & \ldots & \ldots & \ldots & \ldots & \ldots & \ldots \\
		A1228  &  246 $\pm$   23 &  \ldots &  \ldots & -24.62 $\pm$ 0.40 & \ldots & \ldots & \ldots & \ldots & \ldots & \ldots & \ldots \\
		A1257  & 1202 $\pm$   58 &  \ldots &  \ldots & -24.23 $\pm$ 0.43 & \ldots & \ldots & \ldots & \ldots & \ldots & \ldots & \ldots \\
		A1270  &  \ldots &  \ldots &  \ldots & -24.98 $\pm$ 1.33 & \ldots & \ldots & \ldots & \ldots & \ldots & \ldots & \ldots \\
		A1275  &  \ldots &  \ldots &  \ldots & -25.59 $\pm$ 0.42 & \ldots & \ldots & \ldots & \ldots & \ldots & \ldots & \ldots \\
		A1279  &  186 $\pm$   30 &  \ldots &  \ldots & -24.08 $\pm$ 1.05 & \ldots & 35 $\pm$ 42.0 & \ldots & \ldots & \ldots & \ldots & \ldots \\
		A1314  &  648 $\pm$   25 &   76 &  909 $\pm$  100 & -25.14 $\pm$ 0.40 & 14.52 $\pm$ 0.05 & 122 $\pm$ 65.8 & 71 & 5.02 $\pm$ 0.05 & -7.00 $\pm$ 0.10 & -3.41 $\pm$ 0.05 & -15.43 $\pm$ 0.10 \\
		A1324  &  \ldots &  \ldots &  \ldots & -25.81 $\pm$ 1.63 & \ldots & \ldots & \ldots & \ldots & \ldots & \ldots & \ldots \\
		A1356  &  \ldots &  \ldots &  \ldots & -25.95 $\pm$ 1.51 & \ldots & \ldots & \ldots & \ldots & \ldots & \ldots & \ldots \\
		A1365  &  369 $\pm$   61 &  \ldots &  \ldots & -25.45 $\pm$ 1.11 & \ldots & 195 $\pm$ 88.1 & 91 & \ldots & \ldots & \ldots & \ldots \\
		A1367  &  872 $\pm$   42 &  374 & 1176 $\pm$  234 & -23.95 $\pm$ 1.11 & 14.89 $\pm$ 0.09 & 286 $\pm$ 62.6 & 354 & 5.06 $\pm$ 0.09 & -6.64 $\pm$ 0.13 & -3.76 $\pm$ 0.09 & -15.46 $\pm$ 0.13 \\
		A1371  &  577 $\pm$   50 &  \ldots &  \ldots & \ldots & \ldots & \ldots & \ldots & \ldots & \ldots & \ldots & \ldots \\
		A1400  &  332 $\pm$   53 &  \ldots &  \ldots & -25.57 $\pm$ 0.57 & \ldots & \ldots & \ldots & \ldots & \ldots & \ldots & \ldots \\
		A1413  &  \ldots &  142 & 1219 $\pm$  146 & -27.39 $\pm$ 0.53 & \ldots & \ldots & \ldots & \ldots & -7.11 $\pm$ 0.10 & \ldots & \ldots \\
		A1423  &  \ldots &  \ldots &  \ldots & -24.57 $\pm$ 1.26 & \ldots & \ldots & \ldots & \ldots & \ldots & \ldots & \ldots \\
		A1424  &  697 $\pm$   55 &  \ldots &  \ldots & -25.61 $\pm$ 0.85 & \ldots & 442 $\pm$ 78.1 & 168 & \ldots & \ldots & \ldots & \ldots \\
		A1435  &  \ldots &  \ldots &  \ldots & -24.79 $\pm$ 0.51 & \ldots & \ldots & \ldots & \ldots & \ldots & \ldots & \ldots \\
		A1436  &  703 $\pm$   36 &  \ldots &  \ldots & -24.82 $\pm$ 1.37 & \ldots & \ldots & \ldots & \ldots & \ldots & \ldots & \ldots \\
		A1452  &  560 $\pm$   63 &  \ldots &  \ldots & -24.71 $\pm$ 0.62 & \ldots & \ldots & \ldots & \ldots & \ldots & \ldots & \ldots \\
		A1507  &  405 $\pm$   48 &  \ldots &  \ldots & -25.37 $\pm$ 0.40 & \ldots & 414 $\pm$ 58.3 & 139 & \ldots & \ldots & \ldots & \ldots \\
		A1516  &  \ldots &  \ldots &  \ldots & -25.08 $\pm$ 0.88 & \ldots & \ldots & \ldots & \ldots & \ldots & \ldots & \ldots \\
		A1526  &  \ldots &  \ldots &  \ldots & -25.47 $\pm$ 1.12 & \ldots & \ldots & \ldots & \ldots & \ldots & \ldots & \ldots \\
		A1534  &  371 $\pm$   55 &  \ldots &  \ldots & -25.93 $\pm$ 0.40 & \ldots & 25 $\pm$ 64.4 & 157 & \ldots & \ldots & \ldots & \ldots \\
		A1569  &  622 $\pm$ 1314 &   73 & 1002 $\pm$  226 & -25.02 $\pm$ 0.43 & \ldots & \ldots & \ldots & \ldots & -7.14 $\pm$ 0.14 & \ldots & \ldots \\
		A1589  &  899 $\pm$  546 &  \ldots &  \ldots & -26.01 $\pm$ 0.40 & \ldots & 688 $\pm$ 99.0 & \ldots & \ldots & \ldots & \ldots & \ldots \\
		A1610  &  292 $\pm$  403 &  \ldots &  \ldots & -24.82 $\pm$ 0.68 & \ldots & 485 $\pm$ 62.0 & \ldots & \ldots & \ldots & \ldots & \ldots \\
		A1656  & 1035 $\pm$   25 &  \ldots &  \ldots & -25.02 $\pm$ 0.40 & \ldots & \ldots & \ldots & \ldots & \ldots & \ldots & \ldots \\
		A1668  &  \ldots &   75 & 1426 $\pm$  129 & -25.02 $\pm$ 0.86 & \ldots & \ldots & \ldots & \ldots & -7.59 $\pm$ 0.10 & \ldots & \ldots \\
		A1691  &  784 $\pm$   45 &   93 & 1450 $\pm$  202 & -26.06 $\pm$ 0.40 & 14.89 $\pm$ 0.07 & 37 $\pm$ 80.6 & 45 & 4.78 $\pm$ 0.07 & -7.52 $\pm$ 0.11 & -3.90 $\pm$ 0.07 & -16.20 $\pm$ 0.11 \\
		A1749  &  707 $\pm$   66 &   92 & 1238 $\pm$  243 & -24.44 $\pm$ 1.07 & 14.73 $\pm$ 0.10 & 43 $\pm$ 94.3 & 28 & 4.83 $\pm$ 0.10 & -7.31 $\pm$ 0.13 & -3.72 $\pm$ 0.10 & -15.86 $\pm$ 0.13 \\
		A1767  &  887 $\pm$   31 &  \ldots &  \ldots & -25.87 $\pm$ 0.45 & \ldots & 44 $\pm$ 169.1 & 25 & \ldots & \ldots & \ldots & \ldots \\
		A1775  &  568 $\pm$   60 &   79 &  993 $\pm$  197 & -25.59 $\pm$ 0.40 & 14.45 $\pm$ 0.10 & 35 $\pm$ 74.9 & 48 & 4.83 $\pm$ 0.10 & -7.09 $\pm$ 0.13 & -3.43 $\pm$ 0.10 & -15.36 $\pm$ 0.13 \\
		A1795  &  861 $\pm$   56 &  175 & 1011 $\pm$  100 & -26.05 $\pm$ 0.53 & 14.82 $\pm$ 0.05 & 222 $\pm$ 112.1 & 12 & 5.18 $\pm$ 0.05 & -6.77 $\pm$ 0.10 & -3.63 $\pm$ 0.05 & -15.58 $\pm$ 0.10 \\
		A1800  &  767 $\pm$  190 &  \ldots &  \ldots & -25.46 $\pm$ 0.66 & \ldots & 36 $\pm$ 93.9 & 83 & \ldots & \ldots & \ldots & \ldots \\
		A1809  &  745 $\pm$   30 &   77 & 1105 $\pm$  191 & -26.30 $\pm$ 0.40 & 14.73 $\pm$ 0.08 & 204 $\pm$ 81.5 & 47 & 4.98 $\pm$ 0.08 & -7.24 $\pm$ 0.12 & -3.64 $\pm$ 0.08 & -15.86 $\pm$ 0.12 \\
		A1812  &  \ldots &  \ldots &  \ldots & -25.30 $\pm$ 1.25 & \ldots & \ldots & \ldots & \ldots & \ldots & \ldots & \ldots \\
		A1825  & 1024 $\pm$  100 &  \ldots &  \ldots & -25.20 $\pm$ 0.40 & \ldots & 869 $\pm$ 154.3 & \ldots & \ldots & \ldots & \ldots & \ldots \\
		A1828  &  388 $\pm$   84 &  \ldots &  \ldots & -24.52 $\pm$ 1.02 & \ldots & 94 $\pm$ 96.1 & \ldots & \ldots & \ldots & \ldots & \ldots \\
		A1831  & 1176 $\pm$  118 &  \ldots &  \ldots & -26.03 $\pm$ 0.94 & \ldots & 133 $\pm$ 118.3 & 51 & \ldots & \ldots & \ldots & \ldots \\
		A1890  &  550 $\pm$   59 &  \ldots &  \ldots & -25.45 $\pm$ 0.45 & \ldots & 275 $\pm$ 63.1 & 236 & \ldots & \ldots & \ldots & \ldots \\
		A1899  &  646 $\pm$  100 &  \ldots &  \ldots & \ldots & \ldots & 649 $\pm$ 97.2 & \ldots & \ldots & \ldots & \ldots & \ldots \\
		A1904  &  772 $\pm$   31 &  \ldots &  \ldots & -25.77 $\pm$ 0.40 & \ldots & 22 $\pm$ 66.0 & 289 & \ldots & \ldots & \ldots & \ldots \\
		A1913  &  636 $\pm$  130 &   86 & 1136 $\pm$  383 & -25.64 $\pm$ 0.40 & 14.60 $\pm$ 0.18 & \ldots & \ldots & 4.81 $\pm$ 0.18 & -7.23 $\pm$ 0.18 & -3.60 $\pm$ 0.18 & -15.64 $\pm$ 0.21 \\
		A1982  & 1325 $\pm$  100 &  \ldots &  \ldots & -24.11 $\pm$ 1.11 & \ldots & 281 $\pm$ 248.8 & \ldots & \ldots & \ldots & \ldots & \ldots \\
		A1983  &  541 $\pm$   27 &  184 & 1520 $\pm$  234 & -24.07 $\pm$ 0.67 & 14.59 $\pm$ 0.07 & \ldots & \ldots & 4.42 $\pm$ 0.07 & -7.28 $\pm$ 0.11 & -3.78 $\pm$ 0.07 & -15.48 $\pm$ 0.11 \\
		A2022  &  607 $\pm$   74 &  \ldots & \ldots & \ldots & \ldots & 416 $\pm$ 68 & 61 & \ldots & \ldots & \ldots & $\ldots$ \\
		A2029  & 1222 $\pm$   75 &  587 & 1625 $\pm$  323 & -25.63 $\pm$ 0.58 & 15.33 $\pm$ 0.09 & 215 $\pm$ 124.3 & 1 & 5.07 $\pm$ 0.09 & -6.86 $\pm$ 0.13 & -4.19 $\pm$ 0.09 & -16.12 $\pm$ 0.13 \\
		A2052  &  681 $\pm$   41 &  186 &  798 $\pm$  281 & -24.70 $\pm$ 0.47 & 14.51 $\pm$ 0.16 & 92 $\pm$ 66.5 & 0 & 5.18 $\pm$ 0.16 & -6.44 $\pm$ 0.19 & -3.32 $\pm$ 0.16 & -14.94 $\pm$ 0.19 \\
		A2061  &  851 $\pm$   28 &  164 & 1461 $\pm$  217 & -25.99 $\pm$ 0.40 & 14.96 $\pm$ 0.07 & 270 $\pm$ 74.2 & 176 & 4.85 $\pm$ 0.07 & -7.28 $\pm$ 0.11 & -3.94 $\pm$ 0.07 & -16.07 $\pm$ 0.11 \\
		A2063  &  930 $\pm$   57 &  189 & 1237 $\pm$  264 & -24.75 $\pm$ 0.40 & 14.97 $\pm$ 0.10 & 205 $\pm$ 85.9 & 11 & 5.07 $\pm$ 0.10 & -7.00 $\pm$ 0.13 & -3.84 $\pm$ 0.10 & -15.91 $\pm$ 0.13 \\
		A2065  & 1286 $\pm$  140 &  \ldots &  \ldots & -26.46 $\pm$ 0.40 & \ldots & 927 $\pm$ 110.0 & 105 & \ldots & \ldots & \ldots & \ldots \\
		A2107  &  629 $\pm$   46 &  134 & 1301 $\pm$  110 & -24.21 $\pm$ 0.64 & 14.65 $\pm$ 0.05 & 182 $\pm$ 89.2 & 1 & 4.69 $\pm$ 0.05 & -7.22 $\pm$ 0.10 & -3.71 $\pm$ 0.05 & -15.61 $\pm$ 0.10 \\
		A2122  &  \ldots &   99 & 1305 $\pm$  100 & -25.07 $\pm$ 0.40 & \ldots & \ldots & \ldots & \ldots & -7.35 $\pm$ 0.09 & \ldots & \ldots \\
		A2147  & 1033 $\pm$   33 &  362 & 1279 $\pm$  369 & -24.28 $\pm$ 1.55 & 15.08 $\pm$ 0.13 & 361 $\pm$ 68.4 & 0 & 5.13 $\pm$ 0.13 & -6.76 $\pm$ 0.16 & -3.91 $\pm$ 0.13 & -15.80 $\pm$ 0.16 \\
		A2151  &  842 $\pm$   30 &  336 &  474 $\pm$  106 & -24.93 $\pm$ 0.71 & 14.47 $\pm$ 0.10 & 385 $\pm$ 51.4 & 1 & 5.82 $\pm$ 0.10 & -5.50 $\pm$ 0.13 & -2.96 $\pm$ 0.10 & -14.28 $\pm$ 0.14 \\
		A2152  &  456 $\pm$   62 &  \ldots &  \ldots & -24.70 $\pm$ 0.92 & \ldots & 6 $\pm$ 77.3 & 63 & \ldots & \ldots & \ldots & \ldots \\
		A2162  &  435 $\pm$   37 &  \ldots &  \ldots & -23.82 $\pm$ 0.77 & \ldots & 98 $\pm$ 57.2 & 49 & \ldots & \ldots & \ldots & \ldots \\
		A2197  &  615 $\pm$   21 &  \ldots &  \ldots & \ldots & \ldots & 268 $\pm$ 41.8 & 886 & \ldots & \ldots & \ldots & \ldots \\
		A2199  &  819 $\pm$   32 &  714 & 1444 $\pm$  291 & -24.50 $\pm$ 0.87 & 14.93 $\pm$ 0.09 & 229 $\pm$ 39.3 & 1 & 4.83 $\pm$ 0.09 & -6.63 $\pm$ 0.13 & -3.91 $\pm$ 0.09 & -15.37 $\pm$ 0.13 \\
		A2247  &  353 $\pm$   59 &  \ldots &  \ldots & -23.94 $\pm$ 0.96 & \ldots & \ldots & \ldots & \ldots & \ldots & \ldots & \ldots \\
		A2248  & 1224 $\pm$ 1758 &  \ldots &  \ldots & -25.59 $\pm$ 0.40 & \ldots & \ldots & \ldots & \ldots & \ldots & \ldots & \ldots \\
		A2255  &  \ldots &  457 & 1612 $\pm$  260 & -26.66 $\pm$ 0.46 & \ldots & \ldots & \ldots & \ldots & -6.96 $\pm$ 0.11 & \ldots & \ldots \\
		A2256  & 1301 $\pm$   42 &  392 & 1501 $\pm$  109 & -26.59 $\pm$ 0.40 & 15.35 $\pm$ 0.03 & 347 $\pm$ 91.8 & 143 & 5.19 $\pm$ 0.03 & -6.94 $\pm$ 0.09 & -4.15 $\pm$ 0.03 & -16.28 $\pm$ 0.09 \\
		A2271  &  538 $\pm$  135 &  \ldots &  \ldots & -24.96 $\pm$ 0.82 & \ldots & 171 $\pm$ 168.1 & 17 & \ldots & \ldots & \ldots & \ldots \\
		A2293  &  754 $\pm$  100 &  \ldots &  \ldots & \ldots & \ldots & \ldots & \ldots & \ldots & \ldots & \ldots & \ldots \\
		A2308  &  \ldots &  \ldots &  \ldots & -26.39 $\pm$ 0.40 & \ldots & 0 $\pm$ 55.2 & \ldots & \ldots & \ldots & \ldots & \ldots \\
		A2319  &  \ldots &  334 & 1651 $\pm$  100 & -25.63 $\pm$ 0.53 & \ldots & \ldots & \ldots & \ldots & -7.13 $\pm$ 0.09 & \ldots & \ldots \\
		A2388  &  \ldots &  \ldots &  \ldots & -25.05 $\pm$ 1.50 & \ldots & 0 $\pm$ 50.9 & \ldots & \ldots & \ldots & \ldots & \ldots \\
		A2469  &  \ldots &  \ldots &  \ldots & -24.82 $\pm$ 1.15 & \ldots & \ldots & \ldots & \ldots & \ldots & \ldots & \ldots \\
		A2495  &  638 $\pm$  188 &  168 & 1506 $\pm$  267 & -26.23 $\pm$ 0.60 & 14.73 $\pm$ 0.16 & 151 $\pm$ 216.7 & \ldots & 4.57 $\pm$ 0.16 & -7.31 $\pm$ 0.12 & -3.84 $\pm$ 0.16 & -15.72 $\pm$ 0.18 \\
		A2506  &  \ldots &  \ldots &  \ldots & \ldots & \ldots & \ldots & \ldots & \ldots & \ldots & \ldots & \ldots \\
		A2513  &  \ldots &  \ldots &  \ldots & -22.74 $\pm$ 1.22 & \ldots & \ldots & \ldots & \ldots & \ldots & \ldots & \ldots \\
		A2516  &  \ldots &  \ldots &  \ldots & \ldots & \ldots & \ldots & \ldots & \ldots & \ldots & \ldots & \ldots \\
		A2524  &  627 $\pm$  175 &  \ldots &  \ldots & -26.03 $\pm$ 0.43 & \ldots & 402 $\pm$ 176.1 & \ldots & \ldots & \ldots & \ldots & \ldots \\
		A2558  &  \ldots &  \ldots &  \ldots & -24.92 $\pm$ 1.02 & \ldots & 0 $\pm$ 53.7 & \ldots & \ldots & \ldots & \ldots & \ldots \\
		A2572  &  593 $\pm$   36 &  \ldots &  \ldots & -23.58 $\pm$ 1.73 & \ldots & \ldots & \ldots & \ldots & \ldots & \ldots & \ldots \\
		A2589  &  872 $\pm$   60 &  164 & 1123 $\pm$  100 & -24.45 $\pm$ 0.40 & 14.87 $\pm$ 0.05 & 48 $\pm$ 101.5 & 3 & 5.10 $\pm$ 0.05 & -6.94 $\pm$ 0.10 & -3.72 $\pm$ 0.05 & -15.76 $\pm$ 0.10 \\
		A2593  &  644 $\pm$   23 &  182 & 1216 $\pm$  316 & -24.79 $\pm$ 0.40 & 14.64 $\pm$ 0.12 & 25 $\pm$ 51.3 & 10 & 4.77 $\pm$ 0.12 & -7.00 $\pm$ 0.15 & -3.66 $\pm$ 0.12 & -15.42 $\pm$ 0.15 \\
		A2618  &  \ldots &  \ldots &  \ldots & -25.02 $\pm$ 1.88 & \ldots & 0 $\pm$ 50.9 & \ldots & \ldots & \ldots & \ldots & \ldots \\
		A2622  &  860 $\pm$  121 &   59 & 1068 $\pm$  100 & -24.88 $\pm$ 1.04 & 14.84 $\pm$ 0.07 & 121 $\pm$ 120.2 & 55 & 5.13 $\pm$ 0.07 & -7.31 $\pm$ 0.10 & -3.67 $\pm$ 0.07 & -16.12 $\pm$ 0.12 \\
		A2625  & 1506 $\pm$  171 &  \ldots &  \ldots & -23.82 $\pm$ 2.06 & \ldots & 863 $\pm$ 239.5 & 2065 & \ldots & \ldots & \ldots & \ldots \\
		A2626  &  648 $\pm$   53 &  116 & 1471 $\pm$  144 & -24.57 $\pm$ 1.14 & 14.73 $\pm$ 0.06 & 54 $\pm$ 82.3 & 2 & 4.61 $\pm$ 0.06 & -7.44 $\pm$ 0.10 & -3.83 $\pm$ 0.06 & -15.87 $\pm$ 0.11 \\
		A2630  &  420 $\pm$ 1336 &  \ldots &  \ldots & -26.28 $\pm$ 0.40 & \ldots & \ldots & \ldots & \ldots & \ldots & \ldots & \ldots \\
		A2634  &  919 $\pm$   45 &  236 & 1126 $\pm$  147 & -24.34 $\pm$ 0.90 & 14.92 $\pm$ 0.06 & 238 $\pm$ 79.3 & 24 & 5.14 $\pm$ 0.06 & -6.78 $\pm$ 0.11 & -3.75 $\pm$ 0.06 & -15.67 $\pm$ 0.11 \\
		A2637  &  361 $\pm$   59 &  \ldots &  \ldots & -24.63 $\pm$ 1.21 & \ldots & 179 $\pm$ 97.3 & 177 & \ldots & \ldots & \ldots & \ldots \\
		A2657  &  807 $\pm$   52 &   61 & 1046 $\pm$  100 & -24.51 $\pm$ 0.48 & 14.77 $\pm$ 0.05 & \ldots & \ldots & 5.09 $\pm$ 0.05 & -7.27 $\pm$ 0.10 & -3.63 $\pm$ 0.05 & -15.99 $\pm$ 0.10 \\
		A2665  &  \ldots &  118 & 1517 $\pm$  100 & -24.46 $\pm$ 0.70 & \ldots & 0 $\pm$ 80.6 & \ldots & \ldots & -7.47 $\pm$ 0.09 & \ldots & \ldots \\
		A2666  &  377 $\pm$   47 &  104 &  728 $\pm$  467 & -22.54 $\pm$ 1.70 & 13.95 $\pm$ 0.34 & 75 $\pm$ 78.8 & 65 & 4.75 $\pm$ 0.34 & -6.57 $\pm$ 0.35 & -2.98 $\pm$ 0.34 & -14.30 $\pm$ 0.36 \\
		A2675  &  372 $\pm$  156 &  \ldots &  \ldots & -24.55 $\pm$ 1.79 & \ldots & 609 $\pm$ 125.0 & 176 & \ldots & \ldots & \ldots & \ldots \\
		A2678  &  361 $\pm$  156 &  \ldots &  \ldots & \ldots & \ldots & \ldots & \ldots & \ldots & \ldots & \ldots & \ldots \\
		AWM1   &  \ldots &  116 & 1397 $\pm$  225 & -24.11 $\pm$ 0.40 & \ldots & \ldots & \ldots & \ldots & -7.37 $\pm$ 0.11 & \ldots & \ldots \\
		AWM5   &  \ldots &   88 & 1204 $\pm$  100 & -24.32 $\pm$ 1.09 & \ldots & \ldots & \ldots & \ldots & -7.30 $\pm$ 0.10 & \ldots & \ldots \\
		AWM7   &  \ldots &  193 & 1171 $\pm$  208 & -23.66 $\pm$ 0.40 & \ldots & \ldots & \ldots & \ldots & -6.92 $\pm$ 0.12 & \ldots & \ldots \\
		L2027  &  \ldots &  \ldots &  \ldots & -25.29 $\pm$ 0.73 & \ldots & \ldots & \ldots & \ldots & \ldots & \ldots & \ldots \\
		L2030  &  \ldots &  \ldots &  \ldots & -22.97 $\pm$ 1.43 & \ldots & \ldots & \ldots & \ldots & \ldots & \ldots & \ldots \\
		L2069  &  \ldots &  \ldots &  \ldots & -24.48 $\pm$ 1.78 & \ldots & \ldots & \ldots & \ldots & \ldots & \ldots & \ldots \\
		L2093  &  \ldots &  \ldots &  \ldots & -25.59 $\pm$ 0.40 & \ldots & \ldots & \ldots & \ldots & \ldots & \ldots & \ldots \\
		L2211  &  \ldots &  \ldots &  \ldots & -24.48 $\pm$ 0.67 & \ldots & \ldots & \ldots & \ldots & \ldots & \ldots & \ldots \\
		L3009  &  \ldots &  \ldots &  \ldots & \ldots & \ldots & \ldots & \ldots & \ldots & \ldots & \ldots & \ldots \\
		L3055  &  \ldots &  \ldots &  \ldots & -24.25 $\pm$ 1.44 & \ldots & \ldots & \ldots & \ldots & \ldots & \ldots & \ldots \\
		L3152  &  \ldots &  \ldots &  \ldots & -23.68 $\pm$ 0.40 & \ldots & \ldots & \ldots & \ldots & \ldots & \ldots & \ldots \\
		L3186  &  \ldots &  \ldots &  \ldots & \ldots & \ldots & \ldots & \ldots & \ldots & \ldots & \ldots & \ldots \\
		MKW4   &  \ldots &  \ldots &  \ldots & -22.56 $\pm$ 1.09 & \ldots & \ldots & \ldots & \ldots & \ldots & \ldots & \ldots \\
		\enddata
		\tablecomments{Host cluster parameters. The columns show the velocity dispersion of the cluster galaxies $\sigma_{\rm C}$ (2), richness $S$, that is, number of cluster galaxies (3), gravitational radius $r_{\rm g}$ (4), integrated absolute brightness of all cluster galaxies $M_{\rm sat}$ excluding the BCG+ICL (5), gravitational mass $\mathcal{M_{\rm g}}$ (6), systemic velocity offset $v_{\rm syst}$ (7), radial X-ray emission peak offset $r_{\rm syst}$ (8), mass density $\rho$ (9), number density of cluster galaxies $s$ (10), mass phase space density $f_{\mathcal{M_{\rm g}}}$ (11), and number phase space density of the cluster galaxies $f_{\rm s}$ (12).}
	\end{deluxetable}
\end{longrotatetable}

\section{BCG/ICL vs. host cluster correlation plots} \label{sec:corrhost}

\begin{figure*}
	\centering
	\includegraphics[width=\linewidth]{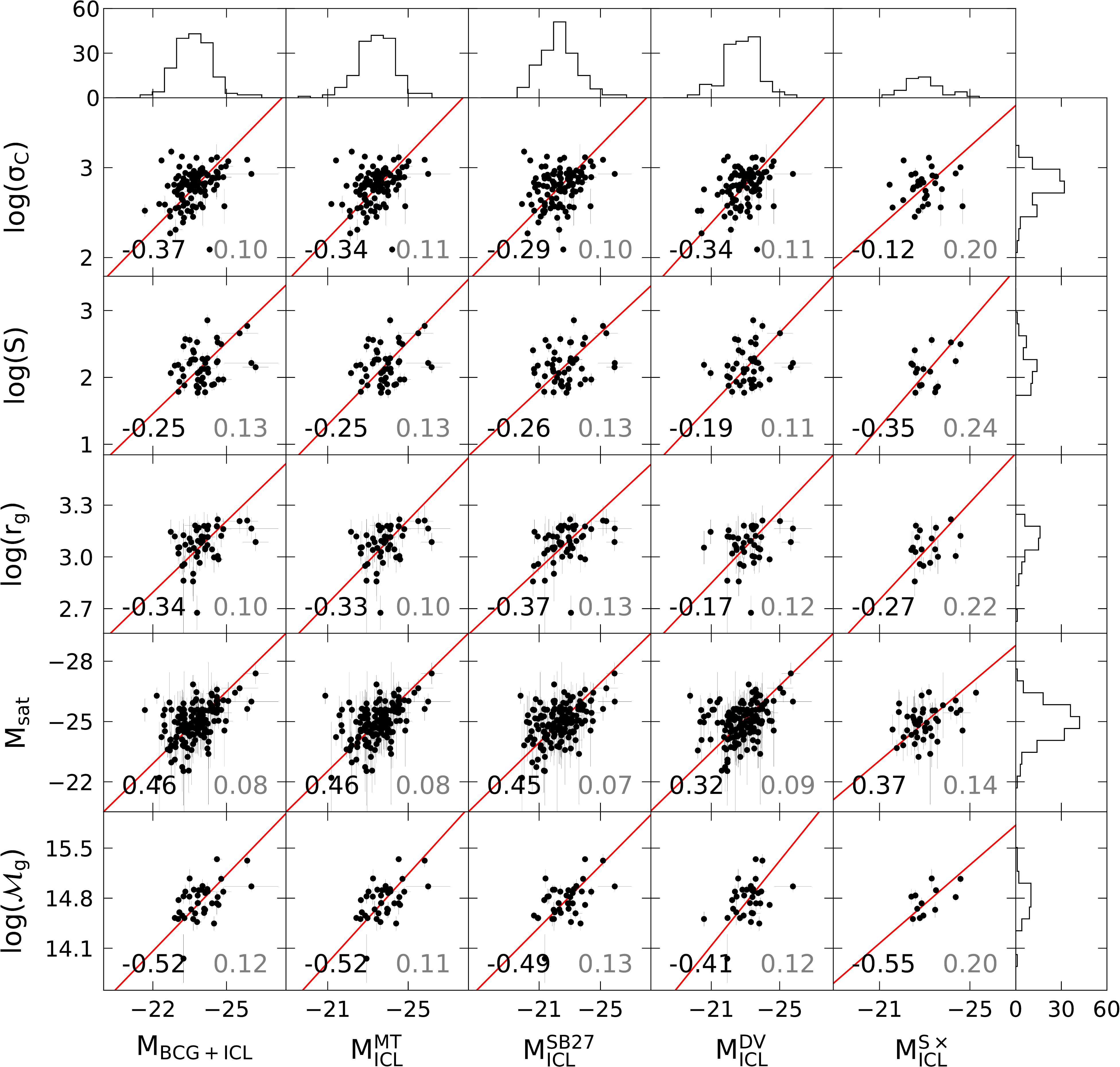}
	\caption{Correlations between BCG/ICL parameters (horizontal) and cluster parameters (vertical). The columns show (1) the absolute brightness of the BCGs+ICL $M_{\rm BCG+ICL}$ [$g'$ mag], the absolute brightness of the ICL only $M_{\rm ICL}$ [$g'$ mag], dissected (2) via the total magnitude threshold of --21.85 $g'$ mag $M_{\rm ICL}^{\rm MT}$, (3) via the surface brightness threshold of 27 $g'$ mag $M_{\rm ICL}^{\rm SB27}$, (4) via the light excess above the inner de Vaucouleurs fit $M_{\rm ICL}^{\rm DV}$, and (5) via the double S\'ersic fit $M_{\rm ICL}^{\rm S\times}$. The methods are explained in Section \ref{sec:iclfrac}. The rows show (1) the velocity dispersion of the satellite galaxies $\sigma_{\rm C}$ [km s$^{-1}$] (taken from \citealt{Lauer2014}), (2) richness $S$, that is, number of satellite galaxies, (3) gravitational radius $r_{\rm g}$ [kpc], (4) integrated absolute brightness of all satellite galaxies (excluding the BCG+ICL) $M_{\rm sat}$ [$g'$ mag], and (5) gravitational mass $\mathcal{M_{\rm g}} [{\rm M}_{\odot}$]. The Pearson coefficient for each correlation is given as a black label and its error (calculated using 10\,000 bootstraps) as a gray label. The histograms show the number of data points in each bin from the subplot containing either $M_{\rm sat}$ or $M_{\rm BCG+ICL}$. \label{fig:correlateall}}
\end{figure*}

\begin{figure*}
	\centering
	\includegraphics[width=\linewidth]{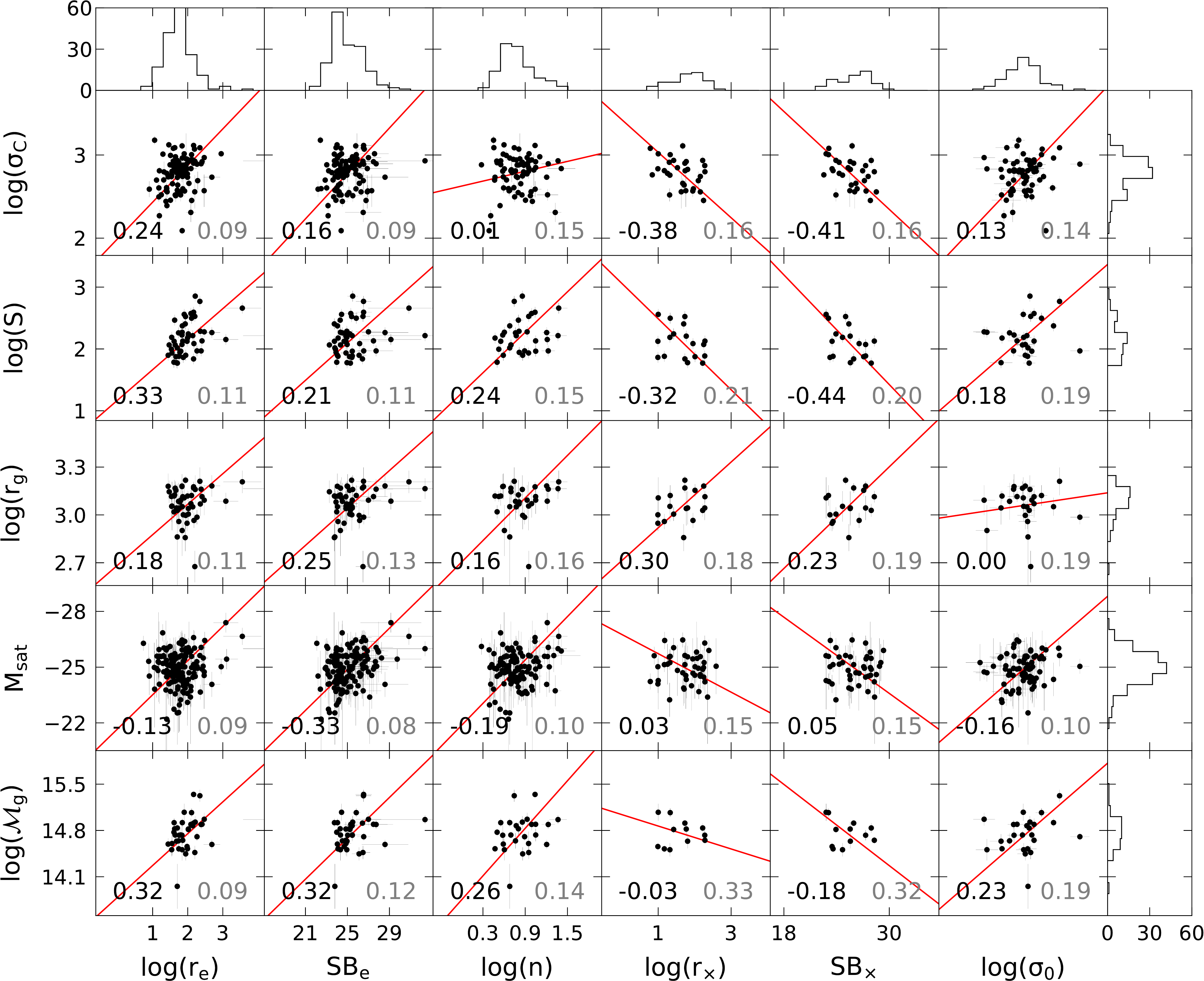}
	\caption{Correlations between BCG/ICL parameters (horizontal) and cluster parameters (vertical). The columns show (1) the effective radius $r_{\rm e}$ [kpc] along the major axis and (2) corresponding effective surface brightnesses ${\rm SB_e}$ [$g'$ mag arcsec$^{-2}$], (3) S\'ersic indices $n$ of the SS BCGs, (4) transition radii $r_{\times}$ [kpc] and (5) transition surface brightnesses ${\rm SB}_{\times}$ [$g'$ mag arcsec$^{-2}$] between the two S\'ersic profiles of the DS BCGs, and (6) central velocity dispersion (data taken from \citealt{Lauer2014}). The rows show (1) the velocity dispersion of the satellite galaxies $\sigma_{\rm C}$ [km s$^{-1}$] (taken from \citealt{Lauer2014}), (2) richness $S$, that is, number of satellite galaxies, (3) gravitational radius $r_{\rm g}$ [kpc], (4) integrated absolute brightness of all satellite galaxies (excluding the BCG+ICL) $M_{\rm sat}$ [$g'$ mag], and (5) gravitational mass $\mathcal{M_{\rm g}} [{\rm M}_{\odot}$]. The Pearson coefficient for each correlation is given as a black label and its error (calculated using 10\,000 bootstraps) as a gray label. The histograms show the number of data points in each bin from the subplot containing either $M_{\rm sat}$ or $M_{\rm BCG+ICL}$ (cf. Figure \ref{fig:correlateall}). \label{fig:correlateall1}}
\end{figure*}

\begin{figure*}
	\centering
	\includegraphics[width=\linewidth]{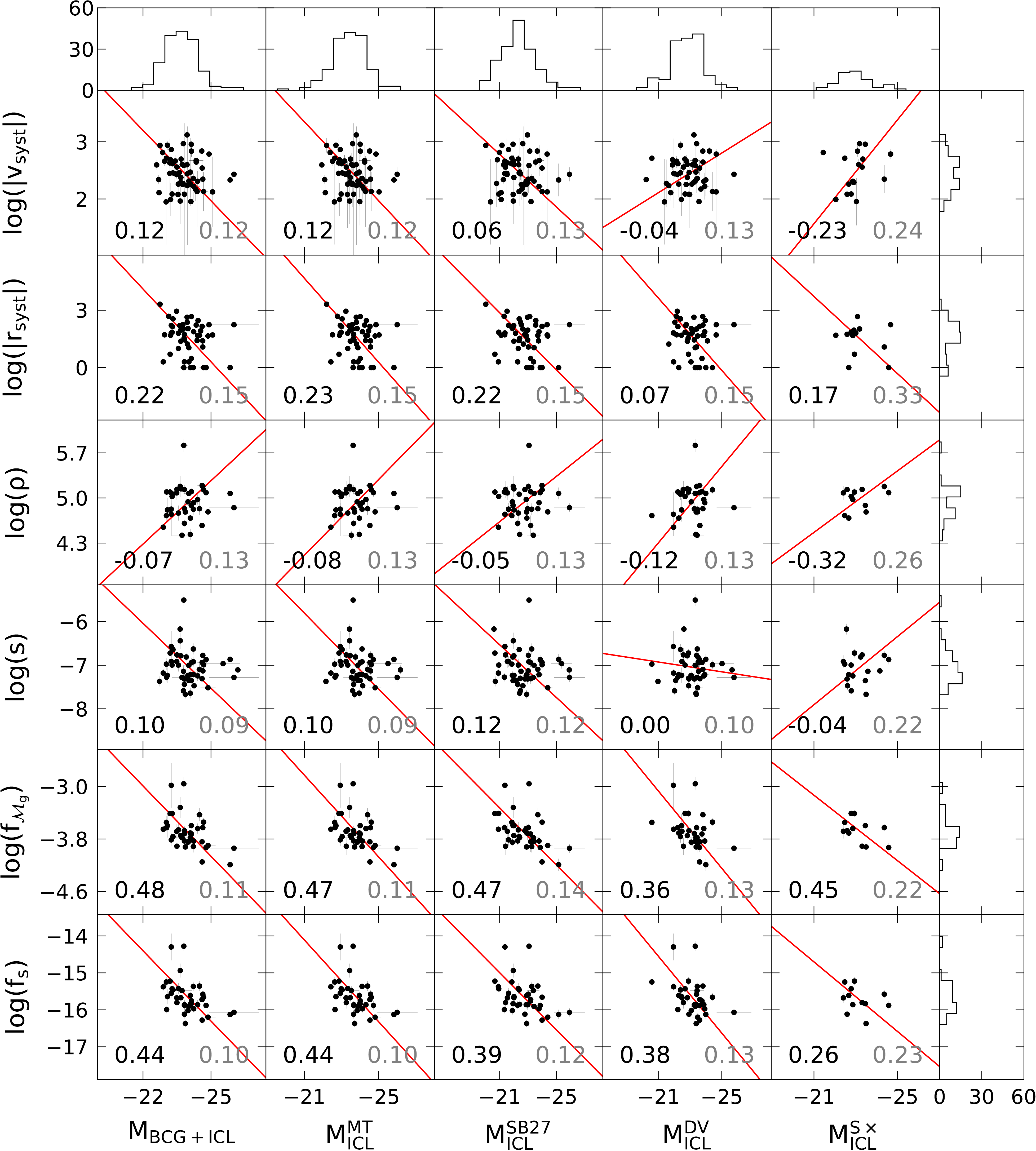}
	\caption{Correlations between BCG/ICL parameters (horizontal) and cluster parameters (vertical). The columns show (1) the absolute brightness of the BCGs+ICL $M_{\rm BCG+ICL}$ [$g'$ mag], the absolute brightnesses of the ICL only $M_{\rm ICL}$ [$g'$ mag], dissected (2) via the total magnitude threshold of --21.85 $g'$ mag $M_{\rm ICL}^{\rm MT}$, (3) via the surface brightness threshold of 27 $g'$ mag $M_{\rm ICL}^{\rm SB27}$, (4) via the light excess above the inner de Vaucouleurs fit $M_{\rm ICL}^{\rm DV}$, and (5) via the double S\'ersic fit $M_{\rm ICL}^{\rm S\times}$. The methods are explained in Section \ref{sec:iclfrac}. The rows show (1) the absolute systemic velocity offset $v_{\rm |syst|}$ (data taken from \citealt{Lauer2014}), (2) radial X-ray emission peak offset $r_{\rm |syst|}$ (data also taken from \citealt{Lauer2014}), (3) mass density $\rho [{\rm M}_{\odot}$ kpc$^{-3}$], (4) number density of satellite galaxies $s$ [kpc$^{-3}$], (5) mass phase space density $f_{\mathcal{M_{\rm g}}} [{\rm M}_{\odot}$ kpc$^{-3}$ km$^{-3}$ s$^3$], and (6) number phase space density of the satellite galaxies $f_{\rm s}$ [kpc$^{-3}$ km$^{-3}$ s$^3$] (6). The Pearson coefficient for each correlation is given as a black label and its error (calculated using 10\,000 bootstraps) as a gray label. The histograms show the number of data points in each bin from the subplot containing either $M_{\rm sat}$ or $M_{\rm BCG+ICL}$ (cf. Figure \ref{fig:correlateall}). \label{fig:correlateall2}}
\end{figure*}

\begin{figure*}
	\centering
	\includegraphics[width=\linewidth]{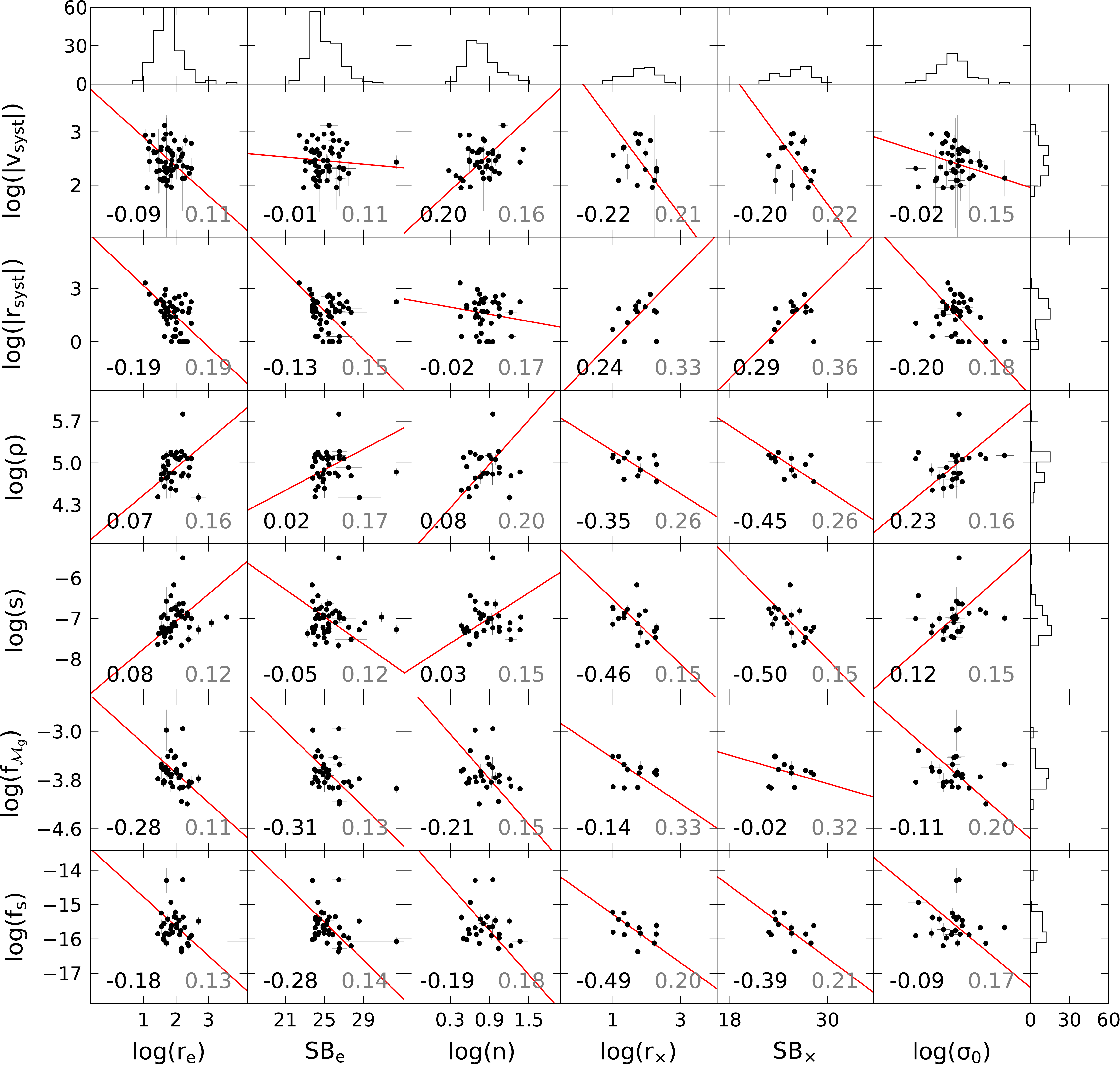}
	\caption{Correlations between BCG/ICL parameters (horizontal) and cluster parameters (vertical). The columns show (1) the effective radius $r_{\rm e}$ [kpc] along the major axis and (2) corresponding effective surface brightnesses ${\rm SB_e}$ [$g'$ mag arcsec$^{-2}$], (3) S\'ersic index $n$ of the SS BCGs, (4) transition radius $r_{\times}$ [kpc] and (5) transition surface brightnesses ${\rm SB}_{\times}$ [$g'$ mag arcsec$^{-2}$] between the two S\'ersic profiles of the DS BCGs, and (6) central velocity dispersion (data taken from \citealt{Lauer2014}). The rows show (1) the absolute systemic velocity offset $v_{\rm |syst|}$ (data taken from \citealt{Lauer2014}), (2) radial X-ray emission peak offset $r_{\rm |syst|}$ (data also taken from \citealt{Lauer2014}), (3) mass density $\rho [{\rm M}_{\odot}$ kpc$^{-3}$], (4) number density of satellite galaxies $s$ [kpc$^{-3}$], (5) mass phase space density $f_{\mathcal{M_{\rm g}}} [{\rm M}_{\odot}$ kpc$^{-3}$ km$^{-3}$ s$^3$], and (6) number phase space density of the satellite galaxies $f_{\rm s}$ [kpc$^{-3}$ km$^{-3}$ s$^3$] (6). The Pearson coefficient for each correlation is given as a black label and its error (calculated using 10\,000 bootstraps) as a gray label. The histograms show the number of data points in each bin from the subplot containing either $M_{\rm sat}$ or $M_{\rm BCG+ICL}$ (cf. Figure \ref{fig:correlateall}). \label{fig:correlateall3}}
\end{figure*}

The correlations with host cluster properties are shown in Figures \ref{fig:correlateall}, \ref{fig:correlateall1}, \ref{fig:correlateall2}, and \ref{fig:correlateall3}. A selection of the strongest correlations between BCG+ICL brightness $M_{\rm BCG+ICL}$ or ICL brightness below an SB of ${\rm SB}>28$ $g'$ mag arcsec$^{-2}$ and host cluster parameters is also presented in Figure \ref{fig:correlatebest}. All slopes, offsets, and correlation strengths are listed in Table \ref{tab:correlateall}. The fitting procedure is described in Section \ref{sec:galdistribution}.

\clearpage
\section{Best-fit parameters of BCG/ICL vs. host cluster correlations}\label{sec:correlateall}

All slopes, offsets, and correlation strengths for the BCG/ICL vs. host cluster correlations are listed in Table \ref{tab:correlateall}.

\startlongtable
\begin{deluxetable}{cccccc}
	\tablewidth{0pt}
	\tabletypesize{\scriptsize}
	
	\tablecaption{BCG/ICL vs.\ host cluster correlations. \label{tab:correlateall}}
	\tablehead{
		\colhead{X} & \colhead{Y} & \colhead{slope $\alpha$} & \colhead{offset $\beta$} & \colhead{$R$} & \colhead{$\Delta R$} \\
	}
	\colnumbers
	\startdata
$\rm{M_{BCG+ICL}}$ &                $\rm{log(\sigma_C)}$ & -0.278 $\pm$ 0.037 & -3.82 & -0.37 & 0.10 \\
$\rm{M_{BCG+ICL}}$ &                       $\rm{log(S)}$ & -0.354 $\pm$ 0.078 & -6.34 & -0.25 & 0.13 \\
$\rm{M_{BCG+ICL}}$ &                     $\rm{log(r_g)}$ & -0.137 $\pm$ 0.027 & -0.23 & -0.34 & 0.10 \\
$\rm{M_{BCG+ICL}}$ &                      $\rm{M_{sat}}$ & 1.19 $\pm$ 0.12 & 3.36 & 0.46 & 0.08 \\
$\rm{M_{BCG+ICL}}$ &           $\rm{log(\mathcal{M}_g)}$ & -0.355 $\pm$ 0.066 & 6.26 & -0.52 & 0.12 \\
$\rm{M_{BCG+ICL}}$ &              $\rm{log(|v_{syst}|)}$ & 0.410 $\pm$ 0.096 & 12.22 & 0.12 & 0.12 \\
$\rm{M_{BCG+ICL}}$ &              $\rm{log(|r_{syst}|)}$ & 1.27 $\pm$ 0.29 & 32.01 & 0.22 & 0.15 \\
$\rm{M_{BCG+ICL}}$ &                    $\rm{log(\rho)}$ & -0.33 $\pm$ 0.10 & -2.93 & -0.07 & 0.13 \\
$\rm{M_{BCG+ICL}}$ &                       $\rm{log(s)}$ & 0.50 $\pm$ 0.13 & 4.91 & 0.10 & 0.09 \\
$\rm{M_{BCG+ICL}}$ &       $\rm{log(f_{\mathcal{M}_g})}$ & 0.357 $\pm$ 0.070 & 4.85 & 0.48 & 0.11 \\
$\rm{M_{BCG+ICL}}$ &                     $\rm{log(f_s)}$ & 0.63 $\pm$ 0.13 & -0.46 & 0.44 & 0.10 \\
$\rm{M_{ICL}^{MT}}$ &                $\rm{log(\sigma_C)}$ & -0.233 $\pm$ 0.032 & -2.68 & -0.34 & 0.11 \\
$\rm{M_{ICL}^{MT}}$ &                       $\rm{log(S)}$ & -0.311 $\pm$ 0.069 & -5.24 & -0.25 & 0.13 \\
$\rm{M_{ICL}^{MT}}$ &                     $\rm{log(r_g)}$ & -0.121 $\pm$ 0.024 & 0.20 & -0.33 & 0.10 \\
$\rm{M_{ICL}^{MT}}$ &                      $\rm{M_{sat}}$ & 0.981 $\pm$ 0.096 & -1.93 & 0.46 & 0.08 \\
$\rm{M_{ICL}^{MT}}$ &           $\rm{log(\mathcal{M}_g)}$ & -0.310 $\pm$ 0.057 & 7.39 & -0.52 & 0.11 \\
$\rm{M_{ICL}^{MT}}$ &              $\rm{log(|v_{syst}|)}$ & 0.345 $\pm$ 0.080 & 10.60 & 0.12 & 0.12 \\
$\rm{M_{ICL}^{MT}}$ &              $\rm{log(|r_{syst}|)}$ & 1.10 $\pm$ 0.25 & 27.75 & 0.23 & 0.15 \\
$\rm{M_{ICL}^{MT}}$ &                    $\rm{log(\rho)}$ & -0.295 $\pm$ 0.090 & -2.08 & -0.08 & 0.13 \\
$\rm{M_{ICL}^{MT}}$ &                       $\rm{log(s)}$ & 0.44 $\pm$ 0.11 & 3.36 & 0.10 & 0.09 \\
$\rm{M_{ICL}^{MT}}$ &       $\rm{log(f_{\mathcal{M}_g})}$ & 0.311 $\pm$ 0.061 & 3.71 & 0.47 & 0.11 \\
$\rm{M_{ICL}^{MT}}$ &                     $\rm{log(f_s)}$ & 0.55 $\pm$ 0.11 & -2.49 & 0.44 & 0.10 \\
$\rm{M_{ICL}^{SB27}}$ &                $\rm{log(\sigma_C)}$ & -0.175 $\pm$ 0.026 & -1.12 & -0.29 & 0.10 \\
$\rm{M_{ICL}^{SB27}}$ &                       $\rm{log(S)}$ & -0.214 $\pm$ 0.047 & -2.68 & -0.26 & 0.13 \\
$\rm{M_{ICL}^{SB27}}$ &                     $\rm{log(r_g)}$ & -0.083 $\pm$ 0.016 & 1.19 & -0.37 & 0.13 \\
$\rm{M_{ICL}^{SB27}}$ &                      $\rm{M_{sat}}$ & 0.746 $\pm$ 0.074 & -8.28 & 0.45 & 0.07 \\
$\rm{M_{ICL}^{SB27}}$ &           $\rm{log(\mathcal{M}_g)}$ & -0.216 $\pm$ 0.041 & 9.86 & -0.49 & 0.13 \\
$\rm{M_{ICL}^{SB27}}$ &              $\rm{log(|v_{syst}|)}$ & 0.230 $\pm$ 0.057 & 7.60 & 0.06 & 0.13 \\
$\rm{M_{ICL}^{SB27}}$ &              $\rm{log(|r_{syst}|)}$ & 0.75 $\pm$ 0.17 & 18.53 & 0.22 & 0.15 \\
$\rm{M_{ICL}^{SB27}}$ &                    $\rm{log(\rho)}$ & -0.176 $\pm$ 0.056 & 0.94 & -0.05 & 0.13 \\
$\rm{M_{ICL}^{SB27}}$ &                       $\rm{log(s)}$ & 0.302 $\pm$ 0.076 & -0.18 & 0.12 & 0.12 \\
$\rm{M_{ICL}^{SB27}}$ &       $\rm{log(f_{\mathcal{M}_g})}$ & 0.217 $\pm$ 0.043 & 1.24 & 0.47 & 0.14 \\
$\rm{M_{ICL}^{SB27}}$ &                     $\rm{log(f_s)}$ & 0.386 $\pm$ 0.084 & -6.89 & 0.39 & 0.12 \\
$\rm{M_{ICL}^{DV}}$ &                $\rm{log(\sigma_C)}$ & -0.210 $\pm$ 0.031 & -2.02 & -0.34 & 0.11 \\
$\rm{M_{ICL}^{DV}}$ &                       $\rm{log(S)}$ & -0.270 $\pm$ 0.068 & -4.07 & -0.19 & 0.11 \\
$\rm{M_{ICL}^{DV}}$ &                     $\rm{log(r_g)}$ & -0.104 $\pm$ 0.027 & 0.66 & -0.17 & 0.12 \\
$\rm{M_{ICL}^{DV}}$ &                      $\rm{M_{sat}}$ & 0.828 $\pm$ 0.099 & -6.09 & 0.32 & 0.09 \\
$\rm{M_{ICL}^{DV}}$ &           $\rm{log(\mathcal{M}_g)}$ & -0.297 $\pm$ 0.067 & 7.90 & -0.41 & 0.12 \\
$\rm{M_{ICL}^{DV}}$ &              $\rm{log(|v_{syst}|)}$ & -0.173 $\pm$ 0.051 & -1.51 & -0.04 & 0.13 \\
$\rm{M_{ICL}^{DV}}$ &              $\rm{log(|r_{syst}|)}$ & 0.94 $\pm$ 0.27 & 23.39 & 0.07 & 0.15 \\
$\rm{M_{ICL}^{DV}}$ &                    $\rm{log(\rho)}$ & -0.295 $\pm$ 0.092 & -1.90 & -0.12 & 0.13 \\
$\rm{M_{ICL}^{DV}}$ &                       $\rm{log(s)}$ & 0.056 $\pm$ 0.060 & -5.75 & 0.00 & 0.10 \\
$\rm{M_{ICL}^{DV}}$ &       $\rm{log(f_{\mathcal{M}_g})}$ & 0.293 $\pm$ 0.071 & 3.08 & 0.36 & 0.13 \\
$\rm{M_{ICL}^{DV}}$ &                     $\rm{log(f_s)}$ & 0.52 $\pm$ 0.12 & -3.68 & 0.38 & 0.13 \\
$\rm{M_{ICL}^{S\times}}$ &                $\rm{log(\sigma_C)}$ & -0.225 $\pm$ 0.074 & -2.39 & -0.12 & 0.20 \\
$\rm{M_{ICL}^{S\times}}$ &                       $\rm{log(S)}$ & -0.39 $\pm$ 0.13 & -6.96 & -0.35 & 0.24 \\
$\rm{M_{ICL}^{S\times}}$ &                     $\rm{log(r_g)}$ & -0.141 $\pm$ 0.052 & -0.20 & -0.27 & 0.22 \\
$\rm{M_{ICL}^{S\times}}$ &                      $\rm{M_{sat}}$ & 0.95 $\pm$ 0.20 & -3.07 & 0.37 & 0.14 \\
$\rm{M_{ICL}^{S\times}}$ &           $\rm{log(\mathcal{M}_g)}$ & -0.274 $\pm$ 0.089 & 8.41 & -0.55 & 0.20 \\
$\rm{M_{ICL}^{S\times}}$ &              $\rm{log(|v_{syst}|)}$ & -0.45 $\pm$ 0.18 & -7.96 & -0.23 & 0.24 \\
$\rm{M_{ICL}^{S\times}}$ &              $\rm{log(|r_{syst}|)}$ & 1.01 $\pm$ 0.49 & 24.90 & 0.17 & 0.33 \\
$\rm{M_{ICL}^{S\times}}$ &                    $\rm{log(\rho)}$ & -0.24 $\pm$ 0.10 & -0.55 & -0.32 & 0.26 \\
$\rm{M_{ICL}^{S\times}}$ &                       $\rm{log(s)}$ & -0.39 $\pm$ 0.19 & -16.12 & -0.04 & 0.22 \\
$\rm{M_{ICL}^{S\times}}$ &       $\rm{log(f_{\mathcal{M}_g})}$ & 0.249 $\pm$ 0.092 & 2.09 & 0.45 & 0.22 \\
$\rm{M_{ICL}^{S\times}}$ &                     $\rm{log(f_s)}$ & 0.47 $\pm$ 0.22 & -4.82 & 0.26 & 0.23 \\
$\rm{log(r_e)}$ &                $\rm{log(\sigma_C)}$ & 0.440 $\pm$ 0.068 & 1.99 & 0.24 & 0.09 \\
$\rm{log(r_e)}$ &                       $\rm{log(S)}$ & 0.493 $\pm$ 0.100 & 1.17 & 0.33 & 0.11 \\
$\rm{log(r_e)}$ &                     $\rm{log(r_g)}$ & 0.191 $\pm$ 0.045 & 2.68 & 0.18 & 0.11 \\
$\rm{log(r_e)}$ &                      $\rm{M_{sat}}$ & -1.84 $\pm$ 0.26 & -21.68 & -0.13 & 0.09 \\
$\rm{log(r_e)}$ &           $\rm{log(\mathcal{M}_g)}$ & 0.48 $\pm$ 0.11 & 13.79 & 0.32 & 0.09 \\
$\rm{log(r_e)}$ &              $\rm{log(|v_{syst}|)}$ & -0.55 $\pm$ 0.13 & 3.45 & -0.09 & 0.11 \\
$\rm{log(r_e)}$ &              $\rm{log(|r_{syst}|)}$ & -1.73 $\pm$ 0.40 & 4.89 & -0.19 & 0.19 \\
$\rm{log(r_e)}$ &                    $\rm{log(\rho)}$ & 0.46 $\pm$ 0.14 & 4.01 & 0.07 & 0.16 \\
$\rm{log(r_e)}$ &                       $\rm{log(s)}$ & 0.68 $\pm$ 0.18 & -8.43 & 0.08 & 0.12 \\
$\rm{log(r_e)}$ &       $\rm{log(f_{\mathcal{M}_g})}$ & -0.48 $\pm$ 0.12 & -2.72 & -0.28 & 0.11 \\
$\rm{log(r_e)}$ &                     $\rm{log(f_s)}$ & -0.86 $\pm$ 0.23 & -13.92 & -0.18 & 0.13 \\
$\rm{SB_e}$ &                $\rm{log(\sigma_C)}$ & 0.137 $\pm$ 0.023 & -0.64 & 0.16 & 0.09 \\
$\rm{SB_e}$ &                       $\rm{log(S)}$ & 0.151 $\pm$ 0.035 & -1.68 & 0.21 & 0.11 \\
$\rm{SB_e}$ &                     $\rm{log(r_g)}$ & 0.059 $\pm$ 0.013 & 1.58 & 0.25 & 0.13 \\
$\rm{SB_e}$ &                      $\rm{M_{sat}}$ & -0.553 $\pm$ 0.063 & -11.07 & -0.33 & 0.08 \\
$\rm{SB_e}$ &           $\rm{log(\mathcal{M}_g)}$ & 0.154 $\pm$ 0.036 & 10.82 & 0.32 & 0.12 \\
$\rm{SB_e}$ &              $\rm{log(|v_{syst}|)}$ & -0.017 $\pm$ 0.024 & 2.87 & -0.01 & 0.11 \\
$\rm{SB_e}$ &              $\rm{log(|r_{syst}|)}$ & -0.55 $\pm$ 0.14 & 15.39 & -0.13 & 0.15 \\
$\rm{SB_e}$ &                    $\rm{log(\rho)}$ & 0.086 $\pm$ 0.033 & 2.74 & 0.02 & 0.17 \\
$\rm{SB_e}$ &                       $\rm{log(s)}$ & -0.169 $\pm$ 0.047 & -2.74 & -0.05 & 0.12 \\
$\rm{SB_e}$ &       $\rm{log(f_{\mathcal{M}_g})}$ & -0.155 $\pm$ 0.037 & 0.27 & -0.31 & 0.13 \\
$\rm{SB_e}$ &                     $\rm{log(f_s)}$ & -0.276 $\pm$ 0.068 & -8.61 & -0.28 & 0.14 \\
$\rm{log(n)}$ &                $\rm{log(\sigma_C)}$ & 0.20 $\pm$ 0.11 & 2.63 & 0.01 & 0.15 \\
$\rm{log(n)}$ &                       $\rm{log(S)}$ & 1.09 $\pm$ 0.31 & 1.29 & 0.24 & 0.15 \\
$\rm{log(n)}$ &                     $\rm{log(r_g)}$ & 0.44 $\pm$ 0.14 & 2.71 & 0.16 & 0.16 \\
$\rm{log(n)}$ &                      $\rm{M_{sat}}$ & -3.89 $\pm$ 0.61 & -21.88 & -0.19 & 0.10 \\
$\rm{log(n)}$ &           $\rm{log(\mathcal{M}_g)}$ & 1.20 $\pm$ 0.37 & 13.75 & 0.26 & 0.14 \\
$\rm{log(n)}$ &              $\rm{log(|v_{syst}|)}$ & 1.14 $\pm$ 0.29 & 1.56 & 0.20 & 0.16 \\
$\rm{log(n)}$ &              $\rm{log(|r_{syst}|)}$ & -0.66 $\pm$ 0.71 & 2.14 & -0.02 & 0.17 \\
$\rm{log(n)}$ &                    $\rm{log(\rho)}$ & 1.22 $\pm$ 0.45 & 3.88 & 0.08 & 0.20 \\
$\rm{log(n)}$ &                       $\rm{log(s)}$ & 1.04 $\pm$ 0.40 & -7.92 & 0.03 & 0.15 \\
$\rm{log(n)}$ &       $\rm{log(f_{\mathcal{M}_g})}$ & -1.22 $\pm$ 0.40 & -2.68 & -0.21 & 0.15 \\
$\rm{log(n)}$ &                     $\rm{log(f_s)}$ & -2.14 $\pm$ 0.71 & -13.82 & -0.19 & 0.18 \\
$\rm{log(r_{\times}})$ &                $\rm{log(\sigma_C)}$ & -0.393 $\pm$ 0.098 & 3.43 & -0.38 & 0.16 \\
$\rm{log(r_{\times}})$ &                       $\rm{log(S)}$ & -0.57 $\pm$ 0.20 & 3.06 & -0.32 & 0.21 \\
$\rm{log(r_{\times}})$ &                     $\rm{log(r_g)}$ & 0.206 $\pm$ 0.074 & 2.71 & 0.30 & 0.18 \\
$\rm{log(r_{\times}})$ &                      $\rm{M_{sat}}$ & 1.04 $\pm$ 0.34 & -26.76 & 0.03 & 0.15 \\
$\rm{log(r_{\times}})$ &           $\rm{log(\mathcal{M}_g)}$ & -0.17 $\pm$ 0.14 & 15.04 & -0.03 & 0.33 \\
$\rm{log(r_{\times}})$ &              $\rm{log(|v_{syst}|)}$ & -0.86 $\pm$ 0.34 & 3.99 & -0.22 & 0.21 \\
$\rm{log(r_{\times}})$ &              $\rm{log(|r_{syst}|)}$ & 1.94 $\pm$ 0.87 & -1.86 & 0.24 & 0.33 \\
$\rm{log(r_{\times}})$ &                    $\rm{log(\rho)}$ & -0.36 $\pm$ 0.15 & 5.56 & -0.35 & 0.26 \\
$\rm{log(r_{\times}})$ &                       $\rm{log(s)}$ & -0.80 $\pm$ 0.23 & -5.73 & -0.46 & 0.15 \\
$\rm{log(r_{\times}})$ &       $\rm{log(f_{\mathcal{M}_g})}$ & -0.37 $\pm$ 0.20 & -3.07 & -0.14 & 0.33 \\
$\rm{log(r_{\times}})$ &                     $\rm{log(f_s)}$ & -0.70 $\pm$ 0.25 & -14.58 & -0.49 & 0.20 \\
$\rm{SB_{\times}}$ &                $\rm{log(\sigma_C)}$ & -0.098 $\pm$ 0.023 & 5.29 & -0.41 & 0.16 \\
$\rm{SB_{\times}}$ &                       $\rm{log(S)}$ & -0.148 $\pm$ 0.045 & 5.86 & -0.44 & 0.20 \\
$\rm{SB_{\times}}$ &                     $\rm{log(r_g)}$ & 0.053 $\pm$ 0.020 & 1.71 & 0.23 & 0.19 \\
$\rm{SB_{\times}}$ &                      $\rm{M_{sat}}$ & 0.34 $\pm$ 0.10 & -33.84 & 0.05 & 0.15 \\
$\rm{SB_{\times}}$ &           $\rm{log(\mathcal{M}_g)}$ & -0.102 $\pm$ 0.052 & 17.34 & -0.18 & 0.32 \\
$\rm{SB_{\times}}$ &              $\rm{log(|v_{syst}|)}$ & -0.209 $\pm$ 0.085 & 7.90 & -0.20 & 0.22 \\
$\rm{SB_{\times}}$ &              $\rm{log(|r_{syst}|)}$ & 0.46 $\pm$ 0.19 & -10.26 & 0.29 & 0.36 \\
$\rm{SB_{\times}}$ &                    $\rm{log(\rho)}$ & -0.089 $\pm$ 0.033 & 7.23 & -0.45 & 0.26 \\
$\rm{SB_{\times}}$ &                       $\rm{log(s)}$ & -0.206 $\pm$ 0.058 & -1.84 & -0.50 & 0.15 \\
$\rm{SB_{\times}}$ &       $\rm{log(f_{\mathcal{M}_g})}$ & -0.039 $\pm$ 0.032 & -2.69 & -0.02 & 0.32 \\
$\rm{SB_{\times}}$ &                     $\rm{log(f_s)}$ & -0.176 $\pm$ 0.070 & -11.29 & -0.39 & 0.21 \\
$\rm{log(\sigma_0)}$ &                $\rm{log(\sigma_C)}$ & 3.30 $\pm$ 0.72 & -5.33 & 0.13 & 0.14 \\
$\rm{log(\sigma_0)}$ &                       $\rm{log(S)}$ & 3.7 $\pm$ 1.2 & -6.77 & 0.18 & 0.19 \\
$\rm{log(\sigma_0)}$ &                     $\rm{log(r_g)}$ & 0.25 $\pm$ 0.30 & 2.46 & 0.00 & 0.19 \\
$\rm{log(\sigma_0)}$ &                      $\rm{M_{sat}}$ & -12.1 $\pm$ 2.4 & 4.82 & -0.16 & 0.10 \\
$\rm{log(\sigma_0)}$ &           $\rm{log(\mathcal{M}_g)}$ & 3.4 $\pm$ 1.1 & 6.36 & 0.23 & 0.19 \\
$\rm{log(\sigma_0)}$ &              $\rm{log(|v_{syst}|)}$ & -1.47 $\pm$ 0.71 & 6.03 & -0.02 & 0.15 \\
$\rm{log(\sigma_0)}$ &              $\rm{log(|r_{syst}|)}$ & -14.8 $\pm$ 3.9 & 37.90 & -0.20 & 0.18 \\
$\rm{log(\sigma_0)}$ &                    $\rm{log(\rho)}$ & 3.3 $\pm$ 1.1 & -3.27 & 0.23 & 0.16 \\
$\rm{log(\sigma_0)}$ &                       $\rm{log(s)}$ & 5.3 $\pm$ 1.9 & -20.04 & 0.12 & 0.15 \\
$\rm{log(\sigma_0)}$ &       $\rm{log(f_{\mathcal{M}_g})}$ & -3.5 $\pm$ 1.3 & 4.86 & -0.11 & 0.20 \\
$\rm{log(\sigma_0)}$ &                     $\rm{log(f_s)}$ & -5.8 $\pm$ 2.2 & -1.24 & -0.09 & 0.17 \\
	\enddata
	\tablecomments{Correlations between BCG/ICL and host cluster parameters. The correlations are in the form of $Y=\alpha X + \beta$. Orthogonal distance regression was applied to find the best-fit parameters. The Pearson coefficient $R$ for each correlation is given in column (5) and its error (calculated using 10\,000 bootstraps) in column (6). The BCG/ICL parameters are given in column (1). They are: absolute brightness of the BCGs+ICL $M_{\rm BCG+ICL}$ [$g'$ mag], absolute brightness of the ICL only $M_{\rm ICL}$ [$g'$ mag], dissected via the total magnitude threshold of --21.85 $g'$ mag $M_{\rm ICL}^{\rm MT}$, via the surface brightness threshold of 27 $g'$ mag $M_{\rm ICL}^{\rm SB27}$, via the light excess above the inner de Vaucouleurs fit $M_{\rm ICL}^{\rm DV}$, and via the double S\'ersic fit $M_{\rm ICL}^{\rm S\times}$. The methods are explained in Section \ref{sec:iclfrac}. The BCG/ICL parameters are measured along the major axis. They are: effective radius $r_{\rm e}$ [kpc], effective surface brightness ${\rm SB_e}$ [$g'$ mag arcsec$^{-2}$], S\'ersic index $n$ (for SS BCGs), transition radius $r_{\times}$ [kpc] and transition surface brightness ${\rm SB}_{\times}$ [$g'$ mag arcsec$^{-2}$] between the two S\'ersic profiles of DS BCGs, and central velocity dispersion (data taken from \citealt{Lauer2014}). The cluster parameters are given in column (2). They are: velocity dispersion of the cluster galaxies $\sigma_{\rm C}$ [km s$^{-1}$] (taken from \citealt{Lauer2014}), richness $S$, that is, number of satellite galaxies, gravitational radius $r_{\rm g}$ [kpc], integrated absolute brightness of all satellite galaxies (excluding the BCG+ICL) $M_{\rm sat}$ [$g'$ mag], gravitational mass $\mathcal{M_{\rm g}} [\rm{M}_{\odot}$], absolute systemic velocity offset $|v_{\rm syst}|$ (data taken from \citealt{Lauer2014}), radial X-ray emission peak offset $r_{\rm syst}$ (data also taken from \citealt{Lauer2014}), mass density $\rho [\rm{M}_{\odot}$ kpc$^{-3}$], number density of satellite galaxies $s$ [kpc$^{-3}$], mass phase space density $f_{\mathcal{M_{\rm g}}} [\rm{M}_{\odot}$ kpc$^{-3}$ km$^{-3}$ s$^3$], and number phase space density of the satellite galaxies $f_{\rm s}$ [kpc$^{-3}$ km$^{-3}$ s$^3$].}
\end{deluxetable}

\newpage

\section{Full-sized images of the clusters}\label{sec:fullscreenshots}

The following Figure \ref{fig:screenshots2} presents the full-sized WWFI observations of all galaxy clusters that are analyzed in this work (left panels in each subfigure). Cut-outs around the BCGs are also shown in Figure 19 in \cite{Kluge2020}. For comparison, far-infrared observations of the same sky regions (\citealt{Planck2014}; 857 GHz band) show only the galactic foreground dust (see also Section 5.2. in \citealt{Kluge2020}). Its distribution is well resembled in unprecedented spatial resolution in the WWFI images.

For the WWFI images, the gradient and a constant are automatically subtracted during basic data reduction by applying the night-sky flat subtraction procedure presented in Section 3.1.5. in \cite{Kluge2020}. For the far-infrared images, the gradient is still present but the median value is subtracted. These images are then flux-scaled by Equation (17) in \cite{Kluge2020} to allow direct comparison with the WWFI images.

\begin{figure*}
	\centering
	\includegraphics[width=\linewidth]{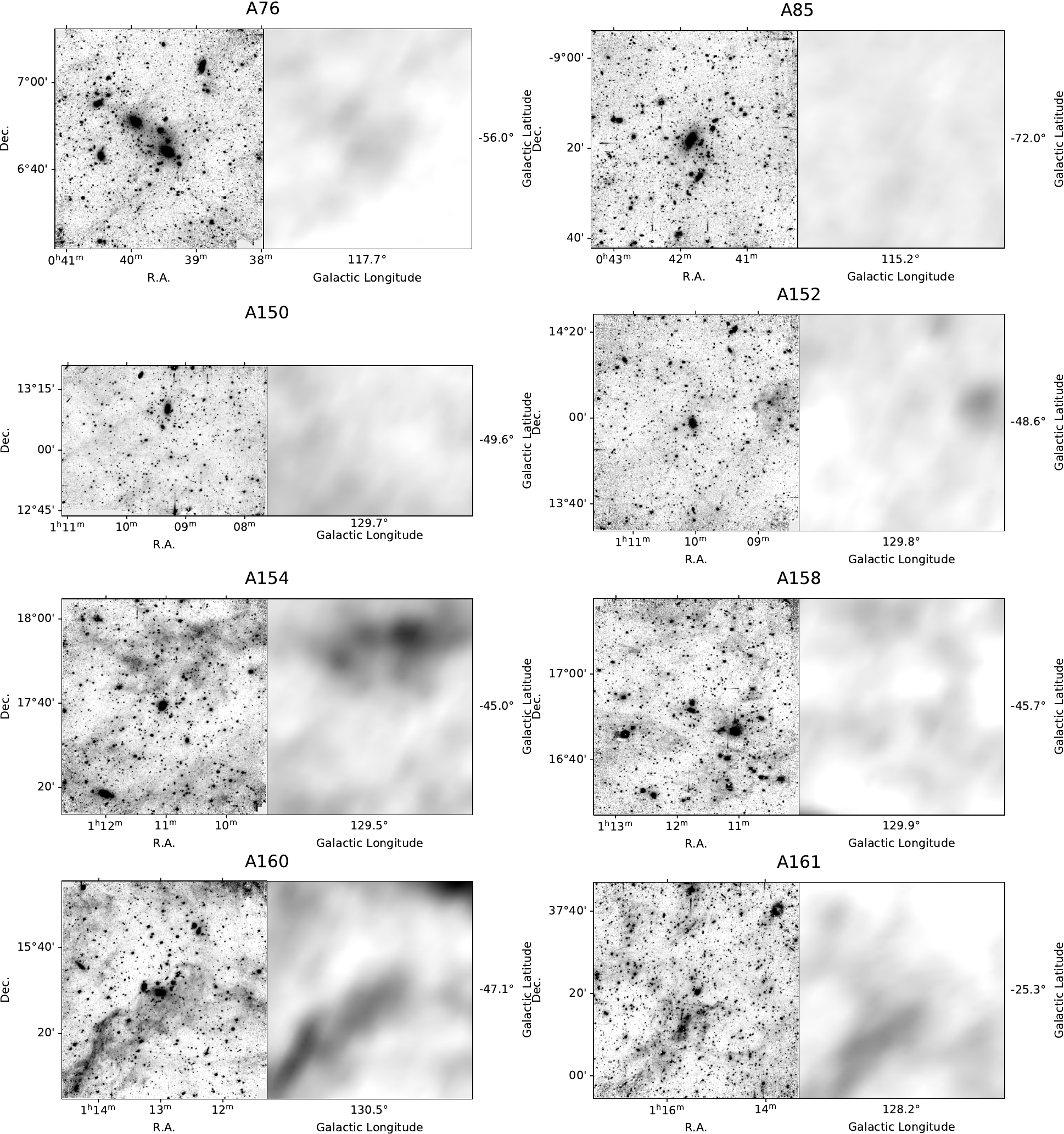} \\~\\~\\
	\includegraphics[width=\linewidth]{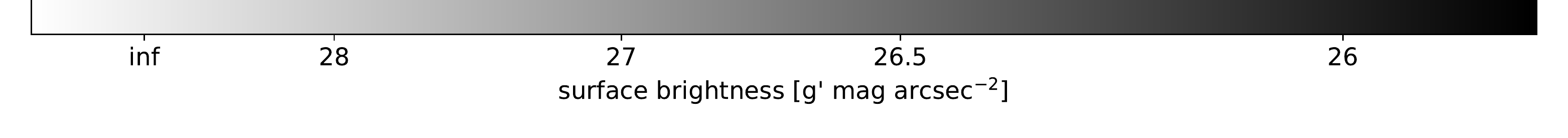}
	\caption{Full-sized images of the observed galaxy clusters. The left panel of the subfigures shows the WWFI $g'$-band stack and the right panel shows the far-infrared 857 GHz Planck image of the same sky region. It is flux scaled using Equation (17) in \cite{Kluge2020} and the median value is subtracted. Galactic coordinates are given for the image center. \label{fig:screenshots2}}
\end{figure*}
\begin{figure*}
	\centering
	\includegraphics[width=\linewidth]{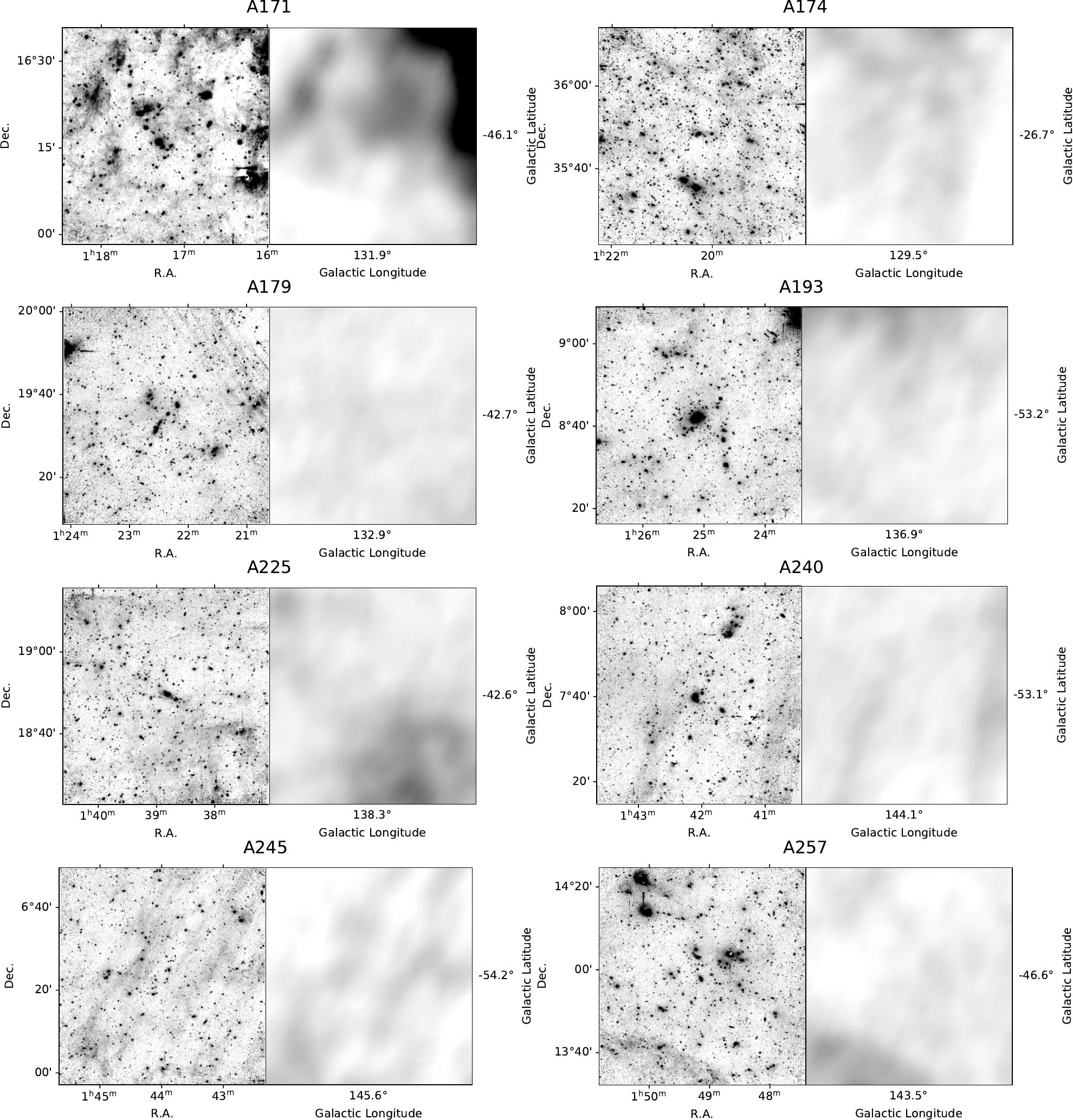} \\~\\~\\
	\includegraphics[width=\linewidth]{fig16bar.pdf}
	\textbf{Figure \ref*{fig:screenshots2}} \textit{(continued)}
\end{figure*}
\begin{figure*}
	\centering
	\includegraphics[width=\linewidth]{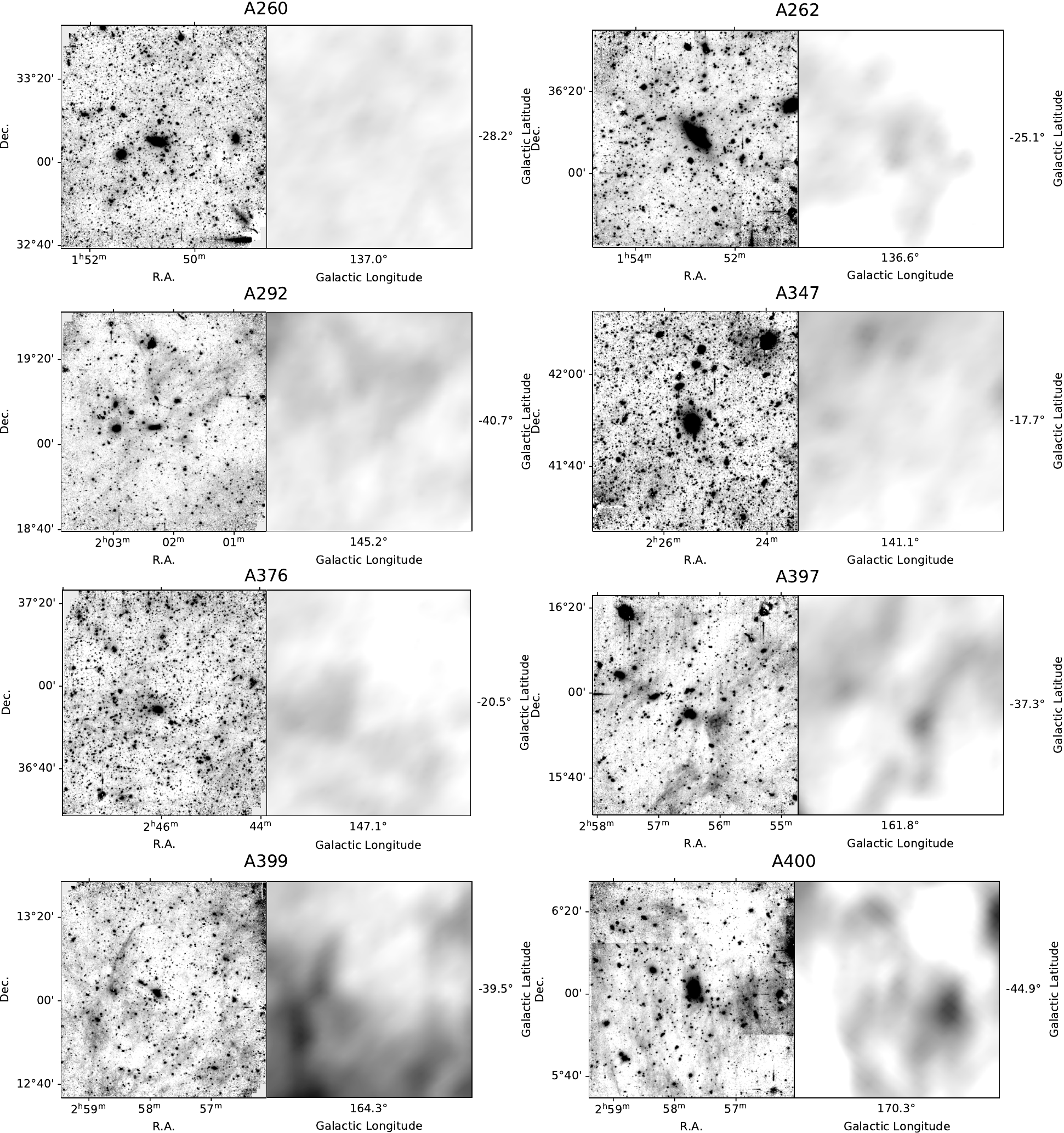} \\~\\~\\
	\includegraphics[width=\linewidth]{fig16bar.pdf}
	\textbf{Figure \ref*{fig:screenshots2}} \textit{(continued)}
\end{figure*}
\begin{figure*}
	\centering
	\includegraphics[width=\linewidth]{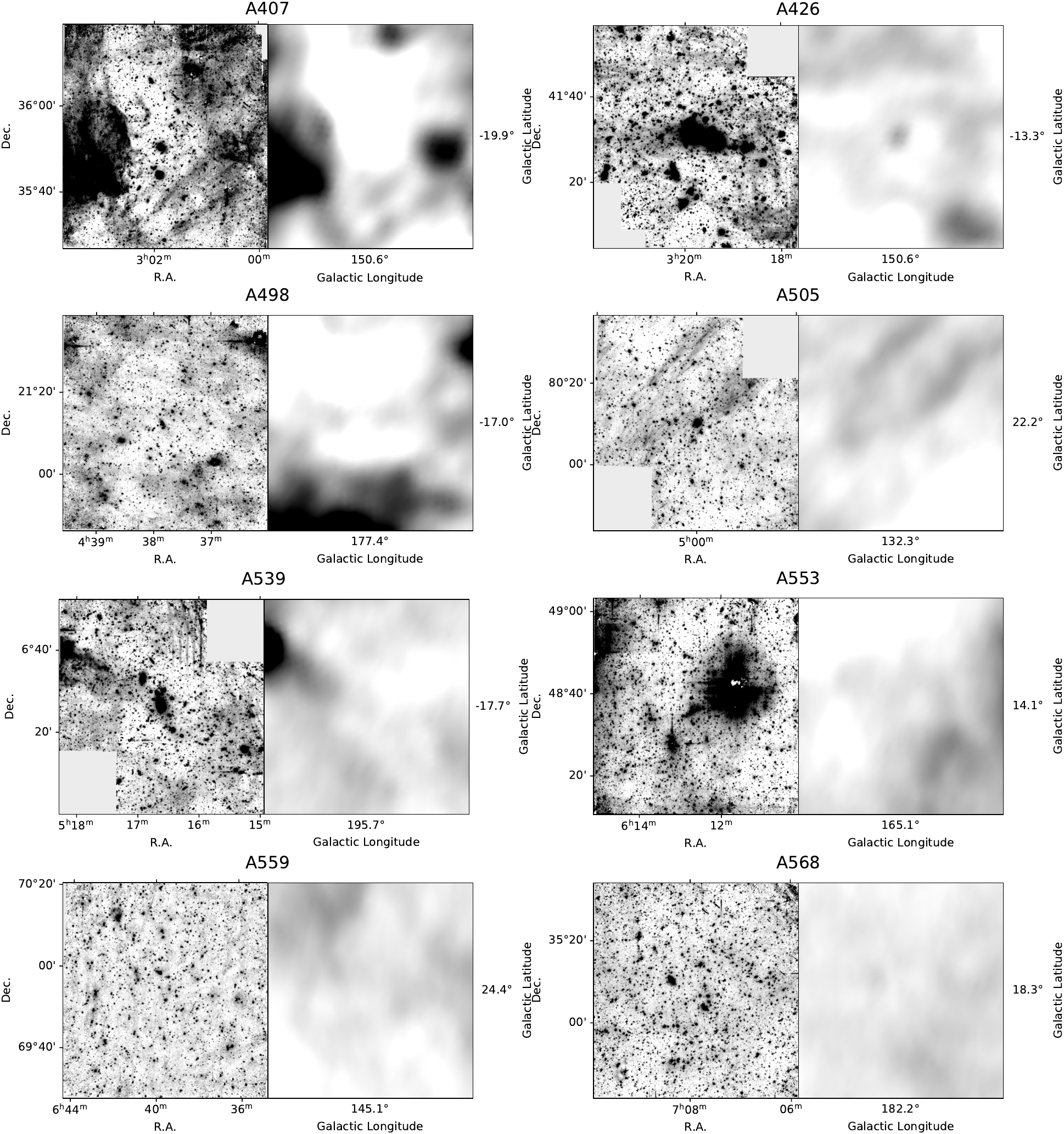} \\~\\~\\
	\includegraphics[width=\linewidth]{fig16bar.pdf}
	\textbf{Figure \ref*{fig:screenshots2}} \textit{(continued)}
\end{figure*}
\begin{figure*}
	\centering
	\includegraphics[width=\linewidth]{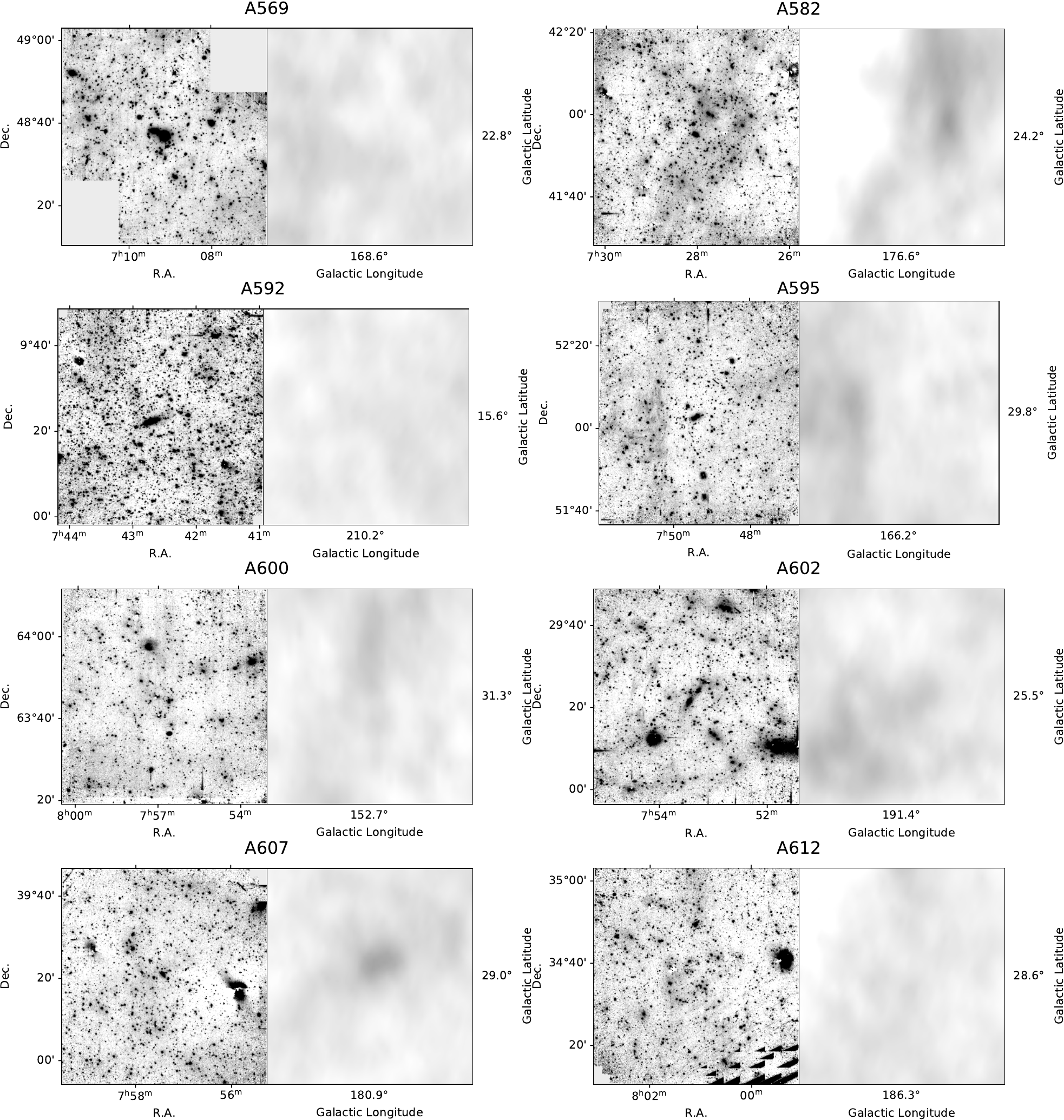} \\~\\~\\
	\includegraphics[width=\linewidth]{fig16bar.pdf}
	\textbf{Figure \ref*{fig:screenshots2}} \textit{(continued)}
\end{figure*}
\begin{figure*}
	\centering
	\includegraphics[width=\linewidth]{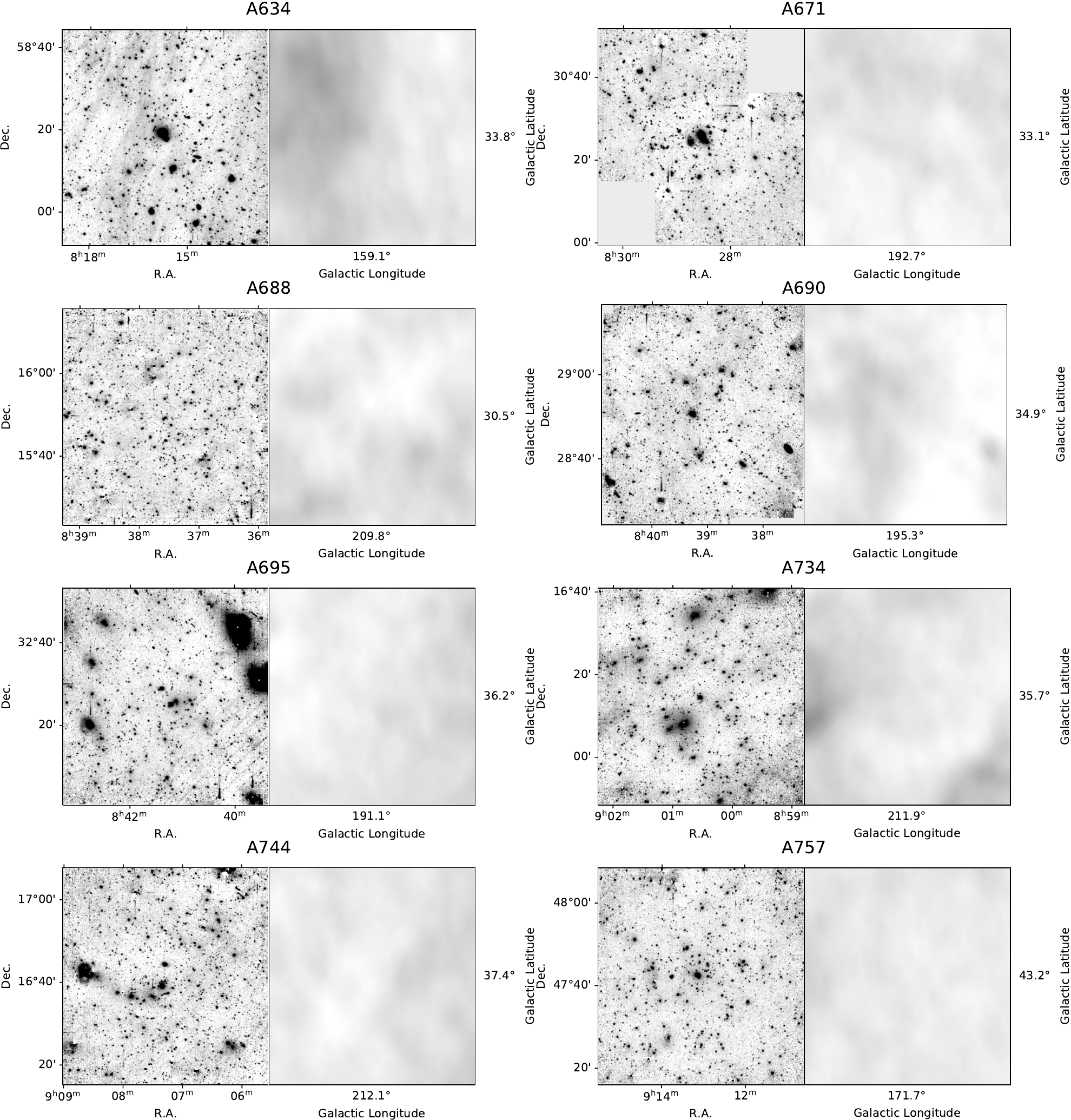} \\~\\~\\
	\includegraphics[width=\linewidth]{fig16bar.pdf}
	\textbf{Figure \ref*{fig:screenshots2}} \textit{(continued)}
\end{figure*}
\begin{figure*}
	\centering
	\includegraphics[width=\linewidth]{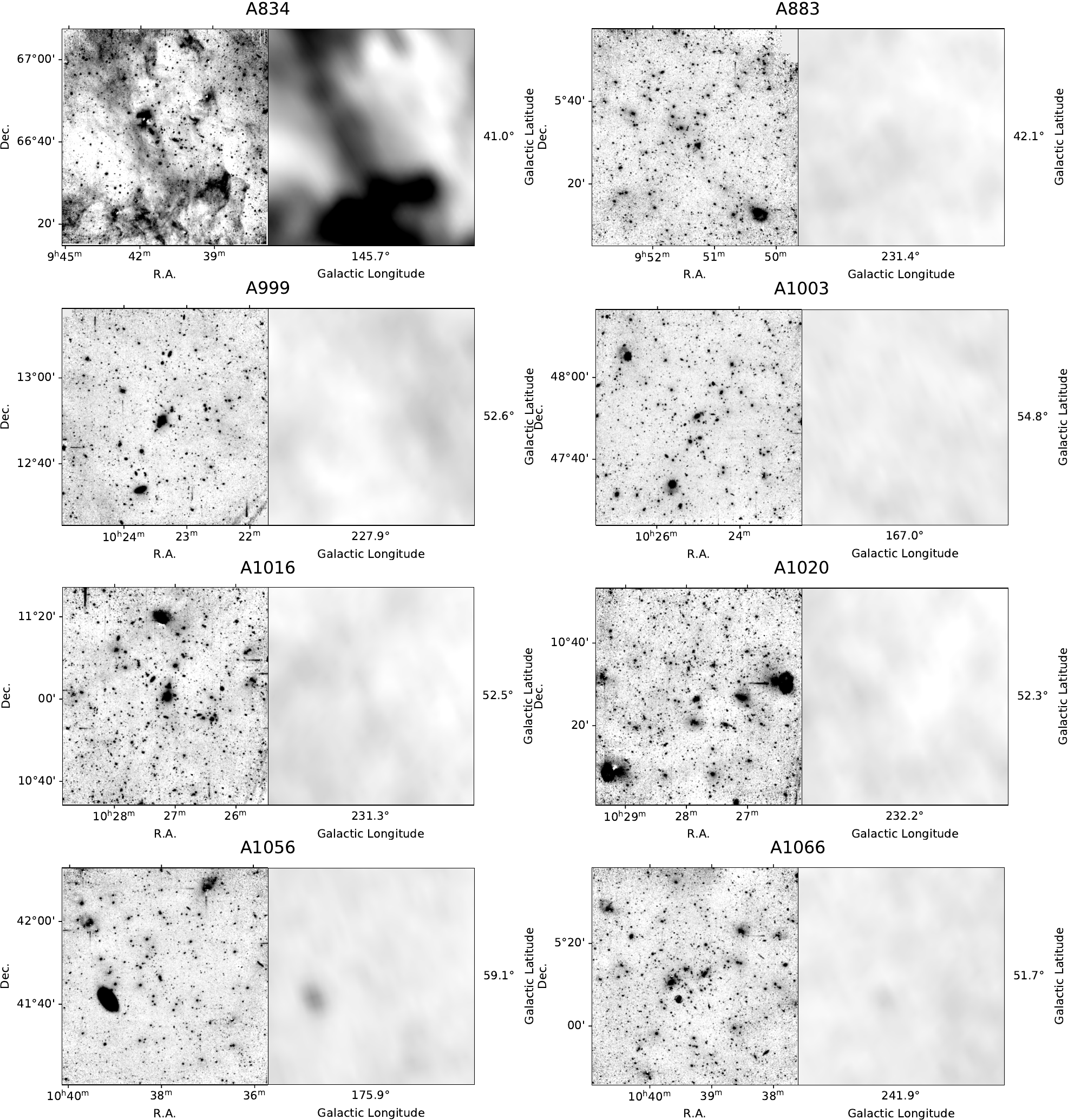} \\~\\~\\
	\includegraphics[width=\linewidth]{fig16bar.pdf}
	\textbf{Figure \ref*{fig:screenshots2}} \textit{(continued)}
\end{figure*}
\begin{figure*}
	\centering
	\includegraphics[width=\linewidth]{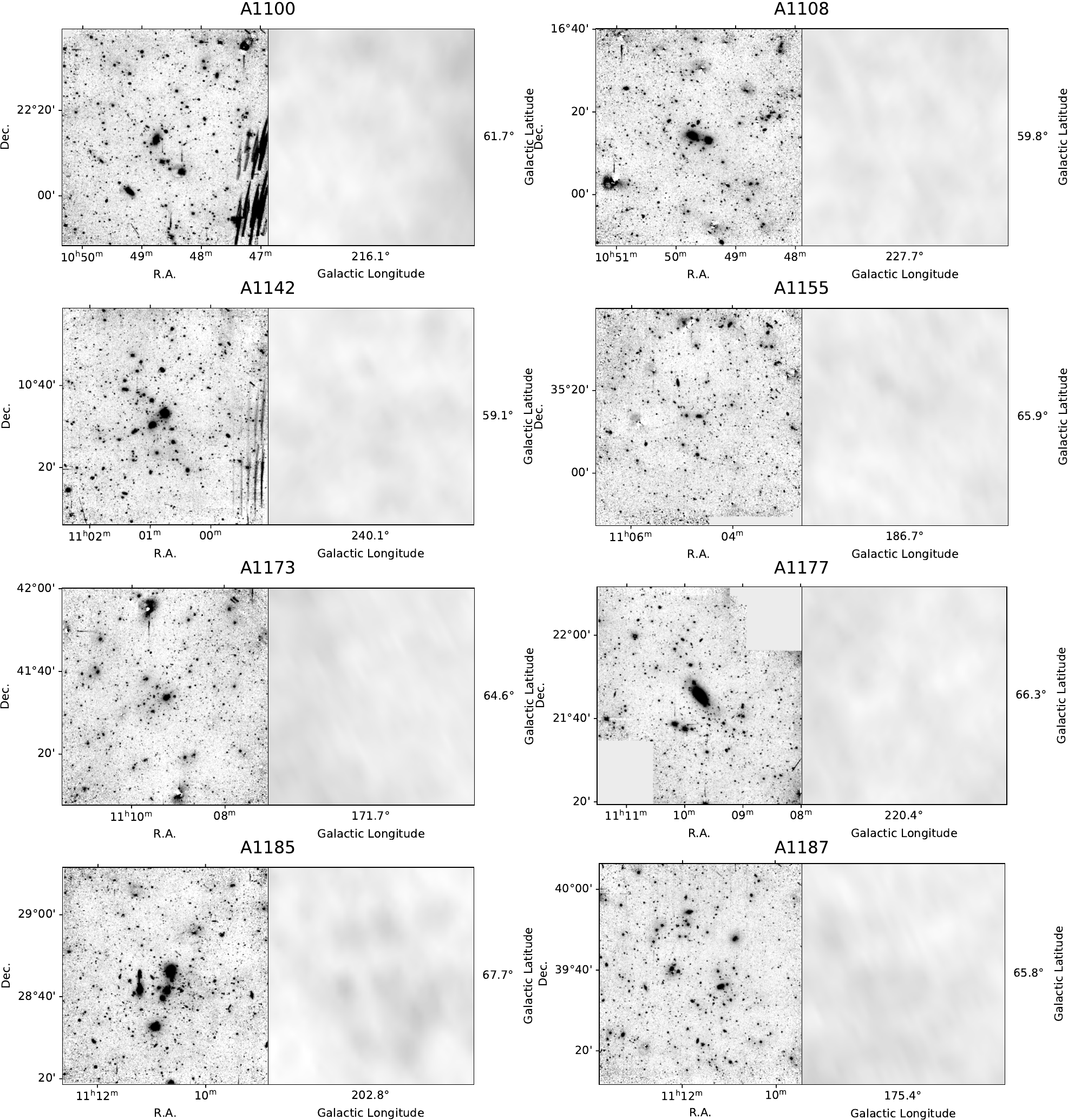} \\~\\~\\
	\includegraphics[width=\linewidth]{fig16bar.pdf}
	\textbf{Figure \ref*{fig:screenshots2}} \textit{(continued)}
\end{figure*}
\begin{figure*}
	\centering
	\includegraphics[width=\linewidth]{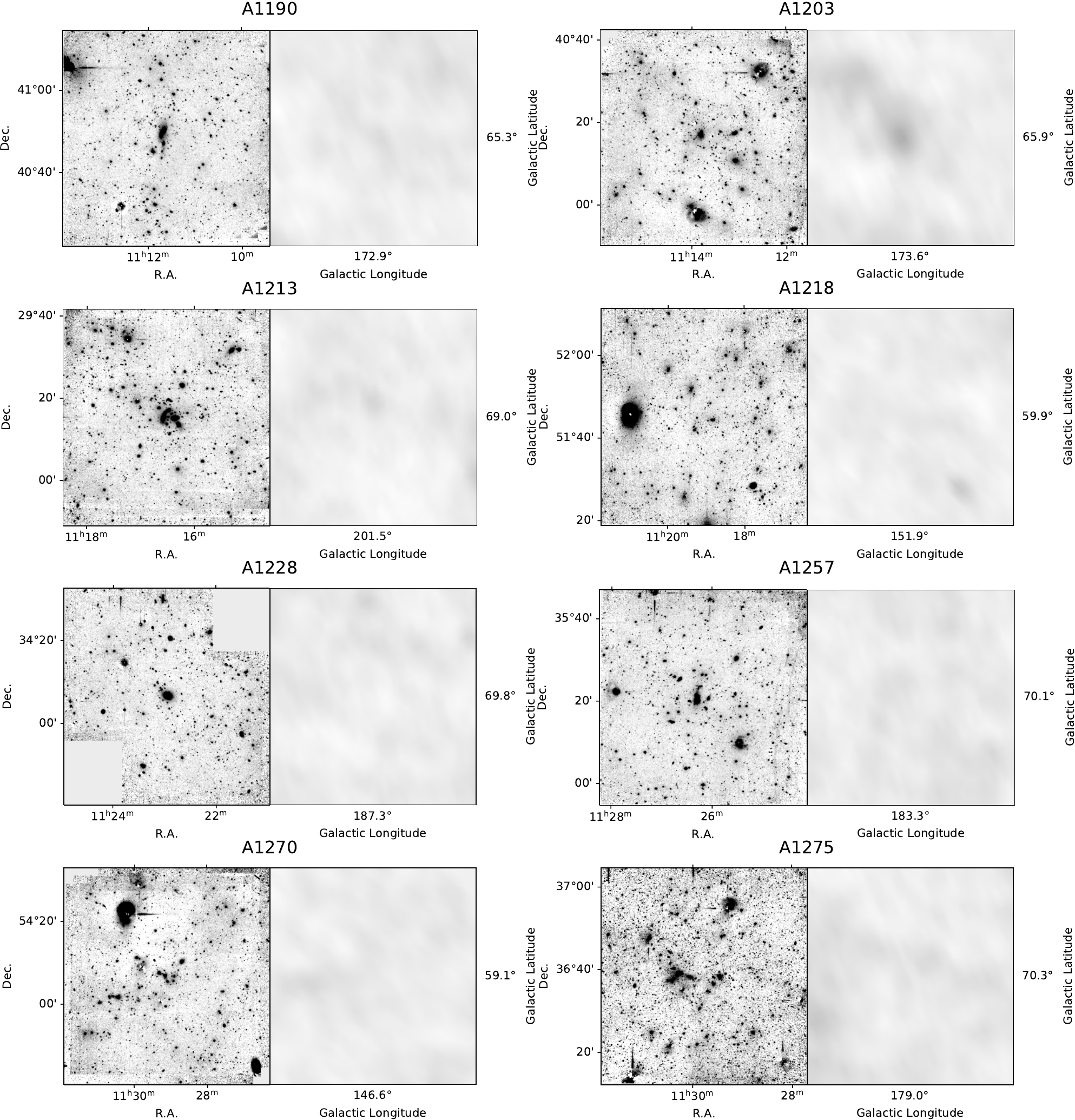} \\~\\~\\
	\includegraphics[width=\linewidth]{fig16bar.pdf}
	\textbf{Figure \ref*{fig:screenshots2}} \textit{(continued)}
\end{figure*}
\begin{figure*}
	\centering
	\includegraphics[width=\linewidth]{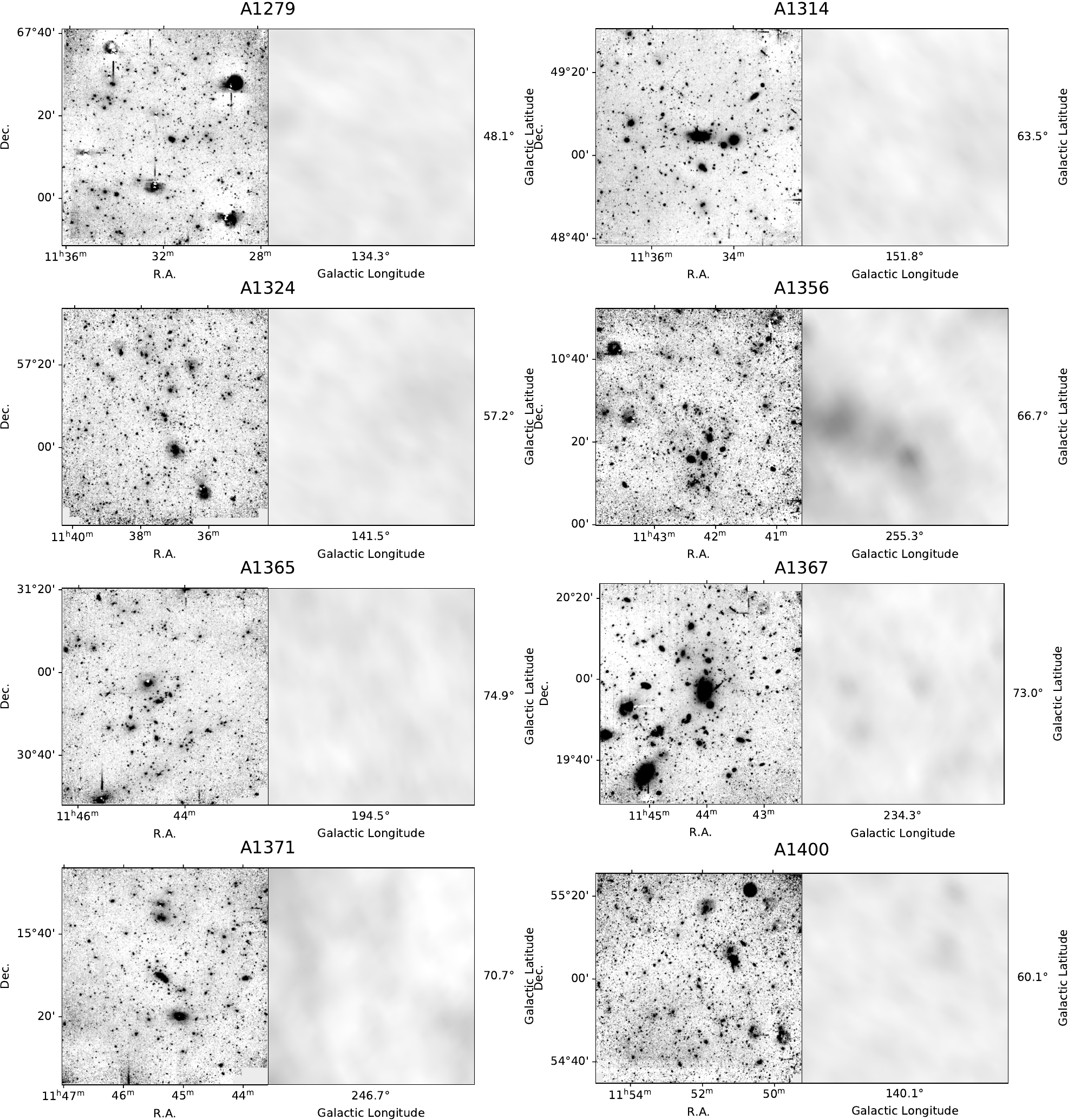} \\~\\~\\
	\includegraphics[width=\linewidth]{fig16bar.pdf}
	\textbf{Figure \ref*{fig:screenshots2}} \textit{(continued)}
\end{figure*}
\begin{figure*}
	\centering
	\includegraphics[width=\linewidth]{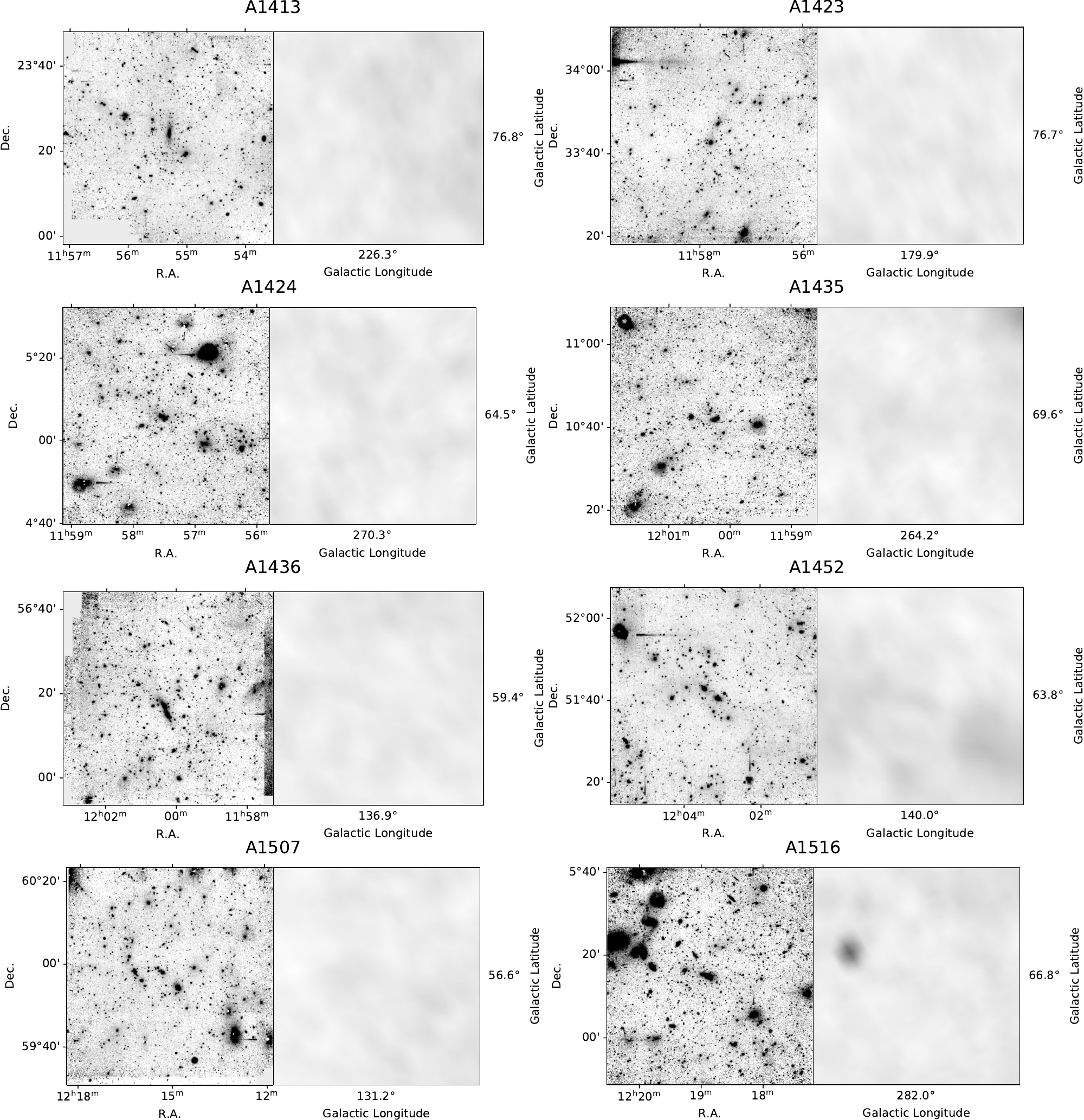} \\~\\~\\
	\includegraphics[width=\linewidth]{fig16bar.pdf}
	\textbf{Figure \ref*{fig:screenshots2}} \textit{(continued)}
\end{figure*}
\begin{figure*}
	\centering
	\includegraphics[width=\linewidth]{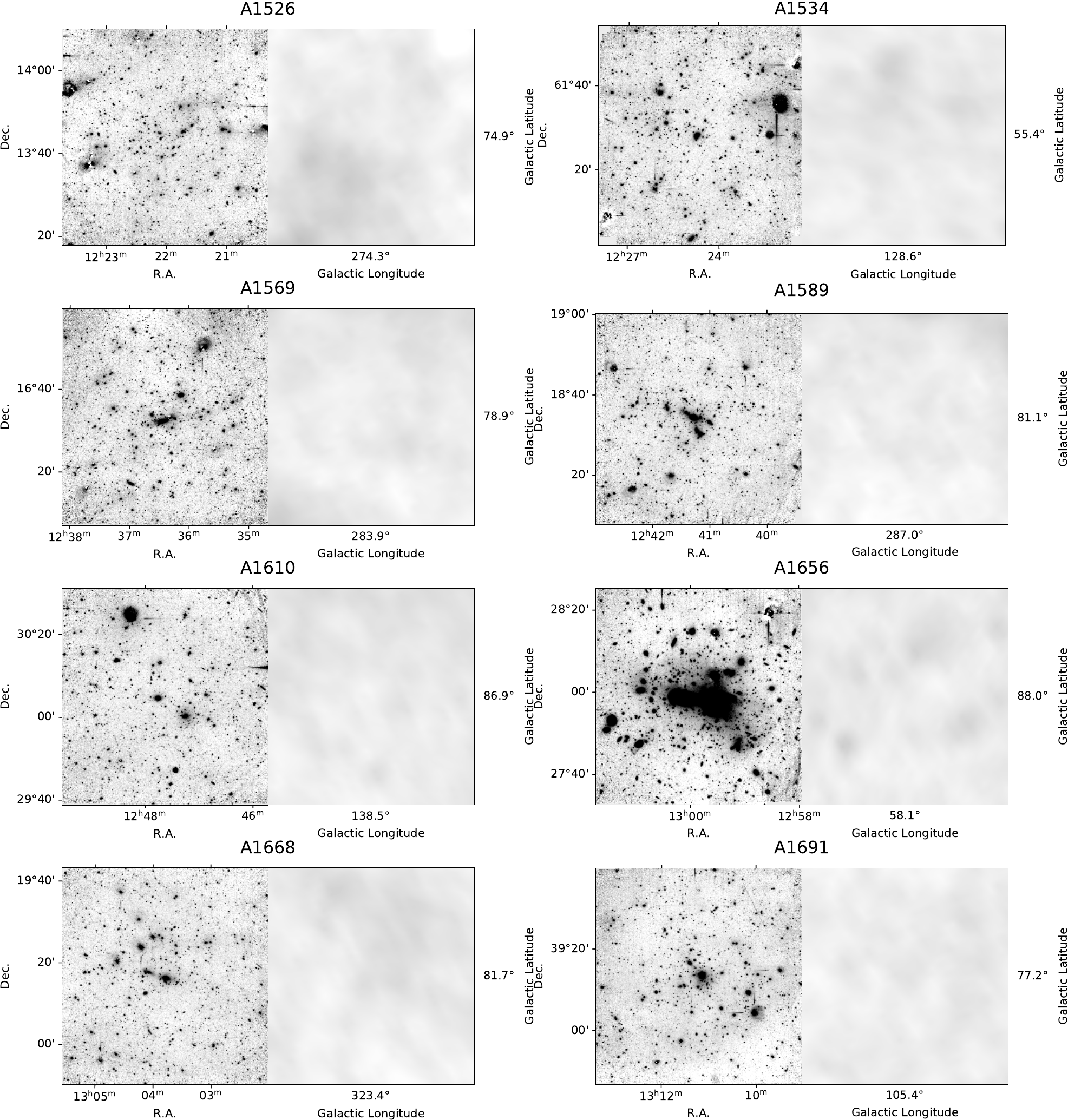} \\~\\~\\
	\includegraphics[width=\linewidth]{fig16bar.pdf}
	\textbf{Figure \ref*{fig:screenshots2}} \textit{(continued)}
\end{figure*}
\begin{figure*}
	\centering
	\includegraphics[width=\linewidth]{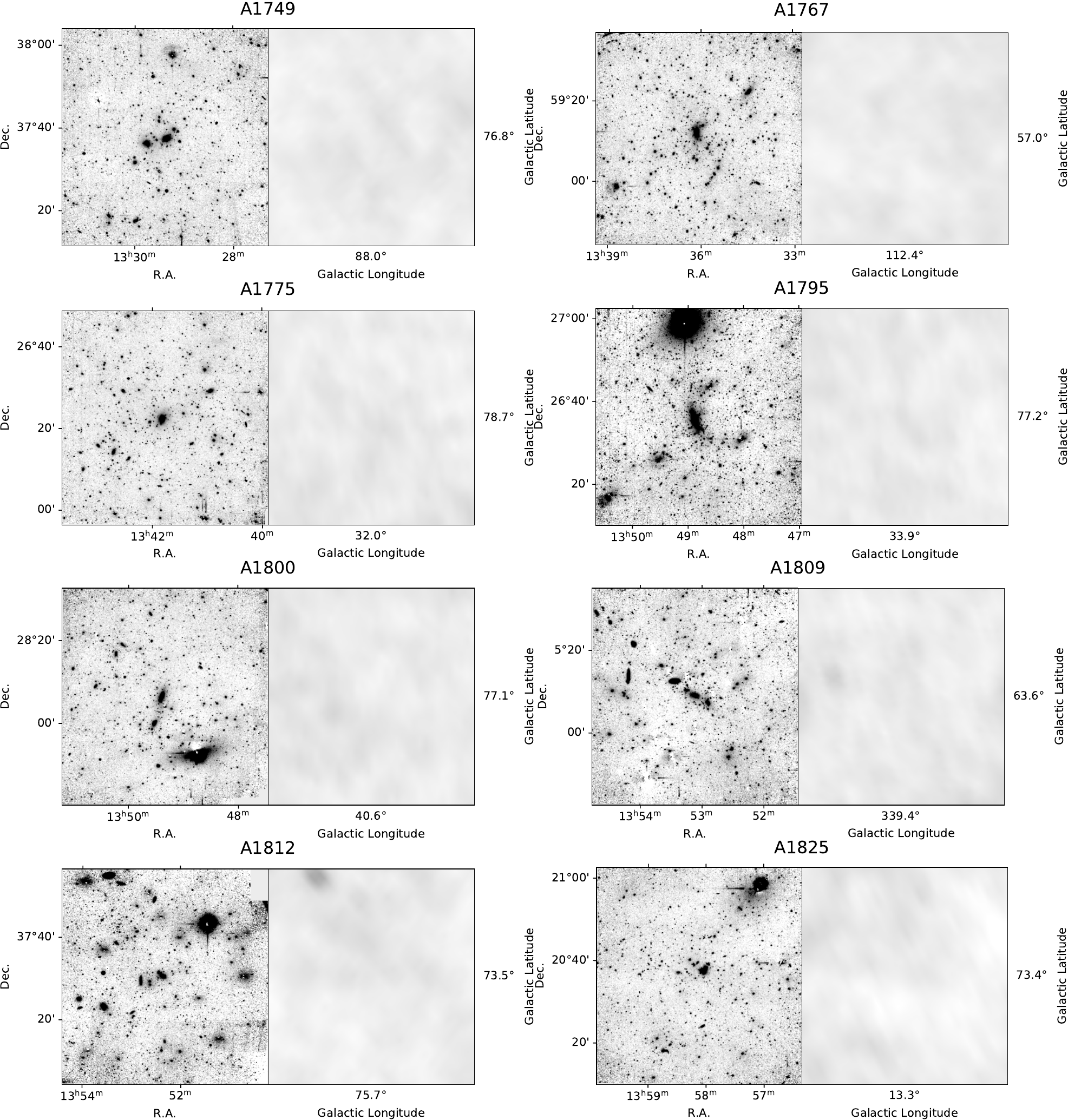} \\~\\~\\
	\includegraphics[width=\linewidth]{fig16bar.pdf}
	\textbf{Figure \ref*{fig:screenshots2}} \textit{(continued)}
\end{figure*}
\begin{figure*}
	\centering
	\includegraphics[width=\linewidth]{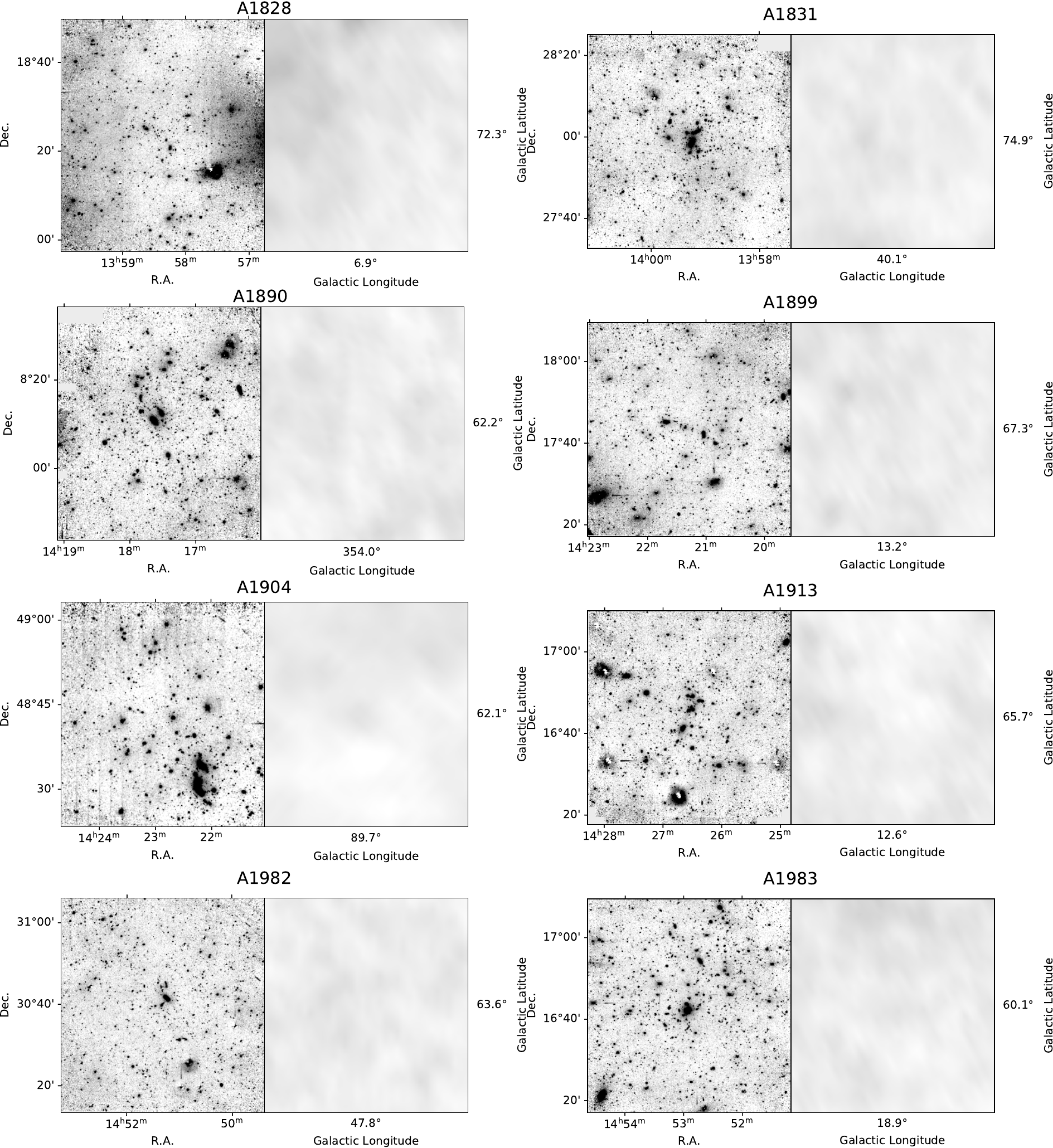} \\~\\~\\
	\includegraphics[width=\linewidth]{fig16bar.pdf}
	\textbf{Figure \ref*{fig:screenshots2}} \textit{(continued)}
\end{figure*}
\begin{figure*}
	\centering
	\includegraphics[width=\linewidth]{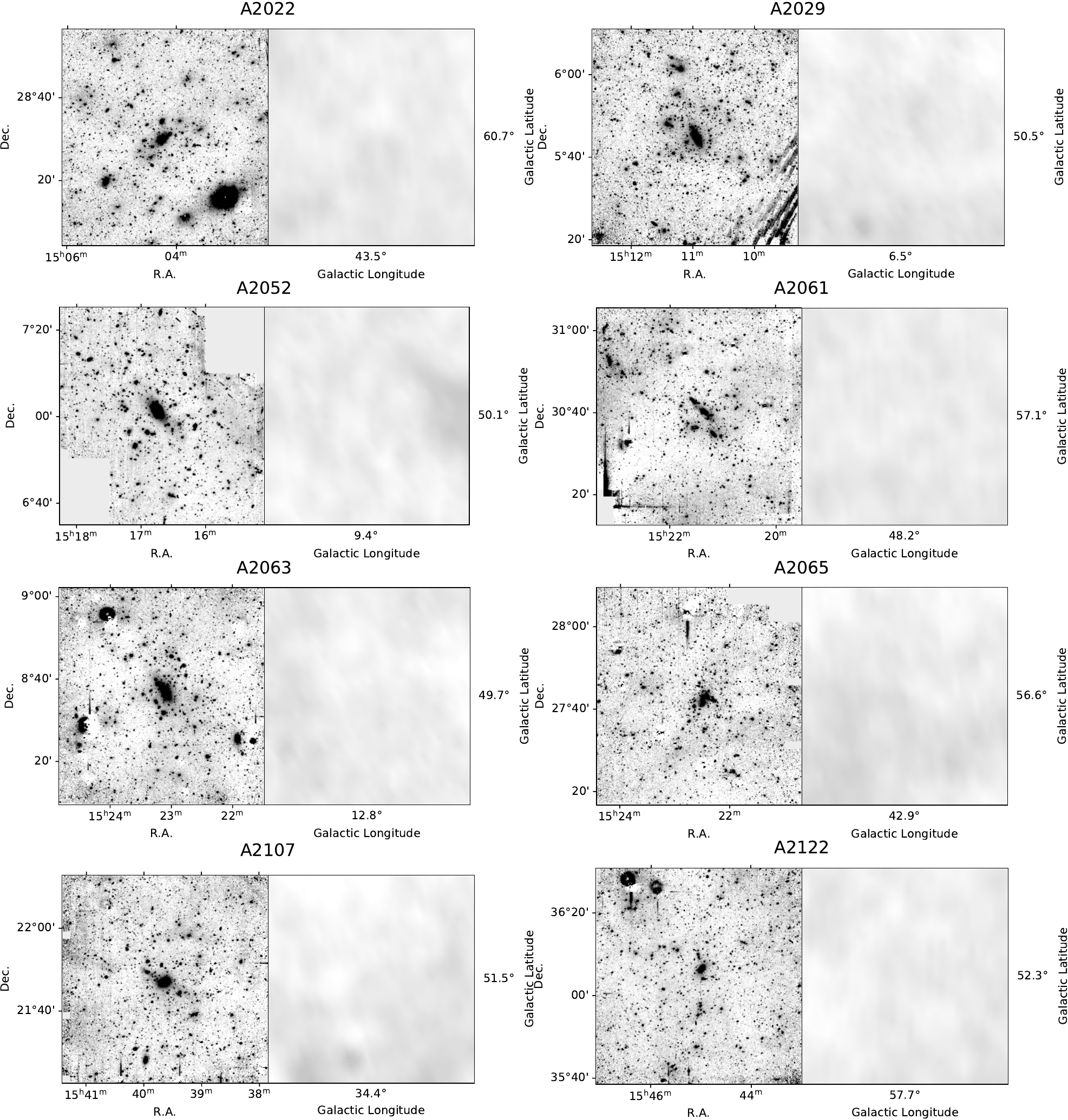} \\~\\~\\
	\includegraphics[width=\linewidth]{fig16bar.pdf}
	\textbf{Figure \ref*{fig:screenshots2}} \textit{(continued)}
\end{figure*}
\begin{figure*}
	\centering
	\includegraphics[width=\linewidth]{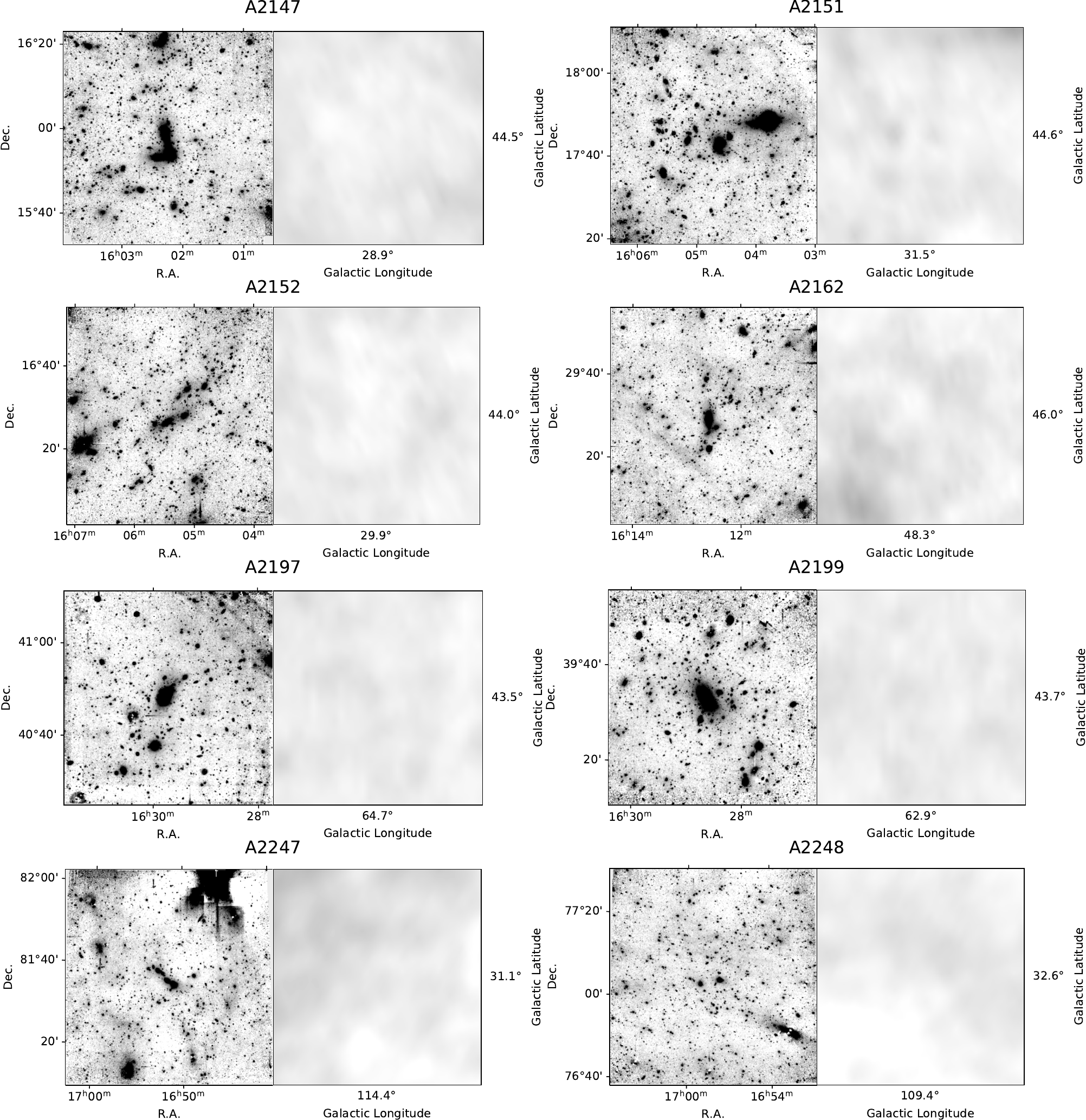} \\~\\~\\
	\includegraphics[width=\linewidth]{fig16bar.pdf}
	\textbf{Figure \ref*{fig:screenshots2}} \textit{(continued)}
\end{figure*}
\begin{figure*}
	\centering
	\includegraphics[width=\linewidth]{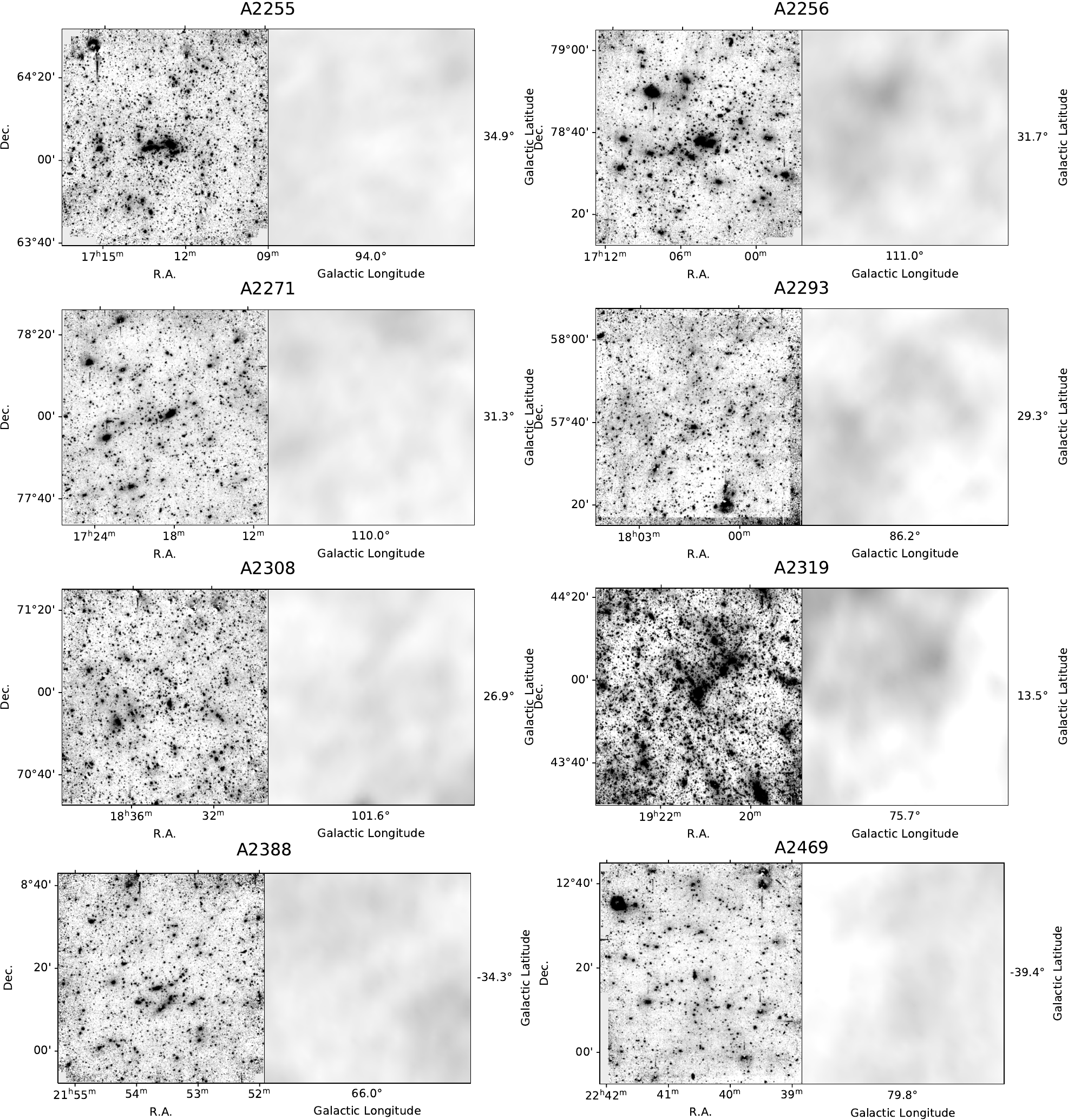} \\~\\~\\
	\includegraphics[width=\linewidth]{fig16bar.pdf}
	\textbf{Figure \ref*{fig:screenshots2}} \textit{(continued)}
\end{figure*}
\begin{figure*}
	\centering
	\includegraphics[width=\linewidth]{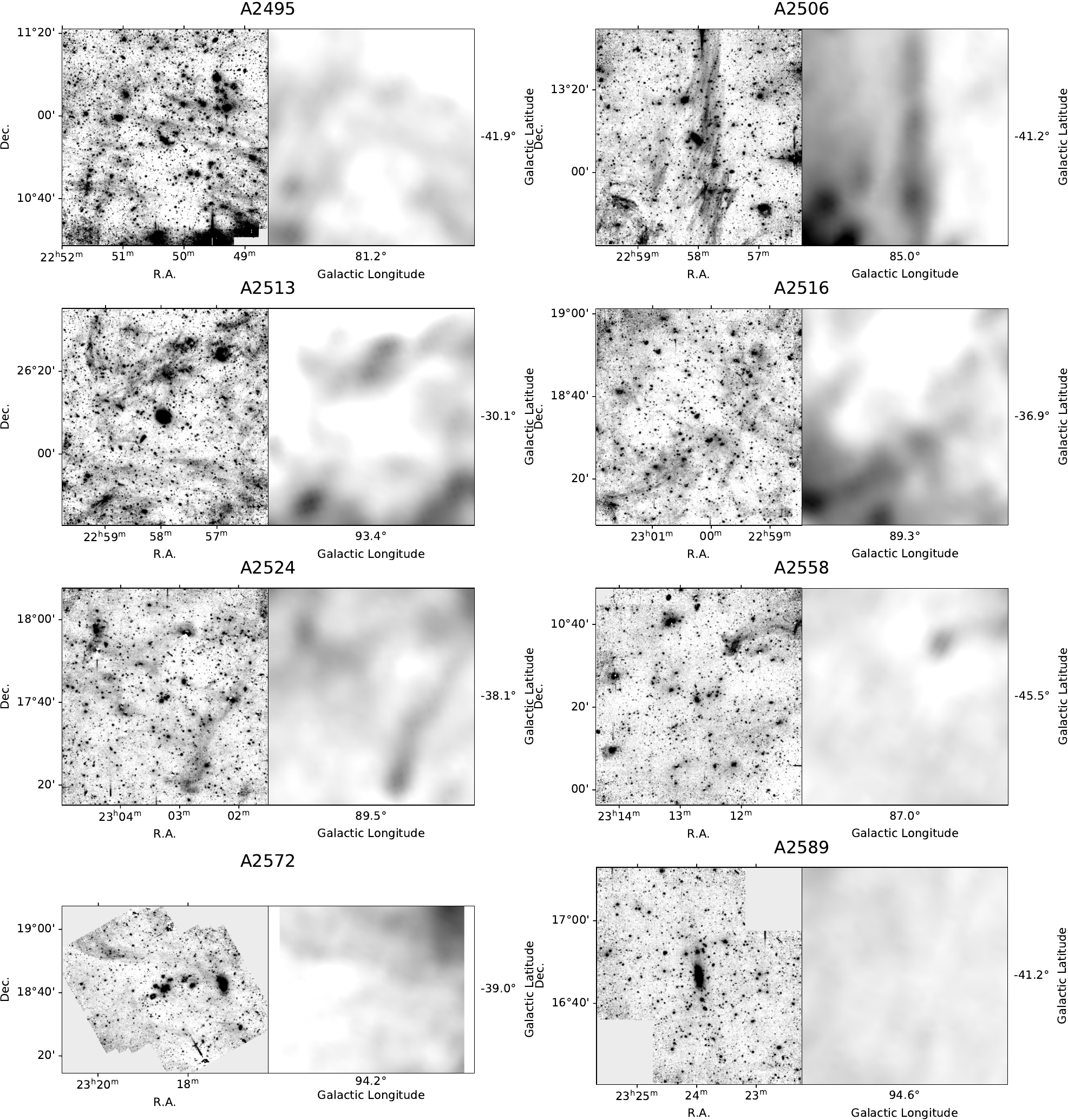} \\~\\~\\
	\includegraphics[width=\linewidth]{fig16bar.pdf}
	\textbf{Figure \ref*{fig:screenshots2}} \textit{(continued)}
\end{figure*}
\begin{figure*}
	\centering
	\includegraphics[width=\linewidth]{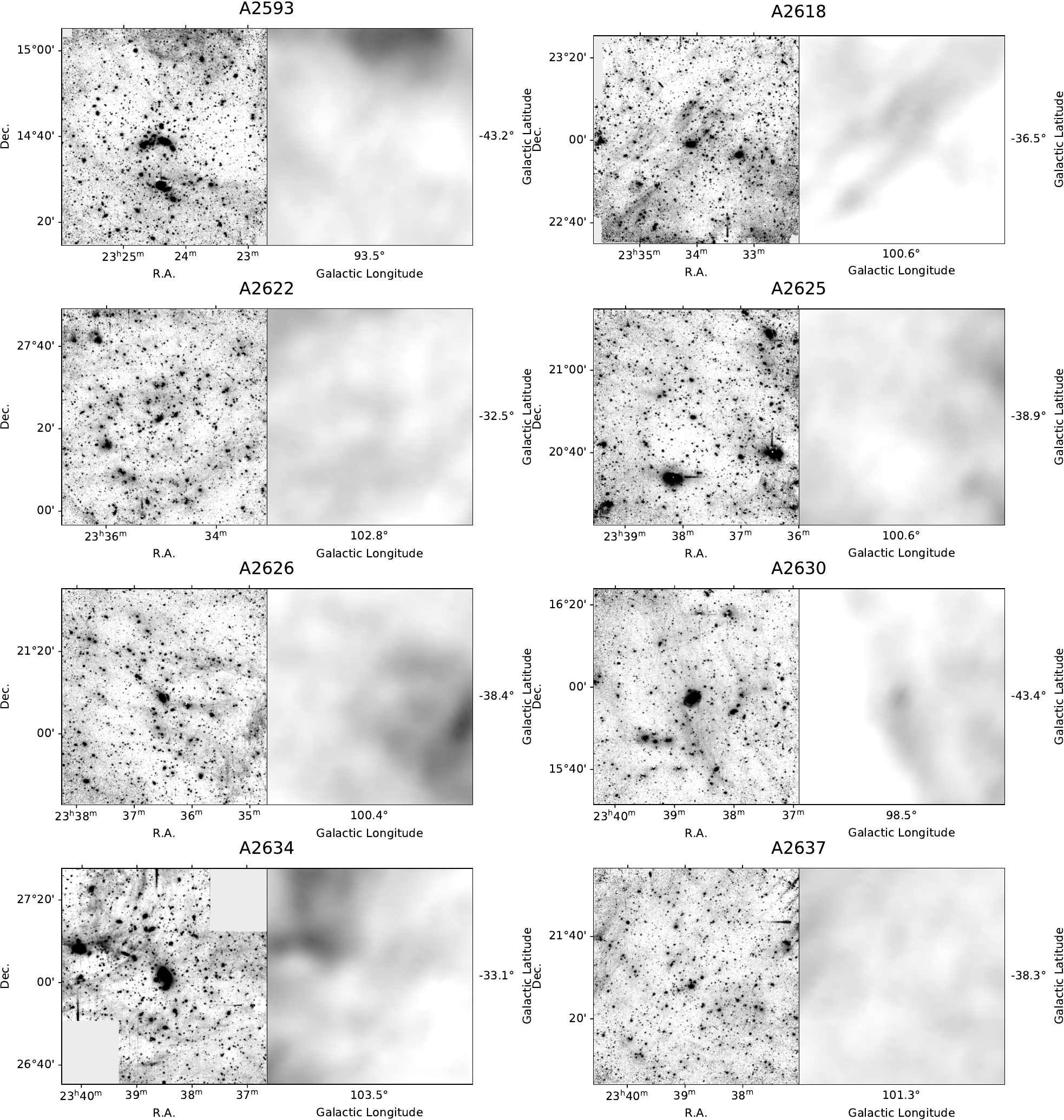} \\~\\~\\
	\includegraphics[width=\linewidth]{fig16bar.pdf}
	\textbf{Figure \ref*{fig:screenshots2}} \textit{(continued)}
\end{figure*}
\begin{figure*}
	\centering
	\includegraphics[width=\linewidth]{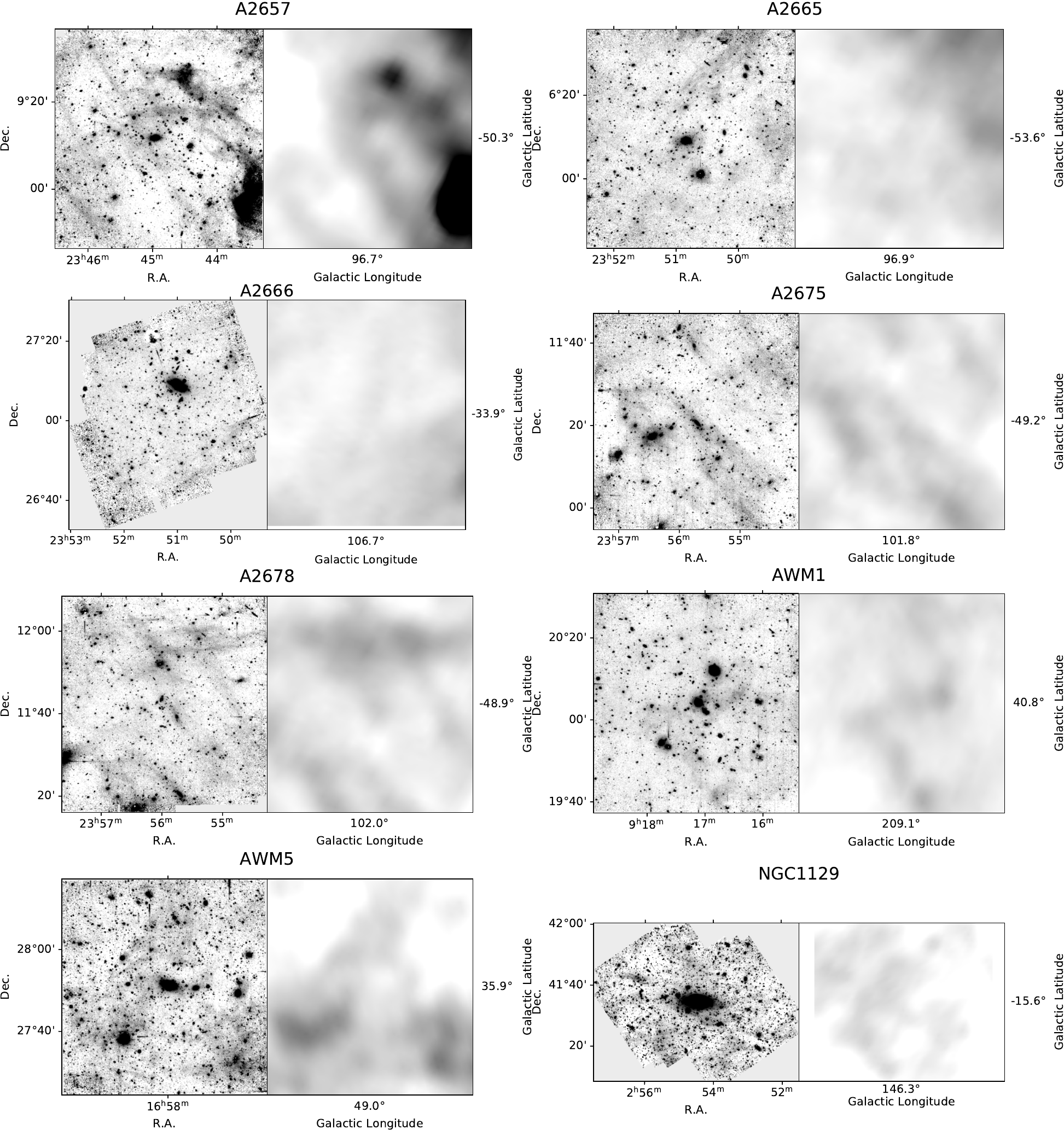} \\~\\~\\
	\includegraphics[width=\linewidth]{fig16bar.pdf}
	\textbf{Figure \ref*{fig:screenshots2}} \textit{(continued)}
\end{figure*}
\begin{figure*}
	\centering
	\includegraphics[width=\linewidth]{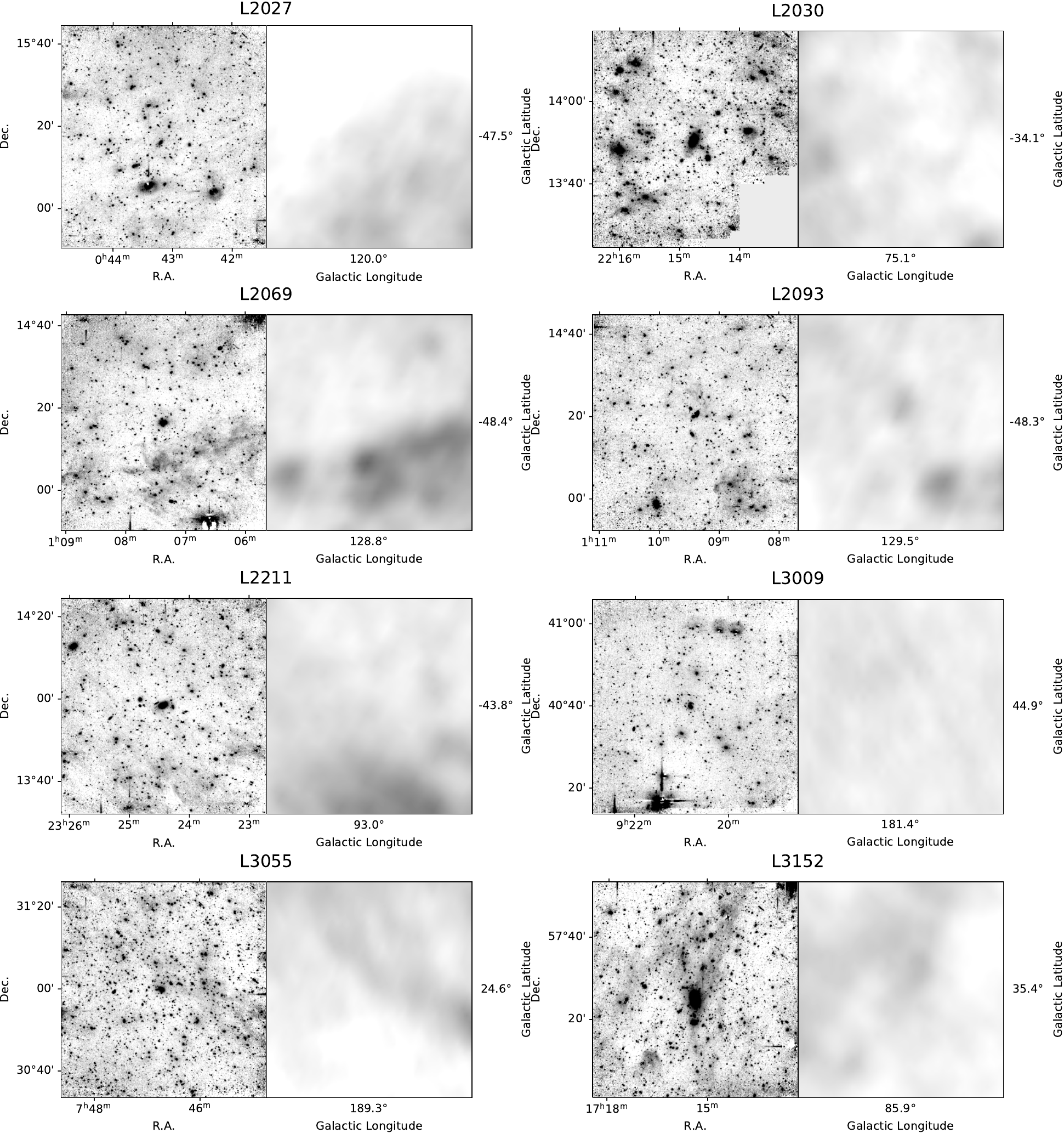} \\~\\~\\
	\includegraphics[width=\linewidth]{fig16bar.pdf}
	\textbf{Figure \ref*{fig:screenshots2}} \textit{(continued)}
\end{figure*}
\begin{figure*}
	\centering
	\includegraphics[width=\linewidth]{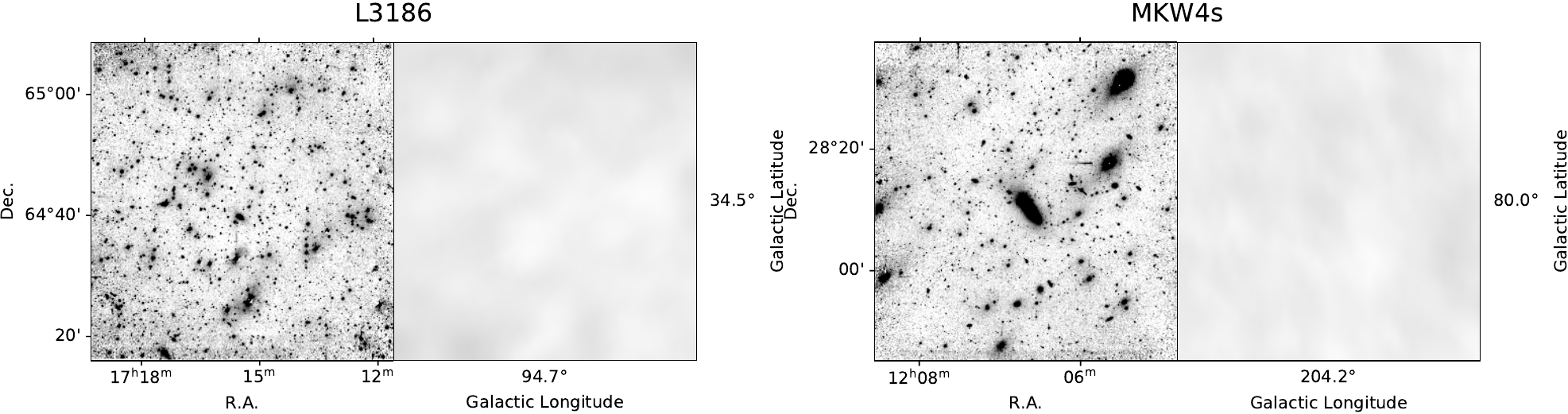} \\~\\~\\
	\includegraphics[width=\linewidth]{fig16bar.pdf}
	\textbf{Figure \ref*{fig:screenshots2}} \textit{(continued)}
\end{figure*}

\clearpage
\bibliography{Paper2}
\bibliographystyle{aasjournal}

\end{document}